\documentclass[aps,pra,twocolumn,footinbib,amsmath,amssymb,amsfonts, superscriptaddress]{revtex4-2} 

\usepackage{color,framed,graphicx}
\usepackage{amsmath, amssymb, amsfonts,mathrsfs, amsthm, dsfont}
\usepackage{physics}
\usepackage{lipsum}
\usepackage[hypertexnames=false]{hyperref}
\usepackage{enumerate}
\usepackage{array}
\usepackage{verbatim}

\usepackage{tikz-cd}
\usepackage{tikz}
\usetikzlibrary{arrows, decorations.pathmorphing, decorations.markings, decorations.pathreplacing, calligraphy}
\usepackage{float}
\usepackage{relsize}
\usepackage{orcidlink}

\usepackage{mathdots}
\usepackage[dvipsnames]{xcolor}

\begin{document}

\title{Disentangling Haldane Phase by Generalized Clifford Circuits}

\author{Minsoo Kim\,\orcidlink{0009-0009-2986-3735}}
\affiliation{Department of Physics, Korea Advanced Institute of Science and Technology, Daejeon 34141, Korea}

\author{Changhun Oh\,\orcidlink{0000-0003-2002-1928}} \email[]{changhun0218@gmail.com}
\affiliation{Department of Physics, Korea Advanced Institute of Science and Technology, Daejeon 34141, Korea}

\author{Donghoon Kim\,\orcidlink{0009-0002-8358-5253}}\email[]{donghoon.kim@riken.jp}
\affiliation{Analytical Quantum Complexity RIKEN Hakubi Research Team, RIKEN Center for Quantum Computing (RQC), Wako, Saitama 351-0198, Japan}

\date{\today}

\begin{abstract}

Disentangling transformations play a central role in the classical simulation of quantum many-body systems, yet their analytic structure and underlying mechanism remain largely unexplored.
Here, we study the structure of the disentangler in the Haldane phase of spin-1 systems using generalized Clifford circuits.
To this end, we extend the Clifford-circuit-augmented matrix product states (CAMPS)-based density-matrix renormalization group (DMRG) method to spin-1 systems.
Within this framework, we find that the local disentanglers optimized for the Haldane phase implement the generalized Kramers--Wannier (KW) transformation, and we analytically verify its optimality for the Affleck--Kennedy--Lieb--Tasaki (AKLT) state.
Beyond reducing entanglement, the KW transformation maps the Haldane phase to a phase with spontaneously broken $\mathbb{Z}_{2}$ symmetry. 
This mapping is distinct from the Kennedy--Tasaki transformation and provides a new unitary route from symmetry-protected topological order to symmetry breaking.

\end{abstract}

\maketitle

\emph{Introduction}---Over the past two decades, quantum information has provided a powerful perspective on the classical simulation of quantum many-body states.
A central lesson is that entanglement plays a key role in classical simulability: states with limited entanglement can be efficiently represented by tensor-network (TN) ansatzes, such as matrix product states (MPS)~\cite{Vidal2003, Vidal2004, Verstraete2006, Hastings2007, Verstraete2008, Schuch2008EntropyScaling, Eisert2010, Schollwock2011, Orus2014}.
Moreover, simulation accuracy can be improved by exploiting the ``disentangler''.
Celebrated methods such as the multiscale entanglement-renormalization ansatz (MERA) and tensor-network renormalization (TNR) employ disentanglers to reduce entanglement in the TN ansatz, thereby lowering computational costs~\cite{Vidal2007EntanglementRenormalization, Vidal2008ClassQuantumManyBodyStates, Evenbly2009AlgorithmsEntanglementRenormalization, Evenbly2009EntanglementRenormalization2D, Pfeifer2009EntanglementRenormalizationScaleInvariance, Vidal2010EntanglementRenormalizationIntroduction, Evenbly2015TensorNetworkRenormalization, Evenbly2015TensorNetworkRenormalizationYieldsMERA, Evenbly2016LocalScaleTransformationsTNR, Evenbly2017AlgorithmsTensorNetworkRenormalization}.

In parallel with entanglement-based TN approaches, Clifford circuits have emerged as another key axis of classical simulability: as exemplified by the Gottesman--Knill theorem, Clifford circuits can generate highly entangled stabilizer states that nevertheless remain efficiently simulable~\cite{Gottesman1999, Raussendorf2001, Hein2004, Aaronson2004, Hostens2005, Farinholt2014}.
These complementary perspectives naturally motivate hybrid Clifford-TN frameworks~\cite{Qian2024, Huang2025, Mello2024, MasotLlima2024, Mello2025, MasotLlima2026, Liu2026}.
A representative example is Clifford circuit-augmented MPS (CAMPS), where local Clifford gates are incorporated into the MPS ansatz as variational disentanglers~\cite{Qian2024, Huang2025, MasotLlima2026, Liu2026, Li2025QutritClifford, Harper2025GCAMPS, Kabir2025}.
This ansatz can be integrated into the density-matrix renormalization group (DMRG)~\cite{White1992, White1993, Schollwock2011, Orus2014}, where the ground-state MPS and the entanglement-minimizing Clifford circuit are optimized together~\cite{Qian2024, Qian2025TDVP, Yosprakob2025GCGMPS}.
Beyond numerical efficiency, CAMPS-based DMRG can provide insight into the underlying physics of phases: the optimized Clifford disentanglers may coincide with duality transformations, thereby encoding physically meaningful information~\cite{Fan2025}.

Despite the numerical success of CAMPS-based DMRG, an analytic understanding of the disentangling mechanism remains limited.
In particular, it remains unclear why the joint MPS--Clifford optimization identifies a nontrivial disentangler, which disentangler is selected as optimal, and how the resulting disentangler realizes a structural transformation associated with the phase.
The Haldane phase of spin-$1$ chains provides an ideal testbed for addressing these questions.
It is a paradigmatic symmetry-protected topological (SPT) phase with hidden nonlocal order~\cite{Haldane1983PRL, Affleck1987, denNijs1989, Kennedy1992CMP, Kennedy1992PRB, Pollmann2010, Chen2011, Pollmann2012}, and it includes the Affleck--Kennedy--Lieb--Tasaki (AKLT) state as an analytically tractable representative, which admits an exact MPS representation with bond dimension 2~\cite{Affleck1987}.

In this work, we study Clifford disentangling of the Haldane phase in spin-$1$ chains.
We extend the CAMPS-based DMRG to qutrit systems using generalized Clifford circuits, and apply it to the spin-1 Heisenberg model and the bilinear-biquadratic chain.
We find that a sequential accumulation of \texttt{SUM} gates serves as an optimal local Clifford disentangler for the Haldane phase in this framework, substantially reducing entanglement and thereby yielding a computational advantage.
For the AKLT state, we analytically demonstrate its optimality and explain its disentangling mechanism.
Moreover, the resulting optimal disentangler coincides with the generalized Kramers--Wannier (KW) transformation, revealing the hidden structure of the Haldane phase.
We show that this transformation maps the Haldane SPT phase to a $\mathbb{Z}_{2}$ spontaneous symmetry-breaking (SSB) phase.
Although this relation is reminiscent of the Kennedy--Tasaki transformation, which connects the Haldane phase to $\mathbb{Z}_{2} \times \mathbb{Z}_{2}$ symmetry breaking~\cite{Kennedy1992PRB, Kennedy1992CMP}, the KW transformation uncovers a different structure: it exposes only a single $\mathbb{Z}_{2}$ component of the hidden symmetry breaking.
This provides a distinct circuit-level route from the SPT order to the SSB, emerging directly from the disentangling circuit.

\emph{Setup and algorithm overview}---For a single qutrit with Hilbert space $\mathcal{H}_{3} = \mathrm{span}\{|0\rangle,|1\rangle,|2\rangle\}$, the generalized Pauli operators are $X|s\rangle = |s+1 \, \mathrm{mod}\, 3\rangle$, $Z|s\rangle = \omega^s |s\rangle$,
where $\omega=e^{2\pi i/3}$. They satisfy $X^{3}=Z^{3}=I$ and $XZ=\omega^{-1}ZX$, and the operators $X^aZ^b$ with $(a,b)\in\mathbb{Z}_3^2$ form a basis of $\mathcal{B}(\mathcal{H}_3)$.
The spin-1 operators $S^x$, $S^y$, and $S^z$ are defined as $S^{z}|s\rangle=(1-s)|s\rangle$, $S^{\pm}|s\rangle=\sqrt{2-(1-s)(1-s \pm 1)}\, |s\mp1\rangle$, with $S^{\pm} = S^{x} \pm iS^{y}$.

For an $N$-qutrit system, the Clifford group is generated by the single-qutrit Pauli operators $X$ and $Z$, the Fourier transform
$H^{(3)} = \frac{1}{\sqrt{3}} \sum_{s,s'=0}^{2} \omega^{ss'} |s\rangle \langle s'|$,
the phase gate
$S^{(3)} = \sum_{s=0}^{2} \omega^{s(s-1)/2} |s\rangle\langle s|$,
and the \texttt{SUM} gate
$U_{j,j+1}^{\texttt{SUM}} |s_{j}, s_{j+1}\rangle = |s_{j}, s_{j}+s_{j+1} \, \pmod 3\rangle$~\cite{Gottesman1999, Hostens2005, Wang2020}.
Under conjugation, these gates map generalized Pauli operators to generalized Pauli operators up to a phase.

\begin{figure}[t]
\centering
\includegraphics[width=\linewidth]{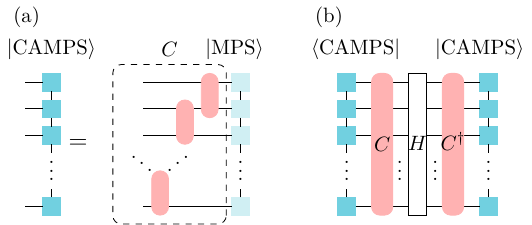}
\caption{
Schematic of the CAMPS-based DMRG method.
(a) The CAMPS is an ansatz $|{\text{CAMPS}}\rangle = C |{\text{MPS}}\rangle$.
Here, $C$ consists of sequential two-site Clifford gates, and each two-site gate is chosen to maximally reduce the entanglement of $|{\text{CAMPS}}\rangle$.
(b) By augmenting the MPS with $C$, the Hamiltonian is transformed accordingly as $\widetilde{H} = C H C^{\dagger}$.
The conjugation preserves the energy expectation between the CAMPS and MPS frames.
}
\label{fig: schematic}
\end{figure}

The MPS provides a standard variational ansatz for one-dimensional systems within the DMRG framework~\cite{Ostlund1995, Schollwock2011, Orus2014, Cirac2021}. For qutrit systems, we write
\begin{equation}
\label{MPS form}
    |\mathrm{MPS}\rangle
    = \sum_{\{s_{j}\}}
    A_{1}^{s_1} A_{2}^{s_{2}} \cdots A_{N}^{s_{N}}
    |s_{1} s_{2} \cdots s_N\rangle ,
\end{equation}
where $s_{j} \in \{0,1,2\}$ and $A_{j}^{s_{j}}$ is a $\chi_{j-1} \times \chi_{j}$ matrix, with bond dimension $\chi = \max_{j}\chi_{j}$.
The CAMPS extends standard DMRG by introducing a variational class $|\mathrm{CAMPS}\rangle = C|\mathrm{MPS}\rangle$, i.e., the CAMPS ansatz, where $C$ is a Clifford circuit built from two-site gates acting as local disentanglers [Fig.~\ref{fig: schematic}(a)]~\cite{Qian2024}.
During the DMRG sweep, one applies a two-site Clifford gate $C_{j,j+1}$ to each bond and selects the one that most effectively reduces the MPS's entanglement across that bond.
The selected two-site gates are then accumulated, in the order in which they are chosen during the sweep, into an optimized Clifford circuit $C_{\mathrm{opt}} = \prod C_{j,j+1}$.
Since the disentangler acts on the state, the Hamiltonian is updated simultaneously [Fig.~\ref{fig: schematic}(b)], and after the sweep, the transformed Hamiltonian is $\widetilde{H} = C_{\mathrm{opt}} H C_{\mathrm{opt}}^{\dagger}$.
For Hamiltonians expressed as sums of generalized Pauli strings, Clifford conjugation preserves the number of Pauli terms, since the Clifford group normalizes the Pauli group.
Moreover, although the full two-qutrit Clifford group is much larger, only $90$ gates act inequivalently on the entanglement and therefore need to be considered in the disentangling step~\cite{Harper2025GCAMPS} (see also the Supplemental Material (SM)~\cite{supple}).

\begin{figure}[t]
    \centering
    \includegraphics[scale=1]{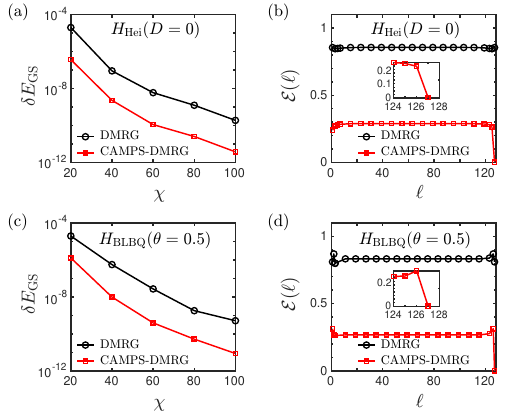}
    \caption{
    The CAMPS-based DMRG results for the Heisenberg and the BLBQ models.
    (a), (c) Relative ground-state energy error $\delta E_{\text{GS}}$ for $N=128$ at (a) $H_{\mathrm{Hei}}(D=0)$ and (c) $H_{\mathrm{BLBQ}}(\theta = 0.5)$.
    The DMRG result with $\chi=1000$ serves as the reference.
    Panels (b) and (d) display the corresponding entanglement entropy $\mathcal{E}(\ell)$ for bipartitions of sizes $\ell$ and $N-\ell$.
    For both models, we observe that $\mathcal{E}(N-1) = 0$ in the CAMPS-based DMRG results; see the insets of (b) and (d).
    }
    \label{fig: CAMPSadv}
\end{figure}

\emph{CAMPS-based DMRG for the Haldane phase}---We apply the qutrit CAMPS-based DMRG method to the Haldane phase of the spin-$1$ Heisenberg model and the bilinear--biquadratic (BLBQ) chain.
The spin-$1$ Heisenberg model Hamiltonian is
\begin{equation}
    H_{\mathrm{Hei}} (D)
    = \sum_{j=1}^{N-1} \mathbf{S}_j\cdot\mathbf{S}_{j+1}
    + D\sum_{j=1}^{N}(S_j^z)^2 ,
\end{equation}
and the BLBQ model is given by
\begin{equation}
    H_{\mathrm{BLBQ}} (\theta)
    =
    \sum_{j=1}^{N-1}
    \left[
        \cos\theta\,(\mathbf{S}_j\cdot\mathbf{S}_{j+1})
        + \sin\theta\,(\mathbf{S}_j\cdot\mathbf{S}_{j+1})^2
    \right],
\end{equation}
where $\mathbf{S}_{j} = (S_{j}^{x}, S_{j}^{y}, S_{j}^{z})$.
For the Heisenberg model, this method improves the accuracy of the ground-state energy over standard DMRG at a fixed bond dimension [Fig.~\ref{fig: CAMPSadv}(a)].
Moreover, this method reduces the entanglement entropy $\mathcal{E}(\ell)$ of the ground-state MPS [Fig.~\ref{fig: CAMPSadv}(b)], with $\mathcal{E}$ vanishing at the rightmost bond (see SM~\cite{supple}).
The same trend persists in the BLBQ model at $\theta=0.5$~\cite{Lauchli2006} [Fig.~\ref{fig: CAMPSadv}(c,d)].

Although our main focus is the Haldane phase, the computational advantage extends to other settings, including the three-state clock model, the trivial phase of $H_{\mathrm{Hei}}$, and the dimerized regimes of $H_{\mathrm{BLBQ}}$.
The supporting numerical results are provided in SM~\cite{supple}.

\emph{Optimal Clifford disentangler for the Haldane phase}---We investigate the underlying mechanism of the computational advantage.
We prove that when local Clifford disentanglers are applied sequentially to the AKLT state, the optimal choice is $U_{\mathrm{KW}}$, defined by
\begin{equation}
\label{KW unitary}
    U_{\mathrm{KW}}
    = \mathcal{U}_{N-1,N} \cdots \mathcal{U}_{2,3} \mathcal{U}_{1,2}, ~~
    \mathcal{U}_{j,j+1} := X_{j+1}^{2} U_{j,j+1}^{\texttt{SUM}}.
\end{equation}
It implies that $U_{\mathrm{KW}}$ is optimal throughout the Haldane phase, as supported by our CAMPS-based DMRG results for $H_{\mathrm{Hei/BLBQ}}$.
Since $U_{\mathrm{KW}}$ matches the conventional generalized Kramers--Wannier (KW) duality transformation up to local unitaries~\cite{Sinha2024, Zhang2025}, we adopt this terminology.
Although the KW duality was originally introduced for the three-state clock model~\cite{Sinha2024, Zhang2025}, our results suggest that its role extends beyond that setting: it also acts as a Clifford disentangler for the Haldane phase.

For the analytical investigation, we focus on the AKLT state $| \mathrm{AKLT}_{ \mathbf{L}, \mathbf{R} } \rangle$, the exact ground state of $H_{\mathrm{BLBQ}}(\tan^{-1}(1/3))$.
In the valence-bond-solid (VBS) construction, each spin-1 degree of freedom is decomposed into two spin-$1/2$'s, with neighboring spin-1/2's forming singlets.
For open boundary conditions (OBC), unpaired spin-$1/2$ degrees of freedom remain at the left and right edges; we denote their states by the 2-dimensional vectors $\mathbf{L}$ and $\mathbf{R}$.
Using the form in Eq.~\eqref{MPS form}, the bond-dimension-$2$ MPS tensors for $| \mathrm{AKLT}_{ \mathbf{L}, \mathbf{R} } \rangle$ are given in the End Matter: the bulk tensors $A_{j}$ are uniform for $2 \leq j \leq N-1$, while the boundary tensors are $\mathbf{L}^{\dagger} A_{1}$ and $A_{N} \mathbf{R}$ at $j=1$ and $j=N$, respectively.

We then show the optimality of $U_{\mathrm{KW}}$ for the AKLT state as follows.
We begin by defining the \emph{$(\mathbf{u}, \mathbf{v}; a_{j})$-canonical form}:
we say that a tensor $B_{j}$ is in this form if its physical components are given by
\begin{equation}
\label{bn form}
    B_{j}^{0}
    = \begin{pmatrix} u_{1} \\ u_{2}
    \end{pmatrix} \begin{pmatrix} 0 & 1 \end{pmatrix}, \quad
    B_{j}^{1}
    = \begin{pmatrix} v_{1} \\ v_{2} \end{pmatrix}
    \begin{pmatrix} 1 & 0 \end{pmatrix}, \quad
    B_{j}^{2} = 0,
\end{equation}
where $\mathbf{u} = \begin{pmatrix} u_{1} & u_{2} \end{pmatrix}^{T}$ and $\mathbf{v} = \begin{pmatrix} v_{1} & v_{2} \end{pmatrix}^{T}$ satisfy
\begin{equation}
\label{condition of uv for B}
    \| \mathbf{u} \|^{2} = a_{j}, \quad
    \| \mathbf{v} \|^{2} = 1-a_{j}, \quad
    \mathbf{u}^{\dagger} \mathbf{v} = - \sqrt{2}/3.
\end{equation}
Using this form, we apply a Clifford gate to the bond between a canonical-form tensor $B_{j}$ and the AKLT tensor $A_{j+1}$, and compute the resulting entanglement via singular value decomposition (SVD).
We prove that applying $\mathcal{U}_{j,j+1}$ followed by SVD maps the tensor at site $(j+1)$ back into the canonical form, now with parameters $(\mathbf{u}', \mathbf{v}'; a_{j+1})$ satisfying 
\begin{equation}
    a_{j+1} = \frac{2-a_{j}}{3}. \label{eq:recurrence}
\end{equation}
The iteration starts from the left boundary, since contracting $\mathbf{L}^{\dagger} A_{1}$ with a constant vector yields a canonical-form tensor with $a_{1} = 2/3$.
Moreover, we show that $\mathcal{U}_{j,j+1}$ minimizes the entanglement among the 90 Clifford-gate classes whenever $a_{j} \in [4/9, 2/3]$.
Since the recurrence relation in Eq.~\eqref{eq:recurrence} keeps $a_{j}$ within $[4/9, 2/3]$, the same argument can be iterated from left to right.
Thus, the sweep yields $U_{\mathrm{KW}}$ in Eq.~\eqref{KW unitary} as an optimal disentangler; the KW-transformed AKLT state is given in the End Matter.
Finally, we establish that no two-qutrit Clifford gate can further reduce the entanglement after this sweep.
The details are provided in SM~\cite{supple}.

The above analytic result explains why numerical CAMPS-based DMRG identifies $U_{\mathrm{KW}}$ as an optimal disentangler.
For $H_{\mathrm{BLBQ}}(\tan^{-1}(1/3))$, we consider a right-to-left sweep initialized with a random MPS, whose tensors are far from the AKLT tensor; hence, the analytic result does not yet apply, and Clifford gates may fail to reduce the entanglement.
After this sweep, the algorithm produces tensors close to the AKLT tensor, so the above analytic result applies to the subsequent left-to-right sweep.
This picture is further supported by the robustness of $\mathcal{U}_{j,j+1}$ against perturbations:
the entanglement gap to the second-best gate is $H(1/2 + 2\sqrt{2}/9) - H(1/2 + \sqrt{2}/3) \approx 0.35$, where $H(x)$ denotes the binary Shannon entropy~\cite{supple}.
Thus, $\mathcal{U}_{j,j+1}$ remains identifiable as the optimal disentangler even under small deviations from the exact AKLT tensors.
Moreover, we observe that the disentangling relies crucially on boundary conditions: no Clifford gate reduces the entanglement between two bulk AKLT tensors $A_{j}$ and $A_{j+1}$~\cite{supple}, so the optimization must start from the boundary tensor $\mathbf{L}^{\dagger} A_{1}$.

\bigskip

\emph{Hidden SSB revealed by Clifford disentangling}---For the Haldane phase, we have demonstrated above that $U_{\mathrm{KW}}$ is an optimal Clifford disentangler.
Beyond its optimality as a disentangler, we now show that it reveals a ``hidden'' $\mathbb{Z}_{2}$ symmetry-breaking structure.

We first analyze the symmetries of the KW-transformed Hamiltonians $\widetilde{H}_{\mathrm{Hei/BLBQ}}$.
Using Eq.~\eqref{KW unitary}, we obtain the KW-transformed local terms
as $U_{\mathrm{KW}} (\mathbf{S}_{j} \cdot \mathbf{S}_{j+1}) U_{\mathrm{KW}}^{\dagger} = \widetilde{T}_{j-1, j, j+1} + (\widetilde{T}_{j-1, j, j+1})^{\dagger}$ and $U_{\mathrm{KW}} (S_{j}^{z})^{2} U_{\mathrm{KW}}^{\dagger}= \widetilde{\Xi}_{j-1, j} + (\widetilde{\Xi}_{j-1, j})^{\dagger}$, where 
\begin{equation}
\begin{gathered}
   \label{SdotS transform}
	\widetilde{T}_{j-1, j, j+1}
	= \sum_{m,n=0}^{2} (\omega^{2} Z_{j-1})^{m} \mathcal{M}_{j}(m,n) (\omega Z_{j+1})^{n}, \\
    \widetilde{\Xi}_{j-1, j}
    = \frac{1}{3} I_{j-1} I_{j} - \frac{1}{3} Z_{j-1}^{2} Z_{j}.
\end{gathered}
\end{equation}
Explicit expressions for $\mathcal{M}_{j}(m,n)$ are provided in the End Matter.
For $(\mathbf{S}_{j} \cdot \mathbf{S}_{j+1})^{2}$, the corresponding term is obtained by replacing $\mathcal{M}_{j}(m,n)$ with $\mathcal{N}_{j}(m,n)$, whose explicit form is also given there.
We note that $\widetilde{H}_{\mathrm{Hei/BLBQ}}$ remain local after the KW transformation; in SM~\cite{supple}, we prove that for a 2-local operator $O_{j,j+1}$, the transformed operator $U_{\mathrm{KW}} O_{j,j+1} U_{\mathrm{KW}}^{\dagger}$ remains local if and only if $[O_{j,j+1}, Z_{j} Z_{j+1}] = 0$.

From Eq.~\eqref{SdotS transform}, we fully classify the symmetries of $\widetilde{H}_{\mathrm{Hei/BLBQ}}$ and prove that both Hamiltonians have a $\mathbb{Z}_{2}$ symmetry.
Let $G_{\mathrm{prod}}(H)$ denote the group of on-site product unitaries commuting with $H$.
Then, we show that every $S \in G_{\mathrm{prod}}(\widetilde{H}_{\mathrm{Hei/BLBQ}})$ takes the form~\cite{supple}
\begin{equation}
\label{symm of KW transformed Hamiltonians}
    S = E_{\mathrm{R}} \mathcal{X}^{\varepsilon},
    \qquad \mathcal{X} = \prod_{j=1}^{N} e^{i \pi S_{j}^{x}},
\end{equation}
where $\varepsilon \in \{0,1\}$, and $E_{\mathrm{R}}$ acts only on the $N$-th qutrit and $[E_{\mathrm{R}}, Z_{N}] = 0$.
The two components, $\mathcal{X} \in \mathbb{Z}_{2}$ and boundary-local symmetry $E_{\mathrm{R}}$, arise as follows.
(1) $\mathcal{X}$: since $[H_{\mathrm{Hei/ BLBQ}}, \mathcal{X}] = 0$, we verify that $[ \mathcal{X}, U_{\mathrm{KW}} ] = 0$.
(2) $E_{\mathrm{R}}$: in Eq.~\eqref{SdotS transform}, the $N$-th site appears in $\widetilde{H}_{\mathrm{Hei/ BLBQ}}$ only through $Z_{N}$ and $Z_{N}^{\dagger}$.
The uniqueness of the form in Eq.~\eqref{symm of KW transformed Hamiltonians} is shown in SM~\cite{supple}.

\begin{figure}[t]
    \centering
    \includegraphics[scale=1]{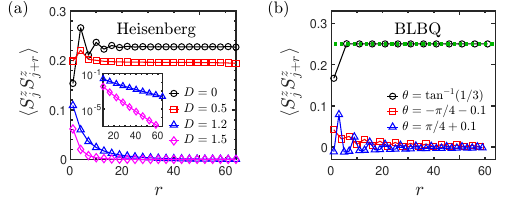}
    \caption{
    Numerical results of the two-point correlation $\langle S_{j}^{z}S_{j+r}^{z} \rangle$ for (a) the Heisenberg model and (b) the BLBQ model, for $j=32$.
    The ground states are obtained using DMRG with $N=128$ and $\chi=300$, and after applying $U_{\mathrm{KW}}$, the corresponding values of $\langle S_{j}^{z}S_{j+r}^{z} \rangle$ are evaluated.
    The green dotted line in (b) represents the analytic value $1/4$ given in Eq.~\eqref{AKLT LRO}.
    Inset (a): In the trivial phase, the correlation decays exponentially.
    }
    \label{fig: corr}
\end{figure}

The bulk $\mathbb{Z}_{2}$ symmetry generated by $\mathcal{X}$ is spontaneously broken in the ground state of $\widetilde{H}_{\mathrm{Hei/BLBQ}}$, whereas no corresponding symmetry breaking occurs in the original models $H_{\mathrm{Hei/BLBQ}}$.
The original models realize the Haldane phase near $D = 0$ for $H_{\mathrm{Hei}}$ and for $-\pi/4 < \theta < \pi/4$ in $H_{\mathrm{BLBQ}}$; this phase is an SPT phase characterized by non-local string order~\cite{Haldane1983PRL, Affleck1987, denNijs1989, Kennedy1992CMP, Kennedy1992PRB, Pollmann2010, Chen2011, Pollmann2012, Lauchli2006}.
In contrast, after the KW transformation, we find that the ground states exhibit long-range order (LRO) in the \emph{local} operator $S_{j}^{z}$.
Specifically, we examine $\langle S_{j}^{z} S_{j+r}^{z} \rangle$ for both models in Fig.~\ref{fig: corr}(a,b).
It remains finite as $r \to \infty$ in the Haldane regime ($D=0,0.5$), including the AKLT point ($\theta = \tan^{-1}(1/3)$), but decays in the trivial ($D=1.2,1.5$), dimerized ($\theta = -\frac{\pi}{4}-0.1$), and gapless ($\theta = \frac{\pi}{4}+0.1$) regimes~\cite{Lauchli2006}.
Since $\mathcal{X} S_{j}^{z} \mathcal{X}^{\dagger} = - S_{j}^{z}$, any finite-size ground state symmetric under $\mathcal{X}$ must satisfy $\langle S_{j}^{z} \rangle = 0$.
Nevertheless, a nonzero long-distance limit of $\langle S_{j}^{z} S_{j+r}^{z} \rangle$ diagnoses spontaneous breaking of the $\mathbb{Z}_{2}$ symmetry in the thermodynamic limit.

To analytically confirm this $\mathbb{Z}_{2}$ SSB, we consider the KW-transformed AKLT state $| \mathrm{AKLT}_{ \mathbf{L}, \mathbf{R} }^{\mathrm{(KW)}} \rangle := U_{\mathrm{KW}} | \mathrm{AKLT}_{ \mathbf{L}, \mathbf{R} } \rangle$.
In particular, we derive that
\begin{equation}
    \label{AKLT LRO}
    \lim_{r \to \infty}
    \lim_{ N \to \infty }
    \langle \mathrm{AKLT}_{ \mathbf{L}, \mathbf{R} }^{\mathrm{(KW)}} |
    S_{j}^{z}S_{j+r}^{z}
    | \mathrm{AKLT}_{ \mathbf{L}, \mathbf{R} }^{\mathrm{(KW)}} \rangle
    = \frac{1}{4},
\end{equation}
consistent with Fig.~\ref{fig: corr}(b).
Furthermore, we prove that
\begin{equation}
\label{AKLT local Sz expectation}
    \lim_{N \to \infty}
    \langle \mathrm{AKLT}_{ \mathbf{L}, \mathbf{R} }^{\mathrm{(KW)}} |
    S_{j}^{z}
    | \mathrm{AKLT}_{ \mathbf{L}, \mathbf{R} }^{\mathrm{(KW)}} \rangle
    =
    \begin{cases}
        \frac{1}{2} & \mathbf{L} = \mathbf{e}_{\uparrow} \\
        -\frac{1}{2} & \mathbf{L} = \mathbf{e}_{\downarrow} \\
    \end{cases},
\end{equation}
where $\mathbf{e}_{\uparrow}$ and $\mathbf{e}_{\downarrow}$ denote the $+1$ and $-1$ eigenvectors of the qubit Pauli-$Z$, $\sigma^{z}$.
Equation~\eqref{AKLT local Sz expectation} clearly shows that $\langle S_{j}^{z} \rangle$ serves as an order parameter characterizing the $\mathbb{Z}_{2}$ SSB phase.
The detailed derivations are given in SM~\cite{supple}.

\emph{Characterization of the KW-transformed Haldane phase}---Under OBC, $H_{\mathrm{Hei/BLBQ}}$ is known to exhibit an approximate fourfold ground-state degeneracy in the Haldane phase~\cite{Kennedy1990}.
Here, we classify the KW-transformed counterparts of these four states:
they are fully labeled by (1) two bulk $\mathbb{Z}_{2}$ symmetry-broken sectors associated with $\mathcal{X}$, and (2) two right-boundary degrees of freedom.
Since the boundary label is distinguished by a local operator and does not define an independent bulk symmetry-broken sector, the KW transformation reveals only the $\mathbb{Z}_{2}$ SSB associated with $\mathcal{X}$.

Since the 4 states in the Haldane phase are represented by the 4 types of AKLT states distinguished by the two edge spins, we employ the VBS framework of the AKLT state.
We then consider the right-boundary $\mathbb{Z}_{2}$ symmetry $\mathcal{Z}_{\mathrm{R}} = e^{i \pi S_{N}^{z}}$, as allowed by Eq.~\eqref{symm of KW transformed Hamiltonians}, and show that
\begin{equation}
\label{AKLT X Z edge transformation}
\begin{gathered}
    \mathcal{X}
    | \mathrm{AKLT}_{ \mathbf{L}, \mathbf{R} }^{\mathrm{(KW)}} \rangle
    = | \mathrm{AKLT}_{ \sigma^{x} \mathbf{L}, \sigma^{x} \mathbf{R} }^{\mathrm{(KW)}} \rangle, \\
    \mathcal{Z}_{\mathrm{R}}
    | \mathrm{AKLT}_{ \mathbf{L}, \mathbf{R} }^{\mathrm{(KW)}} \rangle
    = | \mathrm{AKLT}_{ \sigma^{z} \mathbf{L}, \sigma^{z} \mathbf{R} }^{\mathrm{(KW)}} \rangle.
\end{gathered}
\end{equation}
The derivation is provided in SM~\cite{supple}.
Here, $\mathbf{L} \in \{ \mathbf{e}_{\uparrow}, \mathbf{e}_{\downarrow} \}$ and $\mathbf{R} \in \{ \mathbf{e}_{+}, \mathbf{e}_{-} \}$, where $\mathbf{e}_{+}$ and $\mathbf{e}_{-}$ are the $+1$ and $-1$ eigenstates of the qubit Pauli-$X$, $\sigma^{x}$.
Equation~\eqref{AKLT X Z edge transformation} shows that both $\mathcal{X}$ and $\mathcal{Z}_{\mathrm{R}}$ act on the edge spins, respectively as $\sigma^{x}$ and $\sigma^{z}$.

We emphasize that $\mathcal{X}$ is \emph{global}, whereas $\mathcal{Z}_{\mathrm{R}}$ is \emph{local}, leading to distinct symmetry-breaking structures.
The global symmetry $\mathcal{X}$ exchanges $| \mathrm{AKLT}_{ \mathbf{e}_{\uparrow}, \mathbf{R} }^{\mathrm{(KW)}} \rangle$ and $| \mathrm{AKLT}_{ \mathbf{e}_{\downarrow}, \mathbf{R} }^{\mathrm{(KW)}} \rangle$; thus, in the thermodynamic limit, the two states cannot be connected by a ``finite'' operator.
Consequently, they belong to distinct broken $\mathcal{X}$-symmetry sectors, distinguished by the order parameter $\langle S_{j}^{z} \rangle$ as indicated in Eq.~\eqref{AKLT local Sz expectation}.
Meanwhile, the two states $| \mathrm{AKLT}_{ \mathbf{L}, \mathbf{e}_{+} }^{\mathrm{(KW)}} \rangle$ and $| \mathrm{AKLT}_{ \mathbf{L}, \mathbf{e}_{-} }^{\mathrm{(KW)}} \rangle$ can be connected in finite systems by the boundary-local symmetry $\mathcal{Z}_{\mathrm{R}}$.
Since $\mathcal{Z}_{\mathrm{R}}$ is supported only at the right boundary, every fixed-support bulk local observable commutes with it for sufficiently large $N$ and therefore has the same expectation value in these two states.
Thus, in terms of the bulk quasi-local algebra, the two states define the same bulk state.
In conclusion, the KW transformation reveals $\mathbb{Z}_{2}$ SSB associated with $\mathcal{X}$, but not with $\mathcal{Z}_{\mathrm{R}}$.
Furthermore, Eq.~\eqref{symm of KW transformed Hamiltonians} allows only the bulk symmetry $\mathcal{X}$ and right-boundary symmetries $E_{\mathrm{R}}$; since the latter cannot induce distinct broken-symmetry sectors for the same reason as $\mathcal{Z}_{\mathrm R}$, the $\mathcal{X}$-associated SSB is the only bulk SSB.

Now we compare the actions of the KW and Kennedy–Tasaki (KT) transformations on the Haldane phase.
The KT-transformed Heisenberg model has a $\mathbb{Z}_{2} \times \mathbb{Z}_{2}$ symmetry generated by the set $\{\mathcal{X}, \mathcal{Z}\}$, where $\mathcal{Z} = \prod_{j=1}^{N} e^{i \pi S_{j}^{z}}$.
Unlike the KW transformation, the KT transformation uncovers the ``full'' $\mathbb{Z}_{2} \times \mathbb{Z}_{2}$ SSB~\cite{Kennedy1992PRB, Kennedy1992CMP}.
In detail, after either transformation, $\mathcal{X}$ acts as Pauli-$X$ on the edge spins (see Eq.~\eqref{AKLT X Z edge transformation} and SM~\cite{supple} for the KT case); thus, both transformations reveal the hidden $\mathbb{Z}_{2}$ SSB associated with the ``global'' symmetry $\mathcal{X}$, as established above.
The key contrast is that the KT transformation uncovers an additional hidden $\mathbb{Z}_{2}$ SSB associated with $\mathcal{Z}$, whereas the KW transformation does not.
This contrast follows from which operator realizes the Pauli-$Z$ action on the edges: the global symmetry $\mathcal{Z}$ for the KT case (see SM~\cite{supple}), whereas the right-boundary symmetry $\mathcal{Z}_{\mathrm R}$ for the KW case (see Eq.~\eqref{AKLT X Z edge transformation}).
Analogously to $\mathcal{X}$, the global operator $\mathcal{Z}$ gives rise to a symmetry breaking in the KT case, while the local operator $\mathcal{Z}_{\mathrm R}$ does not play such a role in the KW case.

\emph{Discussion}---In this study, we establish a Clifford-disentangling perspective on the Haldane phase in spin-$1$ systems.
We extend the CAMPS-based DMRG method to qutrit systems and observe a consistent computational advantage.
We then analytically trace the origin of this advantage, which had previously been characterized mainly numerically.
In particular, within this local Clifford-disentangling ansatz, we prove that the KW transformation is an optimal disentangler for the AKLT state.
We further show that, when applied to the Haldane phase, this transformation uncovers hidden $\mathbb{Z}_{2}$ SSB, but in a qualitatively different manner from the KT transformation.
Our finding suggests that an optimal Clifford disentangler, identified from a computational complexity perspective, can reveal nontrivial underlying physics.

Although we focused on the Haldane phase, identifying optimal Clifford gates in other settings is a promising future direction.
In particular, in the subtle regime near $\theta = -3\pi/4$ in the BLBQ model~\cite{Lauchli2006}, the Clifford disentangler may offer new insights.
Furthermore, extending CAMPS-based DMRG to general qudit systems is an interesting direction; the ququart with local dimension 4 is especially appealing, since it involves only 320 relevant local Clifford gates and naturally connects to spinful fermion systems.

Our results show that the KW transformation induces the hidden $\mathbb{Z}_{2}$ SSB while serving as a Clifford disentangler, improving the efficiency of DMRG.
It would be worthwhile to clarify the physical interpretation of its relation to the KT transformation, which also reveals the hidden SSB and is a perfect disentangler for the AKLT state~\cite{Li2023Noninvertible, okunishi2011topological}.
Although the relation between these transformations is understood for qubit cluster states~\cite{Mana2024}, its spin-1 counterpart remains unclear.
In particular, since the KT transformation is non-Clifford~\cite{supple}, it would be interesting to identify the physical role of the non-Clifford ingredient that connects the KW and KT transformations.
More generally, recent work has analyzed duality transformations that induce symmetry breaking and reduce the computational cost of DMRG~\cite{Lootens2025EntanglementDMRG}.
It would be interesting to understand the relation between such dualities and the KW transformation obtained in our Clifford-restricted optimization.

\bigskip

\acknowledgments

We are grateful to Atis Yosprakob for useful discussions and collaborations on related projects.
We also thank Seung-Sup B. Lee for useful discussions.
M.K. and C.O. were supported by the National Research Foundation of Korea Grants (No. RS-2024-00431768 and No. RS-2025-00515456) funded by the Korean government (Ministry of Science and ICT (MSIT)) and the Institute of Information \& Communications Technology Planning \& Evaluation (IITP) Grants funded by the Korean government (MSIT) (No. RS-2024-00437284, No. IITP-2025-RS-2025-02283189 and No. IITP-2025-RS-2025-02263264). This work was supported by Global Partnership Program of Leading Universities in Quantum Science and Technology (RS-2025-08542968) through the National Research Foundation of Korea~(NRF) funded by the Korean government (Ministry of Science and ICT(MSIT)).
D.K. acknowledges support from the RIKEN Special Postdoctoral Researcher Program, the RIKEN Hakubi Project, and JSPS KAKENHI Grant Number JP26K17060.

\bibliography{reference_main}

\section*{End Matter}

\emph{Detailed form of AKLT MPS}
Here, we provide the MPS description of the AKLT state and the KW-transformed AKLT state.
The AKLT state is expressed as 
\begin{equation}
    | \mathrm{AKLT}_{ \mathbf{L}, \mathbf{R} } \rangle
    = \sum_{\{s_{j}\}}
    \mathbf{L}^{\dagger} A_{1}^{s_1} A_{2}^{s_{2}} \cdots A_{N}^{s_{N}} \mathbf{R}
    |s_{1} s_{2} \cdots s_N\rangle ,
\end{equation}
where the tensor $A_{j}$ is given by
\begin{equation}
\begin{aligned}
    A_{j}^{0} &=
    \frac{1}{\sqrt{3}}
    \begin{pmatrix}
        0 & \sqrt{2}\\
        0 & 0 \\
    \end{pmatrix}, \\
    A_{j}^{1} &=
    \frac{1}{\sqrt{3}}
    \begin{pmatrix}
        -1 & 0\\
        0 & 1 \\
    \end{pmatrix}, \\
    A_{j}^{2} &=
    \frac{1}{\sqrt{3}}
    \begin{pmatrix}
        0 & 0\\
        -\sqrt{2} & 0 \\
    \end{pmatrix}.
\end{aligned}
\end{equation}
For $\mathbf{L} = \mathbf{R} = \mathbf{e}_{\uparrow}$, the KW-transformed AKLT state is
\begin{equation}
\begin{aligned}
    U_{\mathrm{KW}} | \mathrm{AKLT}_{ \mathbf{e}_{\uparrow}, \mathbf{e}_{\uparrow} } \rangle
    = \sum_{\{s_{j}\}}
    \mathbf{e}_{\uparrow}^{\dagger} \widetilde{A}_{1}^{s_1} \widetilde{A}_{2}^{s_{2}} \cdots \widetilde{A}_{N}^{s_{N}} \mathbf{e}_{\uparrow}
    |s_{1} s_{2} \cdots s_N\rangle,
\end{aligned}
\end{equation}
where
\begin{equation}
\label{MPS of KW AKLT}
\begin{aligned}
    \widetilde{A}_{j}^{0}
    &= \frac{1}{\sqrt{3}}
    \begin{pmatrix} 0 & \sqrt{2} \\ 0 & 1 \end{pmatrix}, \\
    \widetilde{A}_{j}^{1}
    &= - \frac{1}{\sqrt{3}}
    \begin{pmatrix} 1 & 0 \\ \sqrt{2} & 0 \end{pmatrix}, \\
    \widetilde{A}_{j}^{2} &= 0.
\end{aligned}
\end{equation}

\emph{Detailed forms of $\mathcal{M}_{j}(m,n)$ and $\mathcal{N}_{j}(m,n)$}---Here, the explicit forms of the CAMPS-transformed terms in Eq.~\eqref{SdotS transform} of the main text are provided.
For $j=1$, we find that
$U_{\mathrm{KW}} (\mathbf{S}_{1} \cdot \mathbf{S}_{2}) U_{\mathrm{KW}}^{\dagger}
= \widetilde{T}_{1,2}^{\text{(bd)}} + \left( \widetilde{T}_{1,2}^{\text{(bd)}} \right)^{\dagger}$
and $U_{\mathrm{KW}} (S_{1}^{z})^{2} U_{\mathrm{KW}}^{\dagger}
= \widetilde{\Xi}_{1}^{\text{(bd)}} + \left( \widetilde{\Xi}_{1}^{\text{(bd)}} \right)^{\dagger}$,
where
\begin{equation}
\begin{aligned}
    \widetilde{T}_{1,2}^{\text{(bd)}}
    &= \sum_{m,n=0}^{2} \mathcal{M}_{1}(m,n) (\omega Z_{2})^{n}, \\
    \widetilde{\Xi}_{1}^{\text{(bd)}}
    &= \frac{1}{3} I_{1} - \frac{1}{3}\omega^{2} Z_{1}. \\
\end{aligned}
\end{equation}
Specifically, $\mathcal{M}_{j}(m,n)$ is given by
\begin{equation}
    \label{def of M}
\begin{aligned}
	\mathcal{M}_{j}(0,0) &= \frac{4}{9} X_{j} \\
	\mathcal{M}_{j}(1,0) &= \frac{2}{9} \left(-\omega^{2}X_{j} - X_{j}^{2}\right)Z_{j}^{2} \\
	\mathcal{M}_{j}(0,1) &= \frac{2}{9} \left(-X_{j} -\omega X_{j}^{2}\right)Z_{j}^{2} \\
	\mathcal{M}_{j}(1,1) &= \frac{1}{3} \left(I_{j} + \frac{1}{3}\omega^{2}X_{j} + \frac{1}{3}\omega X_{j}^{2}\right) Z_{j} \\
	\mathcal{M}_{j}(1,2) &= \frac{1}{3}\omega^{2} \left(-I_{j} + \frac{1}{3}X_{j} + \frac{1}{3} X_{j}^{2} \right),
\end{aligned}
\end{equation}
while $\mathcal{M}_{j}(m,n) = 0$ for all other values of $(m,n)$.
In the case of $(\mathbf{S}_{j} \cdot \mathbf{S}_{j+1})^{2}$, the term $\mathcal{M}_{j}(m,n)$ is substituted with $\mathcal{N}_{j}(m,n)$, which is given by
\begin{equation}
    \label{def of N}
\begin{aligned}
	\mathcal{N}_{j}(0,0) &= \frac{1}{3}\left( 2 I_{j} - \frac{1}{6} \left( X_{j} + X_{j}^{\dagger}\right) \right) \\
	\mathcal{N}_{j}(1,0) &= \frac{2}{9} \left(\omega^{2}X_{j} + X_{j}^{2}\right)Z_{j}^{2} \\
	\mathcal{N}_{j}(0,1) &= \frac{2}{9} \left(X_{j} + \omega X_{j}^{2}\right)Z_{j}^{2} \\
	\mathcal{N}_{j}(1,1) &= \frac{2}{9} \left(\omega^{2} X_{j} + \omega X_{j}^{2}\right)Z_{j} \\
	\mathcal{N}_{j}(1,2) &= \frac{1}{3}\omega^{2} \left(I_{j} - \frac{1}{3} \left( X_{j} + X_{j}^{2}\right) \right),
\end{aligned}
\end{equation}
while $\mathcal{N}_{j}(m,n) = 0$ for all other values of $(m,n)$.

\end{document}


\title{Supplemental Material for ``Disentangling Haldane Phase by Generalized Clifford Circuits''}

\author{Minsoo Kim}
\affiliation{Department of Physics, Korea Advanced Institute of Science and Technology, Daejeon 34141, Korea}

\author{Changhun Oh}\email[]{changhun0218@gmail.com}
\affiliation{Department of Physics, Korea Advanced Institute of Science and Technology, Daejeon 34141, Korea}

\author{Donghoon Kim}\email[]{donghoon.kim@riken.jp}
\affiliation{Analytical Quantum Complexity RIKEN Hakubi Research Team, RIKEN Center for Quantum Computing (RQC), Wako, Saitama 351-0198, Japan}

\date{\today}
\maketitle

\newcommand{\beginsup}{
        \setcounter{equation}{0}
        \renewcommand{\theequation}{S\arabic{equation}}
        \setcounter{table}{0}
        \renewcommand{\thetable}{S\arabic{table}}
        \setcounter{figure}{0}
        \renewcommand{\thefigure}{S\arabic{figure}}
     }
\beginsup

\section{Generalized Pauli operators and Clifford group for qutrit systems}

In this section, we define the generalized Pauli operators for qutrit (spin-1) systems.
We also present the definition of the qutrit Clifford group and a detailed description of its generator set.

The local Hilbert space of a qutrit system is given by $\mathcal{H}_3=\mathrm{span}\{ |{0}\rangle, |{1}\rangle, |{2}\rangle\}$, where $|{s}\rangle$ with $s=0,1,2$ are the computational basis states.
By using this computational basis, the generalized Pauli operators $X$ and $Z$ are defined as follows:
\begin{equation}
\begin{aligned}
	X |{s}\rangle := |{s+1~\mathrm{mod}~3}\rangle, \qquad X \doteq \begin{pmatrix} 0 & 0 & 1 \\ 1 & 0 & 0 \\ 0 & 1 & 0 \end{pmatrix}, \qquad
	Z |{s}\rangle := \omega^{s} |{s}\rangle, \qquad Z \doteq \begin{pmatrix} 1 & 0 & 0 \\ 0 & \omega & 0 \\ 0 & 0 & \omega^{2} \end{pmatrix},
\end{aligned}
\end{equation}
where $\omega := \exp(2 \pi i/3)$ and $\doteq$ denotes the matrix form in the $Z$-basis $\{ |{0}\rangle, |{1}\rangle, |{2}\rangle\}$.
These operators satisfy $X^{3} = Z^{3} = I$, $X^{\dagger} = X^{2}$, and $Z^{\dagger} = Z^{2}$, where $I$ denotes the identity operator.
In addition, a phase factor $\omega$ appears in the commutation relation between $X$ and $Z$, namely $ZX = \omega XZ$.
Furthermore, since the operators $X^{a} Z^{b}$ are linearly independent for distinct $(a, b) \in \mathbb{Z}_{3}^{2}$, the set $\mathsf{P}^{(3)} := \{ X^{a} Z^{b} : (a, b) \in \mathbb{Z}_{3}^{2} \}$ constitutes a complete operator basis for $\mathcal{B}(\mathcal{H}_3)$.

The definition of the generalized Pauli operators extends naturally to an $N$-qutrit Hilbert space.
For an $N$-qutrit system, a generalized Pauli string is expressed as $\bigotimes_{j=1}^{N} P_{j}$, where each $P_{j}$ belongs to the set $\mathsf{P}^{(3)}$.
Using these Pauli strings, the generalized $N$-qutrit Pauli group is defined projectively as
\begin{equation}
    \mathbf{P}_{N}^{(3)}
    := \left\{
        \omega^{c}
        \bigotimes_{j=1}^{N} P_{j}
        ~:~
        c \in \mathbb{Z}_{3}, ~ 
        P_{j} \in \mathsf{P}^{(3)}
    \right\} / \langle \omega I \rangle.
\end{equation}
Here and below, we represent projective Pauli classes by their Pauli-string representatives.
By direct analogy with the qubit case, the Clifford group $\mathrm{Cl}_{N}^{(3)}$ for an $N$-qutrit system is defined as the normalizer of the Pauli group $\mathbf{P}_{N}^{(3)}$.
Explicitly, the Clifford group is given by
\begin{equation}
    \mathrm{Cl}_{N}^{(3)}
    := \left\{
        U \in \mathrm{U}(3^{N}) 
        ~:~ U \mathbf{P}_{N}^{(3)} U^{\dagger} = \mathbf{P}_{N}^{(3)}
    \right\} / \mathrm{U}(1),
\end{equation}
where $\mathrm{U}(n)$ denotes the group of $n \times n$ unitary matrices.
The Clifford group is generated by the single-qutrit Pauli operators $X_j$ and $Z_j$, together with the following three gates~\cite{Gottesman1999, Hostens2005, Wang2020}: the single-qutrit Fourier transform $H^{(3)}$, the qutrit phase gate $S^{(3)}$, and the two-qutrit \texttt{SUM} gate $U_{j,j+1}^{\texttt{SUM}}$ acting on the $j$-th and $(j+1)$-th qutrits.
The definitions of each gate are as follows:
\begin{equation}
    H^{(3)} := \frac{1}{\sqrt{3}} \sum_{s,s'=0}^{2} \omega^{ss'} |s\rangle \langle s'| 
    \doteq \frac{1}{\sqrt{3}}
    \begin{pmatrix} 1 & 1 & 1 \\ 1 & \omega & \omega^{2} \\ 1 & \omega^{2} & \omega \end{pmatrix},
    \qquad
    S^{(3)} := \sum_{s=0}^{2} \omega^{s(s-1)/2} |s\rangle\langle s| \doteq \begin{pmatrix} 1 & 0 & 0 \\ 0 & 1 & 0 \\ 0 & 0 & \omega \end{pmatrix},
\end{equation}
\begin{equation}
\label{def of sum gate}
    U_{j,j+1}^{\texttt{SUM}} |{ s_{j}, s_{j+1} }\rangle
    := |{ s_{j}, s_{j} + s_{j+1}~\mathrm{mod}~3 }\rangle, \qquad 
	U_{j,j+1}^{\texttt{SUM}}
	\doteq
    \begin{pmatrix}
		1 & 0 & 0 & 0 & 0 & 0 & 0 & 0 & 0 \\
		0 & 1 & 0 & 0 & 0 & 0 & 0 & 0 & 0 \\
		0 & 0 & 1 & 0 & 0 & 0 & 0 & 0 & 0 \\
		0 & 0 & 0 & 0 & 0 & 1 & 0 & 0 & 0 \\
		0 & 0 & 0 & 1 & 0 & 0 & 0 & 0 & 0 \\
		0 & 0 & 0 & 0 & 1 & 0 & 0 & 0 & 0 \\
		0 & 0 & 0 & 0 & 0 & 0 & 0 & 1 & 0 \\
		0 & 0 & 0 & 0 & 0 & 0 & 0 & 0 & 1 \\
		0 & 0 & 0 & 0 & 0 & 0 & 1 & 0 & 0
	\end{pmatrix}.
\end{equation}
Note that the \texttt{SUM} gate is a natural generalization of the \texttt{CNOT} gate for qubit systems.

Here, we justify why it is sufficient to consider only 90 two-qutrit Clifford gates in the disentangling analysis, rather than the full Clifford group~\cite{Harper2025GCAMPS}.
The key point is that left multiplication by local unitaries does not change the entanglement generated by a two-qutrit gate, e.g., $U_{j,j+1}^{\texttt{SUM}}$ and $X_{j+1}^{2} U_{j,j+1}^{\texttt{SUM}}$ yield the same entanglement.
Thus, we can factor out the Pauli group from the Clifford group; it is enough to consider $\mathrm{Cl}_{2}^{(3)} / \mathbf{P}_{2}^{(3)}$.
The order of this quotient group can be computed using the isomorphism $\mathrm{Cl}_{n}^{(3)} / \mathbf{P}_{n}^{(3)} \cong \mathrm{Sp}(2n, \mathbb{F}_{3})$, where $\mathbb{F}_{3}$ denotes the finite field with three elements, with addition and multiplication defined modulo $3$.
Here, $\mathrm{Sp}(2n, \mathbb{F}_{q})$ is the symplectic group over $\mathbb{F}_{q}$, defined as the group of invertible $2n \times 2n$ matrices that preserve the standard symplectic form.
Using the standard formula for an integer $n$ and a prime $q$,
\begin{equation}
\label{order of Sp}
    \left| \mathrm{Sp}(2n,\mathbb{F}_{q}) \right|
    = q^{n^{2}} \prod_{k=1}^{n} \left( q^{2k}-1 \right),
\end{equation}
we obtain $\left| \mathrm{Sp}(4,\mathbb{F}_{3}) \right| = 3^{4} \left( 3^{2}-1 \right) \left( 3^{4}-1 \right) = 51840$. 
Moreover, we can further identify Clifford gates that differ by left multiplication with local single-qutrit Clifford gates, since such local unitaries also leave the bipartite entanglement invariant.
Under the induced symplectic action, the local single-qutrit Clifford freedom corresponds to $\mathrm{Sp}(2,\mathbb{F}_{3}) \times \mathrm{Sp}(2,\mathbb{F}_{3})$, whose order is $|\mathrm{Sp}(2, \mathbb{F}_{3})|^2 = 24^2$.
Therefore, the number of inequivalent two-qutrit Clifford representatives is the number of left cosets of $\mathrm{Sp}(2,\mathbb{F}_{3}) \times \mathrm{Sp}(2,\mathbb{F}_{3})$ in $\mathrm{Sp}(4,\mathbb{F}_{3})$, namely
\begin{equation}
    \frac{| \mathrm{Sp}(4, \mathbb{F}_{3})|}{|\mathrm{Sp}(2, \mathbb{F}_{3})|^{2}} = \frac{51840}{24^{2}} = 90.
\end{equation}

In addition, on the Hilbert space $\mathcal{H}_{3}$, we define the spin-1 operators $S^{x}$, $S^{y}$, and $S^{z}$ as follows:
\begin{equation}
    S^{x} := \frac{1}{\sqrt{2}} \begin{pmatrix} 0 & 1 & 0 \\ 1 & 0 & 1 \\ 0 & 1 & 0 \end{pmatrix}, \qquad
    S^{y} := \frac{1}{\sqrt{2}} \begin{pmatrix} 0 & -i & 0 \\ i & 0 & -i \\ 0 & i & 0 \end{pmatrix}, \qquad
    S^{z} := \begin{pmatrix} 1 & 0 & 0 \\ 0 & 0 & 0 \\ 0 & 0 & -1 \end{pmatrix},
\end{equation}
and we also define $S^{\pm} := S^{x} \pm i S^{y}$.
Using the generalized Pauli operators, the spin-1 operators can be expressed as
\begin{equation}
    \label{Pauli spinop conversion}
	S^{z} = \frac{1}{3} \br{(1-\omega)Z + (1-\omega^2)Z^{\dagger}}, \qquad
	S^{+} = \frac{\sqrt{2}}{3} X^{\dagger} \br{2I -Z -Z^{\dagger}}, \qquad
	S^{-} = \frac{\sqrt{2}}{3} X \br{2I - \omega Z - \omega^{2} Z^{\dagger}}.
\end{equation}

\section{Details of Kramers--Wannier transformation of Spin-1 Heisenberg and Bilinear-Biquadratic models}

In this section, we analyze details of the generalized Kramers--Wannier (KW) duality transformation for the spin-$1$ Heisenberg and bilinear-biquadratic (BLBQ) models. 
As shown in the main text, for the Haldane phase, we find that the KW transformation is an optimal Clifford disentangler within the Clifford-circuit-augmented matrix product states (CAMPS)-based density-matrix renormalization group (DMRG) framework.
Here, the KW transformation unitary $U_{\mathrm{KW}}$ is defined as~\cite{Sinha2024, Zhang2025}
\begin{equation}
    \label{KW unitary}
    U_{\mathrm{KW}}
    = \mathcal{U}_{N-1,N} \cdots \mathcal{U}_{2,3} \mathcal{U}_{1,2}, \qquad
    \mathcal{U}_{j,j+1} := X_{j+1}^{2} U_{j,j+1}^{\texttt{SUM}}.
\end{equation}
Since $U_{\mathrm{KW}}$ is a sequential application of $\mathcal{U}_{j,j+1}$, to find the KW transformation of generalized Pauli strings, we first compute the  $\mathcal{U}_{j,j+1} (P_{j}P_{j+1}) \mathcal{U}_{j,j+1}^{\dagger}$ for each two-qutrit Pauli string $P_{j}P_{j+1}$.
The resulting transformations are summarized in Table~\ref{KW P transform}.
\begin{table}[h]
    \renewcommand{\arraystretch}{1.7}
    \setlength{\tabcolsep}{8pt}
    \centering
    \begin{tabular}{c | c c c c }
        \hline
        $P_{j}P_{j+1}$
        &  $X_{j} I_{j+1}$  &  $Z_{j} I_{j+1}$  &  $I_{j} X_{j+1}$  & $I_{j} Z_{j+1}$   \\
        \hline
        $\mathcal{U}_{j,j+1} (P_{j}P_{j+1}) (\mathcal{U}_{j,j+1})^{\dagger}$
        &  $X_{j} X_{j+1}$  &  $Z_{j} I_{j+1}$  &  $I_{j} X_{j+1}$  & $\omega Z_{j}^{2}Z_{j+1}$ \\
        \hline
    \end{tabular}
    \caption{
        For the two-qutrit Clifford circuit $\mathcal{U}_{j,j+1}$, the transformation results of each two-qutrit Pauli string $P_{j}P_{j+1}$ are listed in the table.
        The transformations of Pauli strings not explicitly shown can be generated by combining the four listed Pauli strings.
        For example, $\mathcal{U}_{j,j+1} (X_{j}X_{j+1}^{\dagger}) \mathcal{U}_{j,j+1}^{\dagger} = \mathcal{U}_{j,j+1} (X_{j}I_{j+1}) \mathcal{U}_{j,j+1}^{\dagger} \mathcal{U}_{j,j+1} (I_{j}X_{j+1}^{\dagger}) \mathcal{U}_{j,j+1}^{\dagger} = (X_{j}X_{j+1})(I_{j}X_{j+1}^{\dagger}) = X_{j}I_{j+1}$.
    }
    \label{KW P transform}
\end{table}
Consequently, $U_{\mathrm{KW}}$ transforms generalized Pauli strings as follows:
\begin{equation}
	\label{KW transform}
    U_{\mathrm{KW}} X_{j} (U_{\mathrm{KW}})^{\dagger} = \prod_{k=j}^{N} X_{k}, \qquad
    U_{\mathrm{KW}} Z_{j} (U_{\mathrm{KW}})^{\dagger} = 
    \begin{cases}
        \omega Z_{j-1}^{\dagger} Z_{j} & 2 \leq j \leq N \\
        Z_{1} & j = 1
    \end{cases}.
\end{equation}

\subsection{KW-transformed Hamiltonians of spin-1 Heisenberg and BLBQ models}

The Hamiltonian of the Heisenberg model is given by
\begin{equation}
	\label{Heisenberg Hamiltonian}
	H_{\text{Hei}}(D)
	= \underbrace{\sum_{j=1}^{N-1} \mathbf{S}_{j} \cdot \mathbf{S}_{j+1}}_{ H^{(S \cdot S)} }
	+ D\underbrace{\sum_{j=1}^{N} (S_{j}^{z})^{2}}_{ H^{(\text{aniso})} },
\end{equation}
where $H^{(S \cdot S)}$ represents the Heisenberg interaction term, and $H^{(\text{aniso})}$ denotes the single ion anisotropy term.
By using Eq.~\eqref{Pauli spinop conversion}, each term in $H_{\text{Hei}}$ can be written in terms of generalized Pauli operators as follows:
\begin{equation}
    \label{SdotS Pauli decomposition}
	\mathbf{S}_{j} \cdot \mathbf{S}_{j+1}
	= h_{j,j+1}^{(S \cdot S)} + (h_{j,j+1}^{(S \cdot S)})^{\dagger}, \qquad
	(S_{j}^{z})^{2}
	= h_{j}^{(\text{aniso})} + (h_{j}^{(\text{aniso})})^{\dagger},
\end{equation}
where
\begin{equation}
\begin{aligned}
	h_{j,j+1}^{(S \cdot S)}
	& = -\frac{1}{3} \omega Z_{j} Z_{j+1}
	+ \frac{1}{3} Z_{j} Z_{j+1}^{2}
	+ \frac{4}{9} X_{j} X_{j+1}^{2}
	- \frac{2}{9} \omega (XZ)_{j} X_{j+1}^{2}
	- \frac{2}{9} \omega^{2} (XZ^{2})_{j} X_{j+1}^{2} \\
	&\qquad - \frac{2}{9} X_{j} (X^{2}Z)_{j+1}
	+ \frac{1}{9} \omega (XZ)_{j} (X^{2}Z)_{j+1}
	+ \frac{1}{9} \omega^{2} (XZ^{2})_{j} (X^{2}Z)_{j+1} \\
	&\qquad - \frac{2}{9} X_{j} (X^{2}Z^{2})_{j+1}
	+ \frac{1}{9} \omega (XZ)_{j} (X^{2}Z^{2})_{j+1}
	+ \frac{1}{9} \omega^{2} (XZ^{2})_{j} (X^{2}Z^{2})_{j+1},
\end{aligned}
\end{equation}
and
\begin{equation}
	h_{j}^{(\text{aniso})} = \frac{1}{3} I_{j} - \frac{1}{3}\omega^{2} Z_{j}.
\end{equation}

Then, using the results in Eq.~\eqref{KW transform}, we compute $U_{\mathrm{KW}} h_{j,j+1}^{(S \cdot S)} (U_{\mathrm{KW}})^{\dagger}$.
We find that the originally 2-local term $\mathbf{S}_{j} \cdot \mathbf{S}_{j+1}$ becomes a 3-local interaction under the transformation.
More explicitly, we obtain the transformed term as
\begin{equation}
    \label{SdotS transform}
	U_{\mathrm{KW}} (\mathbf{S}_{j} \cdot \mathbf{S}_{j+1}) (U_{\mathrm{KW}})^{\dagger}
	=
    \begin{cases}
        \widetilde{T}_{1,2}^{\text{(bd)}} + \left( \widetilde{T}_{1,2}^{\text{(bd)}} \right)^{\dagger} & j=1 \\
        \widetilde{T}_{j-1, j, j+1} + (\widetilde{T}_{j-1, j, j+1})^{\dagger} & 2 \leq j \leq N-1
    \end{cases},
\end{equation}
where
\begin{equation}
    \label{def of tilde T terms}
\begin{aligned}
    \widetilde{T}_{1,2}^{\text{(bd)}}
    = \sum_{m,n=0}^{2} \mathcal{M}_{1}(m,n) (\omega Z_{2})^{n}, \qquad
	\widetilde{T}_{j-1, j, j+1}
	= \sum_{m,n=0}^{2} (\omega^{2} Z_{j-1})^{m} \mathcal{M}_{j}(m,n) (\omega Z_{j+1})^{n}.
\end{aligned}
\end{equation}
Here, $\mathcal{M}_{j}(m,n)$ is given by
\begin{equation}
    \label{def of M}
\begin{aligned}
	\mathcal{M}_{j}(0,0) &= \frac{4}{9} X_{j} \\
	\mathcal{M}_{j}(1,0) &= \frac{2}{9} \br{-\omega^{2}X_{j} - X_{j}^{2}} Z_{j}^{2} \\
	\mathcal{M}_{j}(0,1) &= \frac{2}{9} \br{-X_{j} -\omega X_{j}^{2}} Z_{j}^{2} \\
	\mathcal{M}_{j}(1,1) &= \frac{1}{3} \br{I_{j} + \frac{1}{3}\omega^{2}X_{j} + \frac{1}{3}\omega X_{j}^{2}} Z_{j} \\
	\mathcal{M}_{j}(1,2) &= \frac{1}{3} \omega^{2} \br{-I_{j} + \frac{1}{3}X_{j} + \frac{1}{3} X_{j}^{2}},
\end{aligned}
\end{equation}
while $\mathcal{M}_{j}(m,n) = 0$ for all other values of $(m,n)$.

Similiarly, we compute $U_{\mathrm{KW}} h_{j}^{(\text{aniso})} (U_{\mathrm{KW}})^{\dagger}$ as follows:
\begin{equation}
	U_{\mathrm{KW}} (S_{j}^{z})^{2} (U_{\mathrm{KW}})^{\dagger}
	=
    \begin{cases}
        \widetilde{\Xi}_{1}^{\text{(bd)}} + \left( \widetilde{\Xi}_{1}^{\text{(bd)}} \right)^{\dagger} & j=1 \\
        \widetilde{\Xi}_{j-1, j} + (\widetilde{\Xi}_{j-1, j})^{\dagger} & 2 \leq j \leq N
    \end{cases},
\end{equation}
where
\begin{equation}
\label{def of tilde Xi terms}
	\widetilde{\Xi}_{1}^{\text{(bd)}} = \frac{1}{3} I_{1} - \frac{1}{3}\omega^{2} Z_{1}, \qquad
	\widetilde{\Xi}_{j-1, j} = \frac{1}{3} I_{j-1} I_{j} - \frac{1}{3} Z_{j-1}^{2} Z_{j}.
\end{equation}

Therefore, using the forms in Eqs.~\eqref{def of tilde T terms},~\eqref{def of M}, and~\eqref{def of tilde Xi terms}, we can write $\widetilde{H}_{\mathrm{Hei}} = U_{\mathrm{KW}} H_{\mathrm{Hei}} U_{\mathrm{KW}}^{\dagger}$ as
\begin{equation}
\label{transformed Hei Ham}
	\widetilde{H}_{\mathrm{Hei}}(D)
    = \left( \widetilde{T}_{1,2}^{\mathrm{(bd)}}+\sum_{j=2}^{N-1}\widetilde{T}_{j-1,j,j+1} \right)
    +D \left( \widetilde{\Xi}_{1}^{\mathrm{(bd)}}+\sum_{j=2}^{N}\widetilde{\Xi}_{j-1,j} \right)+\mathrm{H.c.},
\end{equation}
where H.c. denotes the Hermitian conjugate.

The Hamiltonian of the BLBQ model is given by
\begin{equation}
	H_{\mathrm{BLBQ}}(\theta)
	=
	\sum_{j=1}^{N-1}
	\left[
		\cos\theta \, (\mathbf{S}_{j} \cdot \mathbf{S}_{j+1})
		+
		\sin\theta \left(\mathbf{S}_{j} \cdot \mathbf{S}_{j+1}\right)^{2}
	\right].
\end{equation}
Using the same procedure as in the derivation of Eq.~\eqref{SdotS Pauli decomposition}, we can write $\left(\mathbf{S}_{j} \cdot \mathbf{S}_{j+1}\right)^{2}$ in terms of generalized Pauli operators.
Then, the KW transformation of the term $\left(\mathbf{S}_{j} \cdot \mathbf{S}_{j+1}\right)^{2}$ is as follows:
\begin{equation}
    \label{BLBQ transform}
	U_{\mathrm{KW}} (\mathbf{S}_{j} \cdot \mathbf{S}_{j+1})^{2} (U_{\mathrm{KW}})^{\dagger}
	=
    \begin{cases}
        \widetilde{Q}_{1,2}^{\text{(bd)}} + \left( \widetilde{Q}_{1,2}^{\text{(bd)}} \right)^{\dagger}  & j=1 \\
        \widetilde{Q}_{j-1, j, j+1} + (\widetilde{Q}_{j-1, j, j+1})^{\dagger} & 2 \leq j \leq N-1
    \end{cases},
\end{equation}
where
\begin{equation}
    \label{def of tilde Q terms}
\begin{aligned}
    \widetilde{Q}_{1,2}^{\text{(bd)}}
    = \sum_{m,n=0}^{2} \mathcal{N}_{1}(m,n) (\omega Z_{2})^{n}, \qquad
	\widetilde{Q}_{j-1, j, j+1}
	= \sum_{m,n=0}^{2} (\omega^{2} Z_{j-1})^{m} \mathcal{N}_{j}(m,n) (\omega Z_{j+1})^{n}.
\end{aligned}
\end{equation}
Here, $\mathcal{N}_{j}(m,n)$ is given by
\begin{equation}
    \label{def of N}
\begin{aligned}
	\mathcal{N}_{j}(0,0) &= \frac{1}{3} \br{2 I_{j} - \frac{1}{6} \br{ X_{j} + X_{j}^{\dagger}}} \\
	\mathcal{N}_{j}(1,0) &= \frac{2}{9} \br{\omega^{2}X_{j} + X_{j}^{2}} Z_{j}^{2} \\
	\mathcal{N}_{j}(0,1) &= \frac{2}{9} \br{X_{j} + \omega X_{j}^{2}} Z_{j}^{2} \\
	\mathcal{N}_{j}(1,1) &= \frac{2}{9} \br{\omega^{2} X_{j} + \omega X_{j}^{2}} Z_{j} \\
	\mathcal{N}_{j}(1,2) &= \frac{1}{3} \omega^{2} \br{I_{j} - \frac{1}{3}\br{X_{j} + X_{j}^{2}}},
\end{aligned}
\end{equation}
while $\mathcal{N}_{j}(m,n) = 0$ for all other values of $(m,n)$.
Thus, using Eqs.~\eqref{def of tilde T terms},~\eqref{def of M},~\eqref{def of tilde Q terms}, and~\eqref{def of N}, we can write $\widetilde{H}_{\mathrm{BLBQ}} = U_{\mathrm{KW}} H_{\mathrm{BLBQ}} U_{\mathrm{KW}}^{\dagger}$ as
\begin{equation}
\label{transformed BLBQ Ham}
    \widetilde{H}_{\mathrm{BLBQ}}(\theta)
    = \cos \theta \br{\widetilde{T}_{1,2}^{\mathrm{(bd)}}+\sum_{j=2}^{N-1}\widetilde{T}_{j-1,j,j+1}}
    + \sin \theta \br{\widetilde{Q}_{1,2}^{\mathrm{(bd)}}+\sum_{j=2}^{N-1}\widetilde{Q}_{j-1,j,j+1}} + \mathrm{H.c.}
\end{equation}

\subsection{Vanishing $\mathcal{E}(N-1)$ of KW-transformed Hamiltonians}

Here, a Hamiltonian $H$ on an $N$-qutrit system is considered, satisfying $[H, Z_{N}] = 0$.
After the KW transformation, according to the forms of $\widetilde{T}_{j,j+1,j+2}$, $\widetilde{\Xi}_{j,j+1}$, and $\widetilde{Q}_{j,j+1,j+2}$ in Eqs.~\eqref{def of tilde T terms},~\eqref{def of tilde Xi terms}, and~\eqref{def of tilde Q terms}, the Heisenberg model and the BLBQ model satisfy this condition.
Then, we prove that if $H$ has a unique ground state, the entanglement between the $(N-1)$-th and $N$-th sites vanishes.
\begin{lemma}
\label{lem: ZN commuting Ham unique GS decoupling}
	Let $H$ be a Hamiltonian on an $N$-qutrit system that satisfies $[H, Z_{N}] = 0$.
    Consider a non-degenerate eigenstate $|{\psi}\rangle = a |{\phi_{0}}\rangle \otimes |{0_{N}}\rangle + b |{\phi_{1}}\rangle \otimes |{1_{N}}\rangle + c |{\phi_{2}}\rangle \otimes |{2_{N}}\rangle$, where $|{s_{N}}\rangle$ for $s \in \{0 ,1, 2\}$ denotes the state on the $N$-th qutrit, and $|{\phi_{s}}\rangle \in \mathcal{H}^{(N-1)}$ denotes the state on the first through $(N-1)$-th qutrit.
    Then exactly one of $a$, $b$, and $c$ is nonzero.
\end{lemma}
\begin{proof}
    Since $|{\psi}\rangle$ is a non-degenerate eigenstate and $[H, Z_{N}] = 0$, we find that $Z_{N} |{\psi}\rangle = \lambda |{\psi}\rangle$ for some $\lambda$.
	Equivalently, we have
	\begin{equation}
		a = \lambda a, \qquad \omega b = \lambda b, \qquad \omega^{2} c = \lambda c.
	\end{equation}
	Then, if $a$ is nonzero, $\lambda$ equals 1, and $b = c = 0$.
	When $b$ is nonzero, $\lambda = \omega$, and $a = c = 0$.
	Similarly, if $c \neq 0$, then $a = b = 0$.
\end{proof}
This lemma explains $\mathcal{E}(N-1) = 0$ of the KW-transformed Heisenberg and BLBQ models, as shown in Fig.~2(b) and (d) of the main text.

\subsection{Locality preservation under the generalized KW transformation}

In this section, we characterize when a local operator $O_{j,j+1}$, acting on the $j$-th and $(j+1)$-th qutrits, remains local under the generalized KW transformation $U_{\mathrm{KW}}$ described in Eqs.~\eqref{KW unitary} and~\eqref{KW transform}.

\begin{lemma}
\label{lem: locality preserving}
    Let $O_{j,j+1}$ be a local operator acting on the $j$-th and $(j+1)$-th qutrits, where $1 \leq j < N-2$.
    Then $U_{\mathrm{KW}} O_{j,j+1} (U_{\mathrm{KW}})^{\dagger}$ remains local, with support on at most three qutrits, if and only if $[O_{j,j+1}, Z_{j}Z_{j+1}] = 0$.
\end{lemma}
\begin{proof}

An arbitrary operator $O_{j,j+1}$ can be uniquely expressed as a linear combination of generalized Pauli strings, since the set $\mathsf{P}^{(3)} = \{ X^{a} Z^{b} : (a, b) \in \mathbb{Z}_{3}^{2} \}$ forms a complete basis for the operator space $\mathcal{B}(\mathcal{H}_3)$.
More explicitly, we denote the decomposition as
\begin{equation}
    \label{Pauli string decomposition}
    O_{j,j+1} = \sum_{ \substack{ (a_{j}, b_{j}) \\ (a_{j+1}, b_{j+1}) } }
    c_{(a_{j}, b_{j}), (a_{j+1}, b_{j+1})} ( X^{a_{j}} Z^{b_{j}} )_{j} ( X^{a_{j+1}} Z^{b_{j+1}} )_{j+1},
\end{equation}
where $(a_{j},b_{j})\in\mathbb{Z}_{3}^{2}$ and $c_{(a_{j}, b_{j}), (a_{j+1}, b_{j+1})}$ are the corresponding coefficients.

The next step in the proof is to observe that the KW transformation preserves the locality of $O_{j,j+1}$ if and only if the following condition is satisfied:
\begin{center}
    (\textbf{Condition}) The coefficients $c_{(a_{j}, b_{j}), (a_{j+1}, b_{j+1})}$ are non-zero only for indices satisfying $a_{j} + a_{j+1} \equiv 0 \mod{3}$.
\end{center}
According to Eq.~\eqref{KW transform}, the $Z$-components of an operator always remain local, whereas the $X$-components can be mapped to non-local operators.
Thus, locality is determined solely by the $X$-part, i.e., it only depends on $a_{j}, a_{j+1}$, regardless of $b_{j}, b_{j+1}$.
In other words, a term remains local whenever its $X$-part is proportional to the identity, $X_{j}X_{j+1}^{2}$, or $X_{j}^{2}X_{j+1}$, irrespective of the accompanying $Z$ operators.
In contrast, the other terms that do not satisfy this condition are mapped to non-local operators, since $U_{\mathrm{KW}} X_{j} (U_{\mathrm{KW}})^{\dagger} = \prod_{k=j}^{N} X_{k}$ (see Eq.~\eqref{KW transform}).

Finally, we prove that the above condition is satisfied if and only if $[O_{j,j+1}, Z_{j}Z_{j+1}] = 0$.
The proof is as follows:
From the commutation relation $ZX = \omega XZ$, we have
\begin{equation}
    \label{ZZ conjugation}
    (Z_{j} Z_{j+1}) \left[ ( X^{a_{j}} Z^{b_{j}} )_{j} ( X^{a_{j+1}} Z^{b_{j+1}} )_{j+1} \right] (Z_{j} Z_{j+1})^{\dagger}
    = \omega^{a_{j} + a_{j+1}} \left[ ( X^{a_{j}} Z^{b_{j}} )_{j} ( X^{a_{j+1}} Z^{b_{j+1}} )_{j+1} \right],
\end{equation}
in other words, the coefficients $c_{(a_{j}, b_{j}), (a_{j+1}, b_{j+1})} \rightarrow \omega^{a_{j} + a_{j+1}} c_{(a_{j}, b_{j}), (a_{j+1}, b_{j+1})}$ under the conjugation by $Z_{j}Z_{j+1}$.
Hence, if the condition is satisfied, since $\omega^{3}=1$, all coefficients are invariant under the conjugation, i.e., $[O_{j,j+1}, Z_{j}Z_{j+1}] = 0$.
Conversely, if $(Z_{j}Z_{j+1}) O_{j,j+1} (Z_{j}Z_{j+1})^{\dagger} = O_{j,j+1}$, the condition holds:
This follows because such conjugation multiplies each coefficient by $\omega^{a_{j} + a_{j+1}}$, forcing all coefficients with $\omega^{a_{j} + a_{j+1}} \neq 1$ to vanish. 

\end{proof}
\noindent
As a result of Lemma~\ref{lem: locality preserving}, the KW transformation result of $\mathbf{S}_{j} \cdot \mathbf{S}_{j+1}$ is still local since $[\mathbf{S}_{j} \cdot \mathbf{S}_{j+1}, Z_{j}Z_{j+1}] = 0$.

\section{Benchmark results of the CAMPS-based DMRG for the other systems}

In this section, we present numerical results obtained using the CAMPS-based DMRG method for the three-state clock ($\mathbb{Z}_{3}$ clock) model, the trivial phase of the spin-1 Heisenberg model, and the dimerized regimes of the BLBQ model.

\subsection{$\mathbb{Z}_{3}$ clock model}

\begin{figure}[h]
    \centering
    \includegraphics[scale=1]{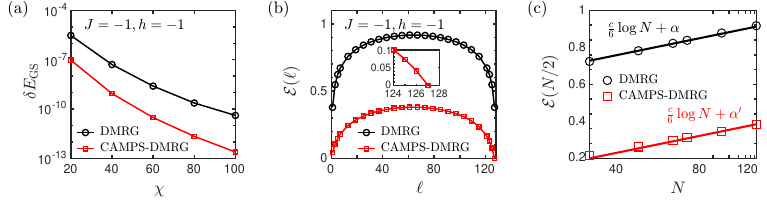}
    \caption{
    The CAMPS-based DMRG results for the $\mathbb{Z}_{3}$ clock model.
    (a) Relative ground-state energy error $\delta E_{\text{GS}}$ for $N=128$ as a function of the bond dimension $\chi$. The DMRG result with $\chi=1000$ is used as a reference.
    (b) For $\chi=100$, entanglement entropy $\mathcal{E}(\ell)$ as a function of $\ell$, where $\ell$ and $N-\ell$ denote the sizes of the two partitions. Inset: the CAMPS-based DMRG results near the right boundary ($\ell = 127$).
    (c) For the half-chain entropy, we fit $\mathcal{E}(N/2) = (c/6) \log N + \alpha$ with central charge $c=4/5$. We numerically find that $\alpha=0.265$ (DMRG) and $\alpha' = -0.265$ (CAMPS).
    }
    \label{Fig: Z3clockdata}
\end{figure}

The $\mathbb{Z}_{3}$ clock model is given by
\begin{equation}
    H_{\mathbb{Z}_{3}\mathrm{clock}}
    = J \sum_{j=1}^{N-1} X_{j} X_{j+1}^\dagger
      + h \sum_{j=1}^{N} Z_{j} + \text{H.c.}
\end{equation}
We focus on the critical point $(J,h)=(-1,-1)$.
As shown in Fig.~\ref{Fig: Z3clockdata}(a,b), the CAMPS-based DMRG consistently improves the ground-state energy accuracy over standard DMRG for all system sizes and bond dimensions considered.
A reduction in the entanglement entropy $\mathcal{E}(\ell)$ is also observed, as shown in Fig.~\ref{Fig: Z3clockdata}(c).

We numerically find that the optimal disentangling circuit is again the generalized KW transformation $U_{\mathrm{KW}}$ in Eq.~\eqref{KW unitary}.
This parallels the qubit Ising case, where CAMPS-based DMRG identifies the qubit KW transformation as the optimal disentangler~\cite{Fan2025}.
The analogy is natural, since the $\mathbb{Z}_{3}$ clock model can be viewed as a minimal three-state extension of the Ising model.

To clearly reveal the duality structure of the KW-transformed $\mathbb{Z}_{3}$ clock model, we introduce a slightly modified unitary $U_{\mathrm{KW}}^{(X)}$, defined by
\begin{equation}
\label{UXKW}
    U_{\mathrm{KW}}^{(X)}
    := \left( \prod_{j=1}^{N} (X_{j})^{j-1} \right) U_{\mathrm{KW}} 
    = U_{N-1,N}^{\texttt{SUM}} U_{N-2,N-1}^{\texttt{SUM}}
    \cdots
    U_{1,2}^{\texttt{SUM}},
\end{equation}
where the second equality follows from the definition $\mathcal{U}_{j,j+1} := X_{j+1}^{2} U_{j,j+1}^{\texttt{SUM}}$ in Eq.~\eqref{KW unitary} and the relation $\mathcal{U}_{j,j+1} X_{j} \mathcal{U}_{j,j+1}^{\dagger} = X_{j} X_{j+1}$ in Table~\ref{KW P transform}.
We emphasize that since $U_{\mathrm{KW}}^{(X)}$ and $U_{\mathrm{KW}}$ are equivalent up to a local unitary, they yield ground states with identical entanglement properties.
Moreover, the numerical optimization can yield $U_{\mathrm{KW}}$ rather than $U_{\mathrm{KW}}^{(X)}$ as the optimal disentangler; this depends on the choice of representative within each local-unitary equivalence class, since two-site Clifford gates differing only by local unitaries yield the same entanglement.
Specifically, if $U_{j,j+1}^{\texttt{SUM}}$ is used instead of $\mathcal{U}_{j,j+1}$, the optimal circuit is represented as $U_{\mathrm{KW}}^{(X)}$ rather than $U_{\mathrm{KW}}$.

It follows from Eq.~\eqref{UXKW} that the KW-transformed Hamiltonian is given by
\begin{equation}
\label{Z3clock CAMPS transformation}
    \widetilde{H}_{\mathbb{Z}_{3}\mathrm{clock}}^{(X)}
    = J \sum_{j=1}^{N-1} X_j + h Z_1
    + h \sum_{j=1}^{N-1} Z_j Z_{j+1}^\dagger
    + \mathrm{H.c.}.
\end{equation}
Equation~\eqref{Z3clock CAMPS transformation} shows that the numerically obtained $U_{\mathrm{KW}}^{(X)}$ coincides with the natural duality transformation, which interchanges the parameters $J$ and $h$.
At the self-dual point, $\widetilde{H}_{\mathbb{Z}_{3}\mathrm{clock}}$ agrees with the original Hamiltonian up to a local basis change and boundary terms, so the bulk critical behavior is unchanged; this explains the same logarithmic scaling of $\mathcal{E}(\ell)$ with central charge $c=4/5$~\cite{Zamolodchikov1985Parafermion, Dotsenko1984Z3PottsJSP, Dotsenko1984ConformalAlgebra}, as shown in Fig.~\ref{Fig: Z3clockdata}(c).
Moreover, the absence of an $X_{N}$-type boundary field term indicates $[\widetilde{H}_{\mathbb{Z}_{3}\mathrm{clock}}, Z_{N}]=0$, and hence $\mathcal{E}(N-1)=0$ for the unique ground state by Lemma~\ref{lem: ZN commuting Ham unique GS decoupling}.
It is consistent with the inset in Fig.~\ref{Fig: Z3clockdata}(b).

\subsection{Trivial phase of Heisenberg model and dimerized phase of BLBQ model}

\begin{figure}[h]
    \centering
    \includegraphics[scale=1]{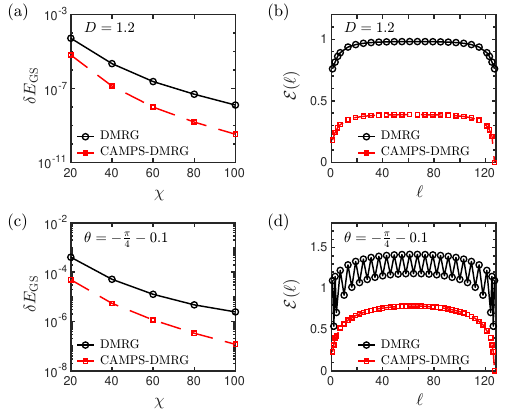}
    \caption{
    The CAMPS-based DMRG results for $H_{\mathrm{Hei}}(D=1.2)$ and $H_{\mathrm{BLBQ}}(\theta=-\frac{\pi}{4}-0.1)$.
    (a), (c) Relative ground-state energy error $\delta E_{\text{GS}}$ for $N=128$ at (a) $H_{\mathrm{Hei}}(D=1.2)$ and (c) $H_{\mathrm{BLBQ}}(\theta = -\pi/4-0.1)$.
    The DMRG result with bond dimension $\chi=1000$ serves as the reference.
    Panels (b) and (d) display the corresponding entanglement entropy $\mathcal{E}(\ell)$.
    In all cases, we observe that $\mathcal{E}(N-1) = 0$.
    }
    \label{Fig: CAMPS_othersystems}
\end{figure}

In Fig.~\ref{Fig: CAMPS_othersystems}, we present the CAMPS-based DMRG results for the trivial phase of the spin-1 Heisenberg model $H_{\mathrm{Hei}}(D=1.2)$ and the dimerized phase of the BLBQ model $H_{\mathrm{BLBQ}}(\theta=-\frac{\pi}{4}-0.1)$.
Across all cases, the CAMPS-based DMRG achieves lower ground-state energy error than standard DMRG at fixed bond dimension, accompanied by a clear reduction of bipartite entanglement entropy.

Our numerical results show that the optimal Clifford circuit is also $U_{\mathrm{KW}}$ in both cases.
Thus, by Lemma~\ref{lem: ZN commuting Ham unique GS decoupling}, we find that $\mathcal{E}(N-1)=0$.
Furthermore, in the dimerized phase of the BLBQ model, $\mathcal{E}(\ell)$ exhibits a period-2 oscillation as shown in Fig.~\ref{Fig: CAMPS_othersystems}(d), which is suppressed after disentangling with $U_{\mathrm{KW}}$.

\section{Optimality of the KW transformation for the AKLT state}

Here, we provide details on the optimality of the KW transformation as a disentangler of the Affleck-Kennedy-Lieb-Tasaki (AKLT) state.

\subsection{Graphical description of AKLT state}

We begin by introducing the AKLT state and the diagrammatic notation.
The AKLT state can be described by using the following valence bond state (VBS) picture~\cite{Affleck1987, Tasaki2020}:
\begin{equation}
	|{\mathrm{AKLT}_{ \mathbf{L}, \mathbf{R} }}\rangle = 
\begin{tikzpicture}[anchor=center, baseline={(0,-0.1)}]
    \def\a{0.35};
	\def\c{0.1};
	\def\h{-1.0}

	\node at (-12*\a, 0.7*\a) {$\textcolor{red}{\mathrm{L}}$};

	\begin{scope}[shift = {(-9.0*\a, 0)}]
        \draw (0,0) ellipse ({2.0*\a} and {0.7*\a});

        \draw (\a,0) -- (0, -2.0*\a);
        \draw (-\a,0) -- (0, -2.0*\a);

        \draw[thick] (1.0*\a,0) -- (2.5*\a,0);
        \fill[black!60] (1.0*\a,0) circle (\c);
        \fill[red] (-1.0*\a,0) circle (\c);
        
        \fill[SkyBlue!80]
        ({1.0*(1+\h/2.0)*\a}, {\h*\a}) -- (0,{-2.0*\a}) -- ({-1.0*(1+\h/2.0)*\a}, {\h*\a}) -- cycle;

        \draw (0, -2.0*\a) -- (0, -2.8*\a);	

        \node at (0,2*\a) {$1$};
	\end{scope}

	\begin{scope}[shift = {(-4.0*\a, 0)}]
        \draw (0,0) ellipse ({2.0*\a} and {0.7*\a});

        \draw (\a,0) -- (0, -2.0*\a);
        \draw (-\a,0) -- (0, -2.0*\a);

        \draw[thick] (1.0*\a,0) -- (2.5*\a,0);
        \draw[thick] (-1.0*\a,0) -- (-2.5*\a,0);
        \fill[black!60] (1.0*\a,0) circle (\c);
        \fill[black!60] (-1.0*\a,0) circle (\c);
        
        \fill[SkyBlue!80]
        ({1.0*(1+\h/2.0)*\a}, {\h*\a}) -- (0,{-2.0*\a}) -- ({-1.0*(1+\h/2.0)*\a}, {\h*\a}) -- cycle;

        \draw (0, -2.0*\a) -- (0, -2.8*\a);	

        \node at (0,2*\a) {$2$};
	\end{scope}

	\node at (0,0) {$\cdots$};

	\begin{scope}[shift = {(4.0*\a, 0)}]
        \draw (0,0) ellipse ({2.0*\a} and {0.7*\a});

        \draw (\a,0) -- (0, -2.0*\a);
        \draw (-\a,0) -- (0, -2.0*\a);

        \draw[thick] (1.0*\a,0) -- (2.5*\a,0);
        \draw[thick] (-1.0*\a,0) -- (-2.5*\a,0);
        \fill[black!60] (1.0*\a,0) circle (\c);
        \fill[black!60] (-1.0*\a,0) circle (\c);
        
        \fill[SkyBlue!80]
        ({1.0*(1+\h/2.0)*\a}, {\h*\a}) -- (0,{-2.0*\a}) -- ({-1.0*(1+\h/2.0)*\a}, {\h*\a}) -- cycle;

        \draw (0, -2.0*\a) -- (0, -2.8*\a);

        \node at (0,2*\a) {$N-1$};
	\end{scope}

	\begin{scope}[shift = {(9.0*\a, 0)}]
        \draw (0,0) ellipse ({2.0*\a} and {0.7*\a});

        \draw (\a,0) -- (0, -2.0*\a);
        \draw (-\a,0) -- (0, -2.0*\a);

        \draw[thick] (1.0*\a,0) -- (2.5*\a,0);
        \draw[thick] (-1.0*\a,0) -- (-2.5*\a,0);
        \fill[black!60] (-1.0*\a,0) circle (\c);
        \fill[black!60] (1.0*\a,0) circle (\c);
        
        \fill[SkyBlue!80]
        ({1.0*(1+\h/2.0)*\a}, {\h*\a}) -- (0,{-2.0*\a}) -- ({-1.0*(1+\h/2.0)*\a}, {\h*\a}) -- cycle;

        \draw (0, -2.0*\a) -- (0, -2.8*\a);

        \node at (0,2*\a) {$N$};
	\end{scope}

    \begin{scope}[shift = {(14.0*\a, 0)}]
        \draw[thick] (-1.0*\a,0) -- (-2.5*\a,0);
        \fill[blue] (-1.0*\a,0) circle (\c);
	\end{scope}

	\node at (14*\a, 0.7*\a) {$\textcolor{blue}{\mathrm{R}}$};
\end{tikzpicture}
\end{equation}
Here, each filled circle denotes a virtual spin-$1/2$, and two spin-$1/2$ particles in the ellipse form a qutrit (i.e., a spin-$1$).
At the left and right boundaries, there are the remaining virtual spin-$1/2$ $\mathrm{L}$ and $\mathrm{R}$.
We denote the corresponding states using row and column vectors in the two-dimensional complex vector space $\mathbb{C}^{2}$; explicitly, 
$\mathbf{L} = L_{1} \mathbf{e}_{\uparrow} + L_{2} \mathbf{e}_{\downarrow}$
and $\mathbf{R} = R_{1} \mathbf{e}_{\uparrow} + R_{2} \mathbf{e}_{\downarrow}$,
where $\mathbf{e}_{\uparrow}, \mathbf{e}_{\downarrow} \in \mathbb{C}^{2}$ are given by $\mathbf{e}_{\uparrow} = \begin{pmatrix} 1 \\ 0 \end{pmatrix}$ and $\mathbf{e}_{\downarrow} = \begin{pmatrix} 0 \\ 1 \end{pmatrix}$.
The matrix representations of the components in the diagram are given by
\begin{equation}
	\begin{tikzpicture}[anchor=center, baseline={(0,-0.1)}]
		\def\a{0.35};
		\def\c{0.1};

        \draw[thick] (-1.0*\a,0) -- (1.0*\a,0);
		\fill[black!60] (1.0*\a,0) circle (\c);
		\fill[black!60] (-1.0*\a,0) circle (\c);

	\end{tikzpicture}
    ~:~\Psi_{\mathrm{singlet}}
	=
    \begin{pmatrix}
        0 & 1/\sqrt{2} \\ -1/\sqrt{2} & 0
    \end{pmatrix}, \qquad
    \begin{tikzpicture}[anchor=center, baseline={(0,-0.1)}]
		\def\c{0.1};
		\fill[red] (0,0) circle (\c);
	\end{tikzpicture}
    ~:~ \mathbf{L}^{\dagger}
    = \begin{pmatrix} L_{1}^{*} & L_{2}^{*} \end{pmatrix}, \qquad
    \begin{tikzpicture}[anchor=center, baseline={(0,-0.1)}]
		\def\c{0.1};
		\fill[blue] (0,0) circle (\c);
	\end{tikzpicture}
    ~:~ \mathbf{R}
    = \begin{pmatrix} R_{1} \\ R_{2} \end{pmatrix}    
\end{equation}
\begin{equation}
	\begin{tikzpicture}[anchor=center, baseline={(0,-0.3)}]
		\def\a{0.35};
		\def\c{0.1};
		\def\h{-1.0}
		
        \draw (\a,0) -- (0, -2.0*\a);
        \draw (-\a,0) -- (0, -2.0*\a);
		
		\fill[SkyBlue!80]
		({1.0*(1+\h/2.0)*\a}, {\h*\a}) -- (0,{-2.0*\a}) -- ({-1.0*(1+\h/2.0)*\a}, {\h*\a}) -- cycle;

		\draw (0, -2.0*\a) -- (0, -2.8*\a);

        \node at (0, -3.7*\a) {$s$};
	\end{tikzpicture}
    ~:~ \Pi^{s}, \qquad
    \Pi^{s=0} = \begin{pmatrix} 1 & 0 \\ 0 & 0 \end{pmatrix}, \qquad
    \Pi^{s=1} = \begin{pmatrix} 0 & 1/\sqrt{2} \\ 1/\sqrt{2} & 0 \end{pmatrix}, \qquad
    \Pi^{s=2} = \begin{pmatrix} 0 & 0 \\ 0 & 1 \end{pmatrix}
\end{equation}
Here, $\Psi_{\mathrm{singlet}}$ denotes the spin singlet state, and the rank-3 tensor $\Pi$ is a projector into the spin-$1$ from two spin-$1/2$.
In the diagram of $\Psi_{\mathrm{singlet}}$, the left and right filled circles indicate the row and column indices, respectively.
Thus, the MPS description of the AKLT state is given by
\begin{equation}
\label{def of AKLT A}
\begin{gathered}
    |\mathrm{AKLT}_{ \mathbf{L}, \mathbf{R} }\rangle
    = \sum_{ \{s_{j}\} }
    \mathbf{L}^{\dagger}
    A_{1}^{s_{1}}
    A_{2}^{s_{2}} A_{3}^{s_{3}} \cdots A_{N-1}^{s_{N-1}}
    A_{N}^{s_{N}}
    \mathbf{R}
    |s_{1} s_{2} \cdots s_{N} \rangle, \\
    A_{j}^{0} =
    \frac{1}{\sqrt{3}}
    \begin{pmatrix}
        0 & \sqrt{2}\\
        0 & 0 \\
    \end{pmatrix}, \qquad
    A_{j}^{1} =
    \frac{1}{\sqrt{3}}
    \begin{pmatrix}
        -1 & 0\\
        0 & 1 \\
    \end{pmatrix}, \qquad
    A_{j}^{2} =
    \frac{1}{\sqrt{3}}
    \begin{pmatrix}
        0 & 0\\
        -\sqrt{2} & 0 \\
    \end{pmatrix},
\end{gathered}
\end{equation}
where it can be seen from the diagrammatic description below:
\begin{equation}
\label{AKLT diagramatic definitions}
    \frac{2}{\sqrt{3}}~
    \begin{tikzpicture}[anchor=center, baseline={(0,-0.3)}]
        \def\a{0.35};
        \def\c{0.1};
        \def\h{-1.0}

        \draw (\a,0) -- (0, -2.0*\a);
        \draw (-\a,0) -- (0, -2.0*\a);

        \draw[thick] (1.0*\a,0) -- (2.5*\a,0);
        \fill[black!60] (1.0*\a,0) circle (\c);
        \fill[black!60] (2.5*\a,0) circle (\c);
        
        \fill[SkyBlue!80]
        ({1.0*(1+\h/2.0)*\a}, {\h*\a}) -- (0,{-2.0*\a}) -- ({-1.0*(1+\h/2.0)*\a}, {\h*\a}) -- cycle;

        \draw (0, -2.0*\a) -- (0, -2.8*\a);	

        \node at (0, -3.7*\a) {$s$};
    \end{tikzpicture}
    =
    \begin{tikzpicture}[anchor=center, baseline={(0,-0.1)}]
        \def\a{0.35};

        \roundedsquare{SkyBlue!40}{\a}{\a}
        \node at (0, 0) {$A_{j}$};    
        
        \draw (1.0*\a, 0) -- (2.0*\a, 0);
        \draw (-1.0*\a, 0) -- (-2.0*\a, 0);
        \draw (0, -1.0*\a) -- (0, -2.0*\a);

        \node at (0, -2.7*\a) {$s$};
    \end{tikzpicture}
    ~:~ \frac{2}{\sqrt{3}} \Pi^{s} \Psi_{\mathrm{singlet}} = A_{j}^{s}, \qquad 
    \frac{2}{\sqrt{3}}~
    \begin{tikzpicture}[anchor=center, baseline={(0,-0.3)}]
        \def\a{0.35};
        \def\c{0.1};
        \def\h{-1.0}

        \draw (\a,0) -- (0, -2.0*\a);
        \draw (-\a,0) -- (0, -2.0*\a);

        \fill[red] (-1.0*\a,0) circle (\c);
        \draw[thick] (1.0*\a,0) -- (2.5*\a,0);
        \fill[black!60] (1.0*\a,0) circle (\c);
        \fill[black!60] (2.5*\a,0) circle (\c);
        
        \fill[SkyBlue!80]
        ({1.0*(1+\h/2.0)*\a}, {\h*\a}) -- (0,{-2.0*\a}) -- ({-1.0*(1+\h/2.0)*\a}, {\h*\a}) -- cycle;

        \draw (0, -2.0*\a) -- (0, -2.8*\a);	

        \node at (0, -3.7*\a) {$s$};
    \end{tikzpicture}
    ~:~ \mathbf{L}^{\dagger} A_{1}^{s}, \qquad
    \frac{2}{\sqrt{3}}~
    \begin{tikzpicture}[anchor=center, baseline={(0,-0.3)}]
        \def\a{0.35};
        \def\c{0.1};
        \def\h{-1.0}

        \draw (\a,0) -- (0, -2.0*\a);
        \draw (-\a,0) -- (0, -2.0*\a);

        \draw[thick] (1.0*\a,0) -- (2.5*\a,0);
        \fill[black!60] (1.0*\a,0) circle (\c);
        \fill[blue] (2.5*\a,0) circle (\c);
        
        \fill[SkyBlue!80]
        ({1.0*(1+\h/2.0)*\a}, {\h*\a}) -- (0,{-2.0*\a}) -- ({-1.0*(1+\h/2.0)*\a}, {\h*\a}) -- cycle;

        \draw (0, -2.0*\a) -- (0, -2.8*\a);	

        \node at (0, -3.7*\a) {$s$};
    \end{tikzpicture}
    ~:~ A_{N}^{s} \mathbf{R}.
\end{equation}
Note that $|\mathrm{AKLT}_{ \mathbf{L}, \mathbf{R} }\rangle$ is unnormalized, and its normalization will be addressed later.

\subsection{Propagation of the canonical form under the action of $\mathcal{U}_{j,j+1}$}

Here, we define \emph{$(\mathbf{u}, \mathbf{v}; a)$-canonical form}, which is a key ingredient for connecting the KW transformation to the AKLT state.
In this section, a vector in $\mathbb{C}^{2}$ is written in boldface, for example, $ \mathbf{v} = v_{1} \mathbf{e}_{\uparrow} + v_{2} \mathbf{e}_{\downarrow}$, while its components $v_{1}$ and $v_{2}$ are denoted by the corresponding nonbold symbols.

\begin{definition}
    \normalfont
    Let $B$ be a tensor where each $B^{p}$ for $p \in \mathbb{Z}_{3}$ is a $2 \times 2$ matrix.
    The tensor $B$ is said to be in \emph{$(\mathbf{u}, \mathbf{v}; a)$-canonical form} if it can be written as
    \begin{equation}
    \label{bn form}
        B^{0} = \mathbf{u} \mathbf{e}_{\downarrow}^{\dagger}
        = \begin{pmatrix} u_{1} \\ u_{2}
        \end{pmatrix} \begin{pmatrix} 0 & 1 \end{pmatrix}, \qquad
        B^{1} = \mathbf{v} \mathbf{e}_{\uparrow}^{\dagger}
        = \begin{pmatrix} v_{1} \\ v_{2} \end{pmatrix}
        \begin{pmatrix} 1 & 0 \end{pmatrix}, \qquad
        B^{2} = 0,
    \end{equation}
    where $\mathbf{u}$ and $\mathbf{v}$ satisfy
    \begin{equation}
    \label{condition of uv for B}
        \| \mathbf{u} \|^{2} = a, \qquad
        \| \mathbf{v} \|^{2} = 1-a, \qquad
        \mathbf{u}^{\dagger} \mathbf{v} = - \frac{\sqrt{2}}{3}.
    \end{equation}
    Here, $\mathbf{u} = u_{1} \mathbf{e}_{\uparrow} + u_{2} \mathbf{e}_{\downarrow}$ and $\mathbf{v} = v_{1} \mathbf{e}_{\uparrow} + v_{2} \mathbf{e}_{\downarrow}$ are vectors in $\mathbb{C}^{2}$.
\end{definition}
\noindent
Note that this canonical form is closely related to the AKLT tensor $A_{j}$ in Eq.~\eqref{def of AKLT A}: it is straightforward to show that for any normalized vector $\mathbf{w} \in \mathbb{C}^{2}$, a tensor $B_{1}$ such that $B_{1}^{s} = \mathbf{w} \mathbf{e}_{\uparrow}^{\dagger} A_{1}^{s}$ is in $(\sqrt{2/3} \mathbf{w}, -\sqrt{1/3} \mathbf{w}; 2/3)$-canonical form.
Diagrammatically,
\begin{equation}
    B_{1}^{0} =
    \begin{tikzpicture}[anchor=center, baseline={(0,-0.1)}]
        \def\a{0.35};

        \roundedsquare{SkyBlue!40}{\a}{\a}
        \node at (0, 0) {$A_{1}$};
        
        \draw (1.0*\a, 0) -- (2.0*\a, 0);
        \draw (-1.0*\a, 0) -- (-2.0*\a, 0);
        \node at (-3.2*\a, 0) {$ \mathbf{w} \mathbf{e}_{\uparrow}^{\dagger} $};
        \draw (0, -1.0*\a) -- (0, -2.0*\a);

        \node at (0, -2.7*\a) {$\ket{0}$};
    \end{tikzpicture} = \sqrt{\frac{2}{3}} \mathbf{w} \mathbf{e}_{\downarrow}^{\dagger}, \qquad 
    B_{1}^{1} =
    \begin{tikzpicture}[anchor=center, baseline={(0,-0.1)}]
        \def\a{0.35};

        \roundedsquare{SkyBlue!40}{\a}{\a}
        \node at (0, 0) {$A_{1}$};
        
        \draw (1.0*\a, 0) -- (2.0*\a, 0);
        \draw (-1.0*\a, 0) -- (-2.0*\a, 0);
        \node at (-3.2*\a, 0) {$ \mathbf{w} \mathbf{e}_{\uparrow}^{\dagger} $};
        \draw (0, -1.0*\a) -- (0, -2.0*\a);

        \node at (0, -2.7*\a) {$\ket{1}$};
    \end{tikzpicture} = - \sqrt{\frac{1}{3}} \mathbf{w} \mathbf{e}_{\uparrow}^{\dagger}, \qquad 
    B_{1}^{2} =
    \begin{tikzpicture}[anchor=center, baseline={(0,-0.1)}]
        \def\a{0.35};

        \roundedsquare{SkyBlue!40}{\a}{\a}
        \node at (0, 0) {$A_{1}$};
        
        \draw (1.0*\a, 0) -- (2.0*\a, 0);
        \draw (-1.0*\a, 0) -- (-2.0*\a, 0);
        \node at (-3.2*\a, 0) {$ \mathbf{w} \mathbf{e}_{\uparrow}^{\dagger} $};
        \draw (0, -1.0*\a) -- (0, -2.0*\a);

        \node at (0, -2.7*\a) {$\ket{2}$};
    \end{tikzpicture} =  0,
\end{equation}
Hence, $B_{1}$ coincides with $\mathbf{L}^{\dagger} A_{1}$ in Eq.~\eqref{AKLT diagramatic definitions} where $\mathbf{L} = \mathbf{e}_{\uparrow}$, apart from the additional column vector $\mathbf{w}$.

We show that the canonical form can be propagated from site $j$ to site $j+1$ by applying the gate $\mathcal{U}_{j,j+1}$ as a disentangler.
Suppose that the tensor $B_{j}$ at the $j$-th site is in $(\mathbf{u}, \mathbf{v}; a)$-canonical form, and define the two-site tensor $\Theta^{p,q} = B_{j}^{p} A_{j+1}^{q}$ where $p,q \in \mathbb{Z}_{3}$.
We apply $\mathcal{U}_{j,j+1}$ to the physical indices (i.e., $p, q$), obtaining the resulting tensor $\widetilde{\Theta}^{p,q}$ as
\begin{equation}
    \widetilde{\Theta}^{p,q}
    : \underbrace{\mathbb{C}^{2}}_{\text{right bond}} \to \underbrace{\mathbb{C}^{2}}_{\text{left bond}}.
\end{equation}
Then, a singular value decomposition (SVD) across the bond between sites $j$ and $j+1$ yields
\begin{equation}
    \widetilde{\Theta}^{p,q}
    = U_{j}^{p} \Sigma V_{j+1}^{q}
    = U_{j}^{p} B_{j+1}^{q}, \qquad 
    B_{j+1}^{q} := \Sigma V_{j+1}^{q}.
\end{equation}
Diagrammatically, 
\begin{equation}
    \widetilde{\Theta}^{p,q}
    ~ = ~
    \begin{tikzpicture}[anchor=center, baseline={(0,-0.1)}]
        \def\a{0.36};

        \lcanon{Green!40}{2*\a}{1.0}{0}
        \node at (0, 0) {$B_{j}$};
        \draw (1.75*\a, 0) -- (2.5*\a, 0);
        \draw (-1.0*\a, 0) -- (-2.0*\a, 0);
        \node[align=center, font=\scriptsize] at (-2.7*\a, 1.2*\a) {left \\ bond};
        \draw (0, -1.0*\a) -- (0, -2.0*\a);

        \begin{scope}[shift = {(4.0*\a, 0)}]
            \roundedsquare{SkyBlue!40}{\a}{\a}
            \node at (0, 0) {$A_{j+1}$};            
            \draw (1.0*\a, 0) -- (2.0*\a, 0);
            \node[align=center, font=\scriptsize] at (2.7*\a, 1.2*\a) {right \\ bond};
            \draw (-1.0*\a, 0) -- (-2.0*\a, 0);
            \draw (0, -1.0*\a) -- (0, -2.0*\a);
        \end{scope}

        \begin{scope}[shift = {(2.0*\a, -2.8*\a)}]
            \roundedsquare{red!40}{3.0*\a}{0.8*\a}
            \node at (0, 0) {$\mathcal{U}_{j,j+1}$};
            \draw (-2.0*\a, -0.8*\a) -- (-2.0*\a, -1.8*\a);
            \draw (2.0*\a, -0.8*\a) -- (2.0*\a, -1.8*\a);
            \node at (-2.0*\a, -2.3*\a) {$p$};
            \node at (2.0*\a, -2.3*\a) {$q$};
        \end{scope}
    \end{tikzpicture}
    ~ \overset{\mathrm{SVD}}{=} ~
    \begin{tikzpicture}[anchor=center, baseline={(0,-0.1)}]
        \def\a{0.36};

        \roundedsquare{Green!40}{\a}{\a}
        \node at (0.1*\a, 0) {$U_{j}$};
        \draw (1.0*\a, 0) -- (2.0*\a, 0);
        \draw (-1.0*\a, 0) -- (-2.0*\a, 0);
        \draw (0, -1.0*\a) -- (0, -2.0*\a);
        \node at (0, -2.5*\a) {$p$};

        \begin{scope}[shift = {(3.0*\a, 0)}]
            \diatensor{Orange!30}{1.0*\a}
            \node at (0, 0) {$\Sigma$};
            \draw (1.3*\a, 0) -- (1.8*\a, 0);
        \end{scope}    

        \begin{scope}[shift = {(6.0*\a, 0)}]
            \roundedsquare{Green!40}{\a}{\a}
            \node at (0.1*\a, 0) {$V_{j+1}$};
            \draw (1.0*\a, 0) -- (2.0*\a, 0);
            \draw (-1.0*\a, 0) -- (-2.0*\a, 0);
            \draw (0, -1.0*\a) -- (0, -2.0*\a);
            \node at (0, -2.5*\a) {$q$};
        \end{scope}
    \end{tikzpicture}
    ~ = ~
    \begin{tikzpicture}[anchor=center, baseline={(0,-0.1)}]
        \def\a{0.36};

        \roundedsquare{Green!40}{\a}{\a}
        \node at (0.1*\a, 0) {$U_{j}$};
        \draw (1.0*\a, 0) -- (2.0*\a, 0);
        \draw (-1.0*\a, 0) -- (-2.0*\a, 0);
        \draw (0, -1.0*\a) -- (0, -2.0*\a);
        \node at (0, -2.5*\a) {$p$};

        \begin{scope}[shift = {(3.0*\a, 0)}]
            \lcanon{Green!40}{2*\a}{1.0}{0}
            \node at (0.2*\a, 0) {$B_{j+1}$};            
            \draw (1.75*\a, 0) -- (2.5*\a, 0);
            \draw (-1.0*\a, 0) -- (-2.0*\a, 0);
            \draw (0, -1.0*\a) -- (0, -2.0*\a);
            \node at (0, -2.5*\a) {$q$};
        \end{scope}
    \end{tikzpicture}
    ~ = ~ U_{j}^{p} B_{j+1}^{q}.
\end{equation}
Here, $\Sigma$ is a diagonal matrix, and $U_{j}$ and $V_{j+1}$ satisfy $\sum_{p} (U_{j}^{p})^{\dagger} U_{j}^{p} = \mathds{1}$ and $ \sum_{q} V_{j+1}^{q} (V_{j+1}^{q})^{\dagger} = \mathds{1}$, where $\mathds{1}$ denotes the identity operator with the same dimension as $\Sigma$.
Below, we establish that $B_{j+1}$ has the canonical form.
\begin{proposition}
\label{prop: propagation of sum gate}
    Let $B_{j}$ be a $(\mathbf{u}, \mathbf{v}; a)$-canonical form tensor.
    For the tensors $B_{j}$ and $A_{j+1}$, after applying $\mathcal{U}_{j,j+1}$ and performing an SVD, the resulting tensor $B_{j+1}$ is a $(\mathbf{u}', \mathbf{v}'; a')$-canonical form tensor such that $a' = (2-a)/3$.
    Furthermore, the SVD yields two nonzero singular values, $\sqrt{\lambda_{+}}$ and $\sqrt{\lambda_{-}}$, where
    \begin{equation}
        \lambda_{\pm} = \frac{1}{2} \left( 1 \pm \frac{1}{3} \sqrt{8 + (2a-1)^{2}} \right).
    \end{equation}
\end{proposition}

\begin{proof}

From the definitions, the tensor $\Theta^{p,q} = B_{j}^{p} A_{j+1}^{q}$ can be directly computed as
\begin{equation}
\label{components of Theta}
    \Theta^{0,1} = \sqrt{\frac{1}{3}} \mathbf{u} \mathbf{e}_{\downarrow}^{\dagger}, \qquad
    \Theta^{0,2} = -\sqrt{\frac{2}{3}} \mathbf{u} \mathbf{e}_{\uparrow}^{\dagger}, \qquad
    \Theta^{1,0} = \sqrt{\frac{2}{3}} \mathbf{v} \mathbf{e}_{\downarrow}^{\dagger}, \qquad
    \Theta^{1,1} = -\sqrt{\frac{1}{3}} \mathbf{v} \mathbf{e}_{\uparrow}^{\dagger}.
\end{equation}
Additionally, $\mathcal{U}_{j,j+1}$ acts as $\mathcal{U}_{j,j+1} |{p, q}\rangle = |{p, p+q+2}\rangle$, as shown in Eq.~\eqref{def of sum gate}.
Thus, we compute the tensor $\widetilde{\Theta}^{p,q}$ as
\begin{equation}
\label{components of tilde Theta}
    \widetilde{\Theta}^{0, 0}
    = \sqrt{\frac{1}{3}} \mathbf{u} \mathbf{e}_{\downarrow}^{\dagger}, \qquad
    \widetilde{\Theta}^{0, 1}
    = -\sqrt{\frac{2}{3}} \mathbf{u} \mathbf{e}_{\uparrow}^{\dagger}, \qquad
    \widetilde{\Theta}^{1, 0}
    = \sqrt{\frac{2}{3}} \mathbf{v} \mathbf{e}_{\downarrow}^{\dagger}, \qquad
    \widetilde{\Theta}^{1, 1}
    = -\sqrt{\frac{1}{3}} \mathbf{v} \mathbf{e}_{\uparrow}^{\dagger},    
\end{equation}
while all remaining $p, q$ components vanish.
We then reshape $\widetilde{\Theta}$ to perform an SVD across the bond.
Explicitly, we use the composite indices $(p, \alpha)$ and $(q, \beta)$ for the rows and columns, respectively, where $\alpha$ and $\beta$ denote the indices of the left and right bonds.
In this way, we regard $\widetilde{\Theta}$ as a $6 \times 6$ matrix.
As shown in Eq.~\eqref{components of tilde Theta}, the tensor $\widetilde{\Theta}$ has non-zero elements only for
$\{ (p=0, \alpha=\hat{\mathbf{u}}), (p=1, \alpha=\hat{\mathbf{v}}) \}$ 
and
$\{ (q=0, \beta=\mathbf{e}_{\downarrow}^{\dagger}), (q=1, \beta=\mathbf{e}_{\uparrow}^{\dagger}) \}$,
where
$\hat{\mathbf{u}} = \frac{1}{\sqrt{a}} \mathbf{u}$ and $\hat{\mathbf{v}} = \frac{1}{\sqrt{1-a}} \mathbf{v}$ denote the normalized vectors.
In this basis, the nonzero part of $\widetilde{\Theta}$ can be written as
\begin{equation}
\label{2 by 2 Theta}
    \widetilde{\Theta}
    ~=~
    \raisebox{0.50em}
    {$
    \begin{array}{@{}c@{\hspace{0.35em}}c@{}}
        &
        \begin{array}{@{}c@{\hspace{1.7em}}c@{}}
            (0, \mathbf{e}_{\downarrow}^{\dagger}) & (1, \mathbf{e}_{\uparrow}^{\dagger})
        \end{array}
        \\[0.1em]
        \begin{array}{@{}c@{}}
            (0, \hat{\mathbf{u}}) \\[0.45em] (1, \hat{\mathbf{v}})
        \end{array}
        &
        \begin{pmatrix}
            \sqrt{\frac{a}{3}} & -\sqrt{\frac{2a}{3}} \\
            \sqrt{\frac{2(1-a)}{3}} & -\sqrt{\frac{1-a}{3}}
        \end{pmatrix}
    \end{array}
    $}.
\end{equation}
Then, the matrix $\widetilde{\Theta}$ in Eq.~\eqref{2 by 2 Theta} admits an SVD $\widetilde{\Theta} = U_{j} \Sigma V_{j+1}$, where the two nonzero squared singular values are given by
\begin{equation}
    \Sigma^{2} = \begin{pmatrix} \lambda_{+} & \\ & \lambda_{-} \end{pmatrix}, \qquad
    \lambda_{\pm} = \frac{1}{2} \left( 1 \pm \frac{1}{3} \sqrt{8 + (2a-1)^{2}} \right).
\end{equation}
Next, we reshape the $2 \times 6$ matrix $\Sigma V_{j+1}$ into the tensor $B_{j+1}$.
Explicitly, $(\Sigma V_{j+1})_{\alpha', (q, \beta)} = (B_{j+1}^{q})_{\alpha', \beta}$, where $\alpha'$ denotes the row index of $\Sigma$, i.e., the new left-bond index.
The nonzero components of $\Sigma V_{j+1}$ are supported only on the columns labeled by
$\{ (q=0, \beta=\mathbf{e}_{\downarrow}^{\dagger}), (q=1, \beta=\mathbf{e}_{\uparrow}^{\dagger}) \}$.
Hence, all columns with $q=2$ vanish; we find that $B_{j+1}^{2} = 0$.
Moreover, for $q=0$ and $q=1$, the nonzero components are supported on the right-bond states $\mathbf{e}_{\downarrow}^{\dagger}$ and $\mathbf{e}_{\uparrow}^{\dagger}$, respectively.
In other words, $B_{j+1}^{0} \mathbf{e}_{\uparrow} = 0$ and $B_{j+1}^{1} \mathbf{e}_{\downarrow} = 0$.
Therefore, $B_{j+1}^{q}$ can be written as
\begin{equation}
    \label{bn form after SVD}
    B_{j+1}^{0} = \mathbf{u}' \mathbf{e}_{\downarrow}^{\dagger}, \qquad
    B_{j+1}^{1} = \mathbf{v}' \mathbf{e}_{\uparrow}^{\dagger}, \qquad
    B_{j+1}^{2} = 0,
\end{equation}
for some $\mathbf{u}', \mathbf{v}' \in \mathbb{C}^{2}$.

To determine $\left\| \mathbf{u}' \right\|$, $\left\| \mathbf{v}' \right\|$, and $\mathbf{u}'^{\dagger} \mathbf{v}'$, we use the identity
\begin{equation}
    \left( \widetilde{\Theta}^{\dagger} \widetilde{\Theta} \right)^{q, q'}
    = \sum_{p} (\widetilde{\Theta}^{p,q})^{\dagger} \widetilde{\Theta}^{p,q'}
    = (B_{j+1}^{q})^{\dagger} U_{j}^{\dagger} U_{j} B_{j+1}^{q'}
    = (B_{j+1}^{q})^{\dagger} B_{j+1}^{q'},
\end{equation}
where the superscripts $q,q'$ indicate the block indices associated with the physical indices.
More explicitly, the nonzero blocks are
\begin{equation}
\label{Theta dg Theta}
    \left( \widetilde{\Theta}^{\dagger} \widetilde{\Theta} \right)^{q, q'}
    =
    \raisebox{0.50em}
    {$
    \begin{array}{@{}c@{\hspace{0.35em}}c@{}}
        &
        \begin{array}{@{}c@{\hspace{2.0em}}c@{}}
            q'=0 & q'=1
        \end{array}
        \\[0.1em]
        \begin{array}{@{}c@{}}
            q=0 \\[0.45em] q=1
        \end{array}
        &
        \begin{pmatrix}
            \frac{2-a}{3} \mathbf{e}_{\downarrow} \mathbf{e}_{\downarrow}^{\dagger}
            & -\frac{\sqrt{2}}{3} \mathbf{e}_{\downarrow} \mathbf{e}_{\uparrow}^{\dagger} \\
            -\frac{\sqrt{2}}{3} \mathbf{e}_{\uparrow} \mathbf{e}_{\downarrow}^{\dagger}
            & \frac{1+a}{3} \mathbf{e}_{\uparrow} \mathbf{e}_{\uparrow}^{\dagger}
        \end{pmatrix}
    \end{array}
    $}
    = \underbrace{\begin{pmatrix}
        \left\| \mathbf{u}' \right\|^{2} \mathbf{e}_{\downarrow} \mathbf{e}_{\downarrow}^{\dagger}
        & (\mathbf{u}'^{\dagger} \mathbf{v}') \mathbf{e}_{\downarrow} \mathbf{e}_{\uparrow}^{\dagger} \\
        (\mathbf{v}'^{\dagger} \mathbf{u}') \mathbf{e}_{\uparrow} \mathbf{e}_{\downarrow}^{\dagger}
        & \left\| \mathbf{v}' \right\|^{2} \mathbf{e}_{\uparrow} \mathbf{e}_{\uparrow}^{\dagger}
    \end{pmatrix}}_{(B_{j+1}^{q})^{\dagger} B_{j+1}^{q'}}.
\end{equation}
Thus, we obtain that
\begin{equation}
    \| \mathbf{u}' \|^{2} = \frac{2-a}{3}, \qquad
    \| \mathbf{v}' \|^{2} = \frac{1+a}{3}, \qquad
    \mathbf{u}'^{\dagger} \mathbf{v}' = - \frac{\sqrt{2}}{3}.
\end{equation}
By comparing this form with Eqs.~\eqref{bn form} and~\eqref{condition of uv for B}, we find that $B_{j+1}$ is in $(\mathbf{u}', \mathbf{v}'; a')$-canonical form.

\end{proof}

\subsection{Optimal Clifford disentangler for the AKLT state under a left-to-right sweep}

Here, we show that $\mathcal{U}_{j,j+1}$ is the optimal Clifford disentangler for the bond between a canonical-form tensor $B_{j}$ and the AKLT tensor $A_{j+1}$.
This provides an analytic verification that the KW transformation $U_{\mathrm{KW}}$ is the optimal disentangler for the AKLT state under a left-to-right sweep.
Since local gates do not change the entanglement, we quotient out this local freedom and restrict the search to $90$ inequivalent representatives, rather than scanning the full two-qutrit Clifford group~\cite{Harper2025GCAMPS}.

For the bond between the $j$-th and $(j+1)$-th sites, we take the two tensors to be a $(\mathbf{u},\mathbf{v}; a)$-canonical form tensor $B_{j}$ and the AKLT tensor $A_{j+1}$.
Following the procedure used to obtain Eqs.~\eqref{components of tilde Theta} and~\eqref{2 by 2 Theta}, we apply a two-qutrit Clifford gate $C_{\nu}$ and construct the $6 \times 6$ matrix $\widetilde{\Theta}(\nu,\mathbf{u},\mathbf{v})$.
Here, $\widetilde{\Theta}$ depends on the Clifford gate $C_{\nu}$ and $(\mathbf{u},\mathbf{v})$ of $B_{j}$.
More explicitly, according to Eq.~\eqref{components of Theta}, $\widetilde{\Theta}(\nu,\mathbf{u},\mathbf{v})$ can be written as
\begin{equation}
\label{tilde theta for general nu}
    \widetilde{\Theta}^{p,q} (\nu,\mathbf{u},\mathbf{v})
    = \mathbf{u} (\mathbf{x}_{\nu}^{p,q})^{\dagger}
    + \mathbf{v} (\mathbf{y}_{\nu}^{p,q})^{\dagger},
\end{equation}
where $\mathbf{x}_{\nu}^{p,q}$ and $\mathbf{y}_{\nu}^{p,q}$ are vectors determined by the Clifford gate $C_{\nu}$, and are independent of $\mathbf{u}$ and $\mathbf{v}$.

To compute its singular values, we define the Hermitian matrix
\begin{equation}
\label{def of 6 by 6 F}
    F(\nu, \mathbf{u}, \mathbf{v})
    := \left( \widetilde{\Theta}(\nu,\mathbf{u},\mathbf{v}) \right)^{\dagger}
    \widetilde{\Theta}(\nu,\mathbf{u},\mathbf{v}),
\end{equation}
whose eigenvalues are the squared singular values of $\widetilde{\Theta}$.
We show that each entry of $F(\nu, \mathbf{u}, \mathbf{v})$ depends only on the parameter $a$, not on the specific choice of $\mathbf{u}$ and $\mathbf{v}$.
Using Eq.~\eqref{tilde theta for general nu}, we obtain
\begin{equation}
    \left( \widetilde{\Theta}^{\dagger} \widetilde{\Theta} \right)^{q, q'}
    = \sum_{p} (\widetilde{\Theta}^{p,q})^{\dagger} \widetilde{\Theta}^{p,q'}
    = \sum_{p}
    \left[
        (\mathbf{u}^{\dagger} \mathbf{u})
        \mathbf{x}_{\nu}^{p,q} (\mathbf{x}_{\nu}^{p,q'})^{\dagger}
        + (\mathbf{u}^{\dagger} \mathbf{v})
        \mathbf{x}_{\nu}^{p,q} (\mathbf{y}_{\nu}^{p,q'})^{\dagger}
        + (\mathbf{v}^{\dagger} \mathbf{u})
        \mathbf{y}_{\nu}^{p,q} (\mathbf{x}_{\nu}^{p,q'})^{\dagger}
        + (\mathbf{v}^{\dagger} \mathbf{v})
        \mathbf{y}_{\nu}^{p,q} (\mathbf{y}_{\nu}^{p,q'})^{\dagger}
    \right].
\end{equation}
Thus, $F(\nu, \mathbf{u}, \mathbf{v})$ depends only on $\| \mathbf{u} \|^{2}$, $\| \mathbf{v} \|^{2}$, $\mathbf{u}^{\dagger}\mathbf{v}$, and $\mathbf{v}^{\dagger}\mathbf{u}$.
Since $B_{j}$ satisfies the condition in Eq.~\eqref{condition of uv for B}, we confirm the above claim; therefore, we can write $F(\nu, a) := F(\nu, \mathbf{u}, \mathbf{v})$.

\begin{table}[h]
    \centering
    \renewcommand{\arraystretch}{1.8}
    \begin{tabular}{c|c|l}
        \hline
        type & mult. & $\chi_{\nu}(x;a) = \det(xI - F(\nu, a))$ \\
        \hline
        $\mathcal{T}_{0}$ & $1$ &
        $x^{4}\left(x^{2}-x+\dfrac{a(1-a)}{9}\right)$
        \\
        $\mathcal{T}_{1}$ & $3$ &
        $x^{4}\left(x^{2}-x+\dfrac{9a(1-a)+10}{81}\right)$
        \\
        $\mathcal{T}_{2}$ & $2$ &
        $x^{4}\left(x^{2}-x+\dfrac{7a(1-a)}{9}\right)$
        \\
        $\mathcal{T}_{3}$ & $3$ &
        $x^{4}\left(x^{2}-x+\dfrac{9a(1-a)+16}{81}\right)$
        \\
        $\mathcal{T}_{4}$ & $6$ &
        $x^{4}(x-a)(x-(1-a))$
        \\
        $\mathcal{T}_{5}$ & $6$ &
        $x^{2}\left(x-\dfrac{a}{3}\right)
        \left(x-\dfrac{1-a}{3}\right)
        \left(x^{2}-\dfrac{2}{3}x+\dfrac{36a(1-a)-8}{81}\right)$
        \\
        $\mathcal{T}_{6}$ & $9$ &
        $x^{2}
        \left(x^{2}-\dfrac{a+1}{3}x+\dfrac{18a(1-a)-4}{81}\right)
        \left(x^{2}-\dfrac{2-a}{3}x+\dfrac{18a(1-a)-4}{81}\right)$
        \\
        $\mathcal{T}_{7}$ & $18$ &
        $\left(x-\dfrac{3a-1}{27}\right)
        \left(x-\dfrac{2-3a}{27}\right)
        \left(x^{2}-\dfrac{13}{27}x+\dfrac{9a(1-a)-2}{729}\right)^{2}$
        \\
        $\mathcal{T}_{8}$ & $36$ &
        $\left(x^{2}-\dfrac{13}{27}x+\dfrac{63a(1-a)-14}{729}\right)
        \left(x^{2}-\dfrac{7}{27}x+\dfrac{63a(1-a)-14}{729}\right)^{2}$
        \\
        $\mathcal{T}_{9}$ & $6$ &
        $x^{2}\left(x-\dfrac{2a}{3}\right)
        \left(x-\dfrac{2(1-a)}{3}\right)
        \left(x^{2}-\dfrac{1}{3}x+\dfrac{9a(1-a)-2}{81}\right)$
        \\
        \hline
    \end{tabular}
    \caption{
        Characteristic polynomials of $F(\nu, a)$ for each type of Clifford gates.
        Here, ``mult.'' denotes the multiplicity, i.e., the number of Clifford gates in each type.
    }
    \label{characteristic polynomial}
\end{table}

Next, we compute the characteristic polynomials of the matrices $F(\nu, a)$ for the 90 Clifford gates $C_{\nu}$.
We denote the characteristic polynomial as a function of $x$ by $\chi_{\nu}(x; a)$.
Then, we observe that the 90 values of $\nu$ are partitioned into 10 types according to $\chi_{\nu}(x;a)$, so that only 10 distinct characteristic polynomials appear.
We label the 10 types by $\mathcal{T}_{t}$ for $t = 0,1,\ldots,9$, choosing $\mathcal{T}_{0}$ to be the type that contains $\mathcal{U}_{j,j+1}$, i.e., the \texttt{SUM} gate type.
In Table~\ref{characteristic polynomial}, we provide the explicit characteristic polynomials for all Clifford-gate types.

Using characteristic polynomials in Table~\ref{characteristic polynomial}, we derive the following result:

\begin{lemma}
\label{lem: sum is optimal}
    For a given $a \in [4/9, 2/3]$, we consider a $(\mathbf{u}, \mathbf{v}; a)$-canonical form tensor $B_{j}$.
    Then, $\mathcal{U}_{j,j+1}$ is the optimal disentangler between $B_{j}$ and the AKLT tensor $A_{j+1}$, namely, the gate minimizes the entanglement entropy across the bond.
\end{lemma}

\begin{proof}
For each type $\mathcal{T}_{t}$, we denote the largest root of $\chi_{\nu \in \mathcal{T}_{t}}(x,a)$ as $\lambda_{\max}^{t}(a)$.
Then, in the interval $a \in [4/9, 2/3]$, we compute the infimum of $\lambda_{\max}^{t=0}(a)$ as
\begin{equation}
    \inf_{a \in [4/9,2/3]} \lambda_{\max}^{t=0}(a)
    = \frac{1}{2} \left( 1+ \frac{2\sqrt{2}}{3} \right).
\end{equation}
For the 9 non-\texttt{SUM} types, the supremum is calculated as
\begin{equation}
\renewcommand{\arraystretch}{1.6}
\begin{array}{c|ccccccccc}
\text{type}
& \mathcal{T}_{1}
& \mathcal{T}_{2}
& \mathcal{T}_{3}
& \mathcal{T}_{4}
& \mathcal{T}_{5}
& \mathcal{T}_{6}
& \mathcal{T}_{7}
& \mathcal{T}_{8}
& \mathcal{T}_{9}
\\
\hline
\displaystyle \sup_{a \in [4/9,2/3]} \lambda_{\max}^{t}(a)
& \displaystyle \frac{1}{2}\left(1+\frac{\sqrt{33}}{9}\right)
& \displaystyle \frac{7}{9}
& \displaystyle \frac{2}{3}
& \displaystyle \frac{2}{3}
& \displaystyle \frac{2}{3}
& \displaystyle \frac{5}{9}
& \displaystyle \frac{13}{27}
& \displaystyle \frac{13}{27}
& \displaystyle \frac{4}{9}
\end{array}.
\end{equation}

To compute the entanglement entropy, for each type $\mathcal{T}_{t}$, we denote the Schmidt probability vector by $p_{\mathrm{Sch}}^{t}$.
Here, $p_{\mathrm{Sch}}^{t}$ is defined as the vector of roots of $\chi_{\nu \in \mathcal{T}_{t}}(x,a)$, arranged in non-increasing order.
Then, we find that all $p_{\mathrm{Sch}}^{t}$ for $1\leq t \leq 9$ are majorized by $p_{\mathrm{Sch}}^{t=0}$; explicitly, for any $k \geq 1$,
\begin{equation}
\label{majorization condition}
    \sum_{r=1}^{k} (p_{\mathrm{Sch}}^{t=0})_{r}
    \geq \sum_{r=1}^{k} (p_{\mathrm{Sch}}^{t \neq 0})_{r},
\end{equation}
where $(p_{\mathrm{Sch}}^{t})_{r}$ denotes the $r$-th component of $p_{\mathrm{Sch}}^{t}$.
For $k=1$, it follows from
\begin{equation}
    (p_{\mathrm{Sch}}^{t=0})_{1}
    \geq \inf_{a \in [4/9,2/3]} \lambda_{\max}^{t=0}(a)
    = \frac{1}{2} \left( 1+ \frac{2\sqrt{2}}{3} \right)
    \geq \frac{1}{2}\left(1+\frac{\sqrt{33}}{9}\right)
    = \sup_{a \in [4/9,2/3]} \lambda_{\max}^{t \neq 0}(a)
    \geq (p_{\mathrm{Sch}}^{t \neq 0})_{1}.
\end{equation}
Furthermore, by Proposition~\ref{prop: propagation of sum gate}, $(p_{\mathrm{Sch}}^{t=0})_{2} = 1 - (p_{\mathrm{Sch}}^{t=0})_{1}$, so the left-hand side of Eq.~\eqref{majorization condition} equals 1 for $k \geq 2$, and the inequality follows from the normalization of $p_{\mathrm{Sch}}^{t}$.
Therefore, by the majorization condition in Eq.~\eqref{majorization condition}, a Clifford gate of type $\mathcal{T}_{0}$ minimizes the entanglement entropy for any $a \in [4/9,2/3]$.

\end{proof}

Therefore, the KW transformation $U_{\mathrm{KW}}$ is optimal for disentangling the AKLT state under a left-to-right sweep.

\begin{theorem}
    For the AKLT state $|\mathrm{AKLT}_{ \mathbf{L}, \mathbf{R} }\rangle$ with $\mathbf{L} = \mathbf{e}_{\uparrow}$ and $\mathbf{R} = \mathbf{e}_{\uparrow}$, we sequentially determine the optimal two-qutrit disentangling gate on the bond between the $j$-th and $(j+1)$-th sites, from $j=1$ to $j=N-1$.
    Then, for each $j$, the optimal choice is $\mathcal{U}_{j,j+1}$.
    Equivalently, the KW transformation $U_{\mathrm{KW}}$ in Eq.~\eqref{KW unitary} is the optimal Clifford circuit under this left-to-right sweep.
\end{theorem}

\begin{proof}

We first show that Lemma~\ref{lem: sum is optimal} also applies to the bond between the first and second sites.
As shown above, by using a normalized vector $\mathbf{w} \in \mathbb{C}^{2}$, the left-boundary tensor can be written as $\mathbf{e}_{\uparrow}^{\dagger} A_{1} = \mathbf{w}^{\dagger} B_{1}$,
where $B_{1} = \mathbf{w} \mathbf{e}_{\uparrow}^{\dagger} A_{1}$.
Here, $B_{1}$ is in $(\mathbf{u} = \sqrt{2/3} \mathbf{w}, \mathbf{v} = -\sqrt{1/3} \mathbf{w}; a_{1} = 2/3)$-canonical form, i.e., $\mathbf{e}_{\uparrow}^{\dagger} A_{1}$ is obtained from $B_{1}$ by applying the row vector $\mathbf{w}^{\dagger}$ from the left.
Hence, as in the proof of Lemma~\ref{lem: sum is optimal}, we compute the singular values across the bond and find that $\mathcal{U}_{1,2}$ minimizes the entanglement.
Consequently, Proposition~\ref{prop: propagation of sum gate} gives the canonical form tensor $B_{2}$, which again satisfies the condition in Eq.~\eqref{condition of uv for B} with the parameter $a_{2} = \frac{2-a_{1}}{3} = 4/9$.

Next, we show that Lemma~\ref{lem: sum is optimal} and Proposition~\ref{prop: propagation of sum gate} can be recursively applied for $1 \leq j < N-1$.
Suppose that the $j$-th site tensor is a canonical form tensor $B_{j}$ with parameter $a_{j}$ satisfying $4/9 \leq a_{j} \leq 2/3$.
Then, by these Lemma and Proposition, $\mathcal{U}_{j,j+1}$ is optimal, and the resulting tensor $B_{j+1}$ is again in canonical form with parameter $a_{j+1} = \frac{2-a_{j}}{3}$.
Therefore, it remains to show that the sequence defined by $a_{j+1} = \frac{2-a_{j}}{3}$ with initial condition $a_{1} = 2/3$ stays in the interval $[4/9, 2/3]$ for all $j \geq 1$.
This follows from the explicit solution of the recurrence relation,
\begin{equation}
    a_{j} = \frac{1}{2} + \frac{(-1)^{j-1}}{2 \cdot 3^{j}}.
\end{equation}

It remains to determine the optimal disentangler for the bond between the $(N-1)$-th and $N$-th sites.
In particular, we need to find the optimal disentangler between $B_{N-1}$ and $A_{N} \mathbf{e}_{\uparrow}$.
Following the same procedure as in the proof of Lemma~\ref{lem: sum is optimal}, the resulting singular value spectrum shows that $\mathcal{U}_{N-1, N}$ gives the minimum entanglement.

\end{proof}

\begin{figure}[h]
    \centering
    \includegraphics[scale=1]{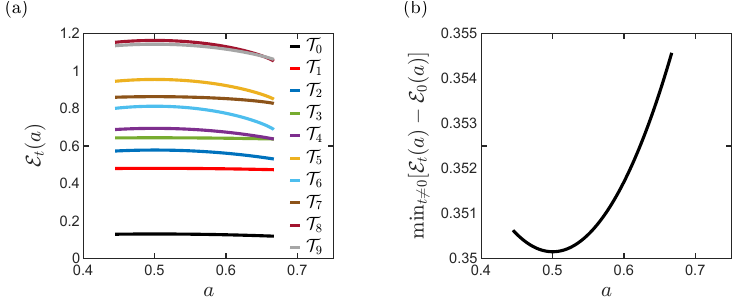}
    \caption{
    Entanglement spectra generated by the Clifford equivalence classes.
    (a) Entanglement entropy $\mathcal{E}_{t}(a)$ obtained from the roots of the characteristic polynomial for the 10 types $\mathcal{T}_{t}$ over $4/9 \leq a \leq 2/3$.
    (b) Entanglement gap $\min_{t \neq 0}[\mathcal{E}_{t}(a)-\mathcal{E}_{0}(a)]$.
    }
    \label{Fig: disentangling_gap}
\end{figure}

We check the optimality of $\mathcal{U}_{j,j+1}$ in a more direct way.
Using all the roots of $\chi_{\nu \in \mathcal{T}_{t}}(x;a)$ listed in Table~\ref{characteristic polynomial}, we compute the entanglement entropy $\mathcal{E}_{t}(a)$ for each type $\mathcal{T}_{t}$.
Figure~\ref{Fig: disentangling_gap}(a) shows $\mathcal{E}_{t}(a)$ in the range $4/9 \leq a \leq 2/3$.
It clearly shows that the type $\mathcal{T}_{0}$, which includes $\mathcal{U}_{j,j+1}$, yields the minimal entanglement throughout the entire regime.

In Fig.~\ref{Fig: disentangling_gap}(b), we plot the entanglement gap between the optimal type $\mathcal{T}_{0}$ and the next-best type.
As shown in Fig.~\ref{Fig: disentangling_gap}(a), the next-best type is $\mathcal{T}_{1}$.
Throughout the range $4/9 \leq a \leq 2/3$, we obtain the minimum value of the gap as
\begin{equation}
    \min_{a \in [4/9, 2/3]} \min_{t \neq 0} [\mathcal{E}_{t}(a) - \mathcal{E}_{0}(a)]
    = \mathcal{E}_{1}(1/2) - \mathcal{E}_{0}(1/2)
    = H\left( \frac{1}{2} + \frac{2\sqrt{2}}{9} \right)
    - H\left( \frac{1}{2} + \frac{\sqrt{2}}{3} \right)
    \approx 0.35,
\end{equation}
where $H(x)$ denotes the binary Shannon entropy.
Compared with $\log 2 \approx 0.69$, the maximum possible entanglement for bond dimension $2$, this gap is substantial.
This significant gap indicates that the optimality of $\mathcal{U}_{j,j+1}$ is robust against small perturbations.
In the numerical CAMPS-based DMRG process, even if $B_{j}$ slightly deviates from the condition in Eq.~\eqref{condition of uv for B} or $A_{j+1}$ deviates from the exact AKLT tensor, the optimal disentangler can still be identified as $\mathcal{U}_{j,j+1}$.

Furthermore, we verify that applying a Clifford gate to two adjacent bulk AKLT tensors $A_{j}$ and $A_{j+1}$ cannot reduce the entanglement.
For all 90 two-qutrit Clifford gates $C_{\nu}$, we compute the resulting entanglement $\mathcal{E}_{\nu}^{\mathrm{bulk}}$.
In the bulk case, the identity gate $\nu = \mathrm{id}$ minimizes the entanglement with $\mathcal{E}_{\mathrm{id}}^{\mathrm{bulk}}=\log 2$, and the entanglement gap to the second-smallest value is evaluated as
\begin{equation}
    \min_{\nu \neq \mathrm{id}}
    \left[
        \mathcal{E}_{\nu}^{\mathrm{bulk}}
        - \mathcal{E}_{\mathrm{id}}^{\mathrm{bulk}}
    \right]
    = H\left( \frac{1}{2} + \frac{\sqrt{73}}{18} \right)
    \approx 0.12.
\end{equation}
This behavior differs from the boundary case in Lemma~\ref{lem: sum is optimal}: the bond between $\mathbf{e}_{\uparrow}^{\dagger} A_{1}$ and $A_{2}$ can be disentangled, whereas a bulk bond cannot.
This observation shows that the boundary effect is essential for disentangling; the disentangling procedure must start from the boundaries.

\subsection{Right-to-left update after applying the KW unitary}

Finally, we examine the right-to-left update of the KW-transformed AKLT state under Clifford-gate disentangling.
By showing that no $2$-local Clifford gate can further reduce the entanglement, we confirm that the KW unitary $U_{\mathrm{KW}}$ is the optimal Clifford circuit for this iterative disentangling method.

After disentangling with $U_{\mathrm{KW}}$, we obtain the following MPS representation of the KW-transformed AKLT state:
\begin{equation}
\label{def of KW AKLT MPS}
\begin{gathered}
    U_{\mathrm{KW}} |\mathrm{AKLT}_{ \mathbf{e}_{\uparrow}, \mathbf{e}_{\uparrow} }\rangle
    = \sum_{ \{s_{j}\} }
    \mathbf{e}_{\uparrow}^{\dagger}
    \widetilde{A}_{1}^{s_{1}}
    \widetilde{A}_{2}^{s_{2}} \widetilde{A}_{3}^{s_{3}} \cdots \widetilde{A}_{N-1}^{s_{N-1}}
    \widetilde{A}_{N}^{s_{N}}
    \mathbf{e}_{\uparrow}
    |s_{1} s_{2} \cdots s_{N} \rangle, \\
    \widetilde{A}_{j}^{0} =
    \frac{1}{\sqrt{3}}
    \begin{pmatrix} \sqrt{2} \\ 1 \end{pmatrix}
    \begin{pmatrix} 0 & 1 \end{pmatrix}, \qquad
    \widetilde{A}_{j}^{1} =
    - \frac{1}{\sqrt{3}}
    \begin{pmatrix} 1 \\ \sqrt{2} \end{pmatrix}
    \begin{pmatrix} 1 & 0 \end{pmatrix}, \qquad
    \widetilde{A}_{j}^{2} = 0.
\end{gathered}
\end{equation}
Then, as in the previous subsection, we apply the disentangling procedure to the bond between the two tensors $\widetilde{A}_{j}$ and $\widetilde{A}_{j+1}$.
For each Clifford gate $C_{\nu}$, we construct the corresponding $6 \times 6$ Hermitian matrix $\widetilde{F}(\nu)$, as in Eq.~\eqref{def of 6 by 6 F}.
In this case, since $\Tr[\widetilde{F}(\nu)] = 2$, we consider $\widetilde{F}_{\mathrm{norm}}(\nu) := \frac{1}{2} \widetilde{F}(\nu)$ and compute its characteristic polynomial $\widetilde{\chi}_{\nu}(x)$.
Analogous to Table~\ref{characteristic polynomial}, all $90$ Clifford gates are classified into 10 types $\widetilde{\mathcal{T}}_{t}$.
Below, Table~\ref{characteristic polynomial for tilde} lists the explicit forms of $\widetilde{\chi}_{\nu}(x)$ for these types.
The type $\widetilde{\mathcal{T}}_{0}$ contains the identity gate.
\begin{table}[h]
    \centering
    \renewcommand{\arraystretch}{1.8}
    \begin{tabular}{c|c|l}
        \hline
        type & mult. & $\widetilde{\chi}_{\nu}(x) = \det(xI - \widetilde{F}_{\mathrm{norm}}(\nu))$ \\
        \hline
        $\widetilde{\mathcal{T}}_{0}$ & $1$ &
        $x^{4}\left(x^{2}-x+\dfrac{1}{36}\right)$
        \\
        $\widetilde{\mathcal{T}}_{1}$ & $3$ &
        $x^{4}\left(x^{2}-x+\dfrac{49}{324}\right)$
        \\
        $\widetilde{\mathcal{T}}_{2}$ & $2$ &
        $x^{4}\left(x^{2}-x+\dfrac{7}{36}\right)$
        \\
        $\widetilde{\mathcal{T}}_{3}$ & $3$ &
        $x^{4}\left(x^{2}-x+\dfrac{73}{324}\right)$
        \\
        $\widetilde{\mathcal{T}}_{4}$ & $6$ &
        $x^{4}\left(x-\dfrac{1}{2}\right)^{2}$
        \\
        $\widetilde{\mathcal{T}}_{5}$ & $6$ &
        $x^{2}\left(x-\dfrac{1}{6}\right)^{2}
        \left(x^{2}-\dfrac{2}{3}x+\dfrac{1}{81}\right)$
        \\
        $\widetilde{\mathcal{T}}_{6}$ & $9$ &
        $x^{2}\left(x^{2}-\dfrac{1}{2}x+\dfrac{1}{162}\right)^{2}$
        \\
        $\widetilde{\mathcal{T}}_{7}$ & $18$ &
        $\left(x-\dfrac{1}{54}\right)^{2}
        \left(x^{2}-\dfrac{13}{27}x+\dfrac{1}{2916}\right)^{2}$
        \\
        $\widetilde{\mathcal{T}}_{8}$ & $36$ &
        $\left(x^{2}-\dfrac{13}{27}x+\dfrac{7}{2916}\right)
        \left(x^{2}-\dfrac{7}{27}x+\dfrac{7}{2916}\right)^{2}$
        \\
        $\widetilde{\mathcal{T}}_{9}$ & $6$ &
        $x^{2}\left(x-\dfrac{1}{3}\right)^{2}
        \left(x^{2}-\dfrac{1}{3}x+\dfrac{1}{324}\right)$
        \\
        \hline
    \end{tabular}
    \caption{
        For the bond between $\widetilde{A}_{j}$ and $\widetilde{A}_{j+1}$, the characteristic polynomials of $\widetilde{F}_{\mathrm{norm}}(\nu)$ for each type of Clifford gates.
    }
    \label{characteristic polynomial for tilde}
\end{table}

Moreover, for each type $\widetilde{\mathcal{T}}_{t}$, we compute $\widetilde{\lambda}_{\max}^{t}$, defined as the largest root of the corresponding characteristic polynomial $\widetilde{\chi}_{\nu \in \widetilde{\mathcal{T}}_{t}}(x)$, as follows:
\begin{equation}
\renewcommand{\arraystretch}{1.6}
\begin{array}{c|cccccccccc}
\text{type}
& \widetilde{\mathcal{T}}_{0}
& \widetilde{\mathcal{T}}_{1}
& \widetilde{\mathcal{T}}_{2}
& \widetilde{\mathcal{T}}_{3}
& \widetilde{\mathcal{T}}_{4}
& \widetilde{\mathcal{T}}_{5}
& \widetilde{\mathcal{T}}_{6}
& \widetilde{\mathcal{T}}_{7}
& \widetilde{\mathcal{T}}_{8}
& \widetilde{\mathcal{T}}_{9}
\\
\hline
\widetilde{\lambda}_{\max}^{t}
& \displaystyle \frac{1}{2}\left(1+\frac{2\sqrt{2}}{3}\right)
& \displaystyle \frac{1}{2}\left(1+\frac{4\sqrt{2}}{9}\right)
& \displaystyle \frac{1}{2}\left(1+\frac{\sqrt{2}}{3}\right)
& \displaystyle \frac{2}{3}
& \displaystyle \frac{1}{2}
& \displaystyle \frac{2}{3}
& \displaystyle \frac{5}{9}
& \displaystyle \frac{13}{27}
& \displaystyle \frac{13}{27}
& \displaystyle \frac{4}{9}
\end{array}.
\end{equation}
The maximum value is attained by the class $\widetilde{\mathcal{T}}_{0}$, which corresponds to the identity gate.
To compare the entanglement entropy, we examine the Schmidt probability vector $\widetilde{p}_{\mathrm{Sch}}^{t}$.
From the facts (1) only non-zero entries of $\widetilde{p}_{\mathrm{Sch}}^{0}$ are $\widetilde{\lambda}_{\max}^{0}$ and $1-\widetilde{\lambda}_{\max}^{0}$ and (2) $\widetilde{\lambda}_{\max}^{0} > \widetilde{\lambda}_{\max}^{t \neq 0}$, we find that $\widetilde{p}_{\mathrm{Sch}}^{0}$ majorizes $\widetilde{p}_{\mathrm{Sch}}^{t \neq 0}$.
Therefore, the identity gate minimizes the entanglement entropy among the Clifford gates, so no Clifford gate can reduce the entanglement further.

Finally, we examine the disentangling at the left and right boundaries.
Specifically, at the right boundary, only the $s_{N}=1$ component survives, since $\widetilde{A}_{N}^{s_{N} = 0} \mathbf{e}_{\uparrow} = \widetilde{A}_{N}^{s_{N} = 2} \mathbf{e}_{\uparrow} = 0$.
Thus, the $N$-th site decouples completely from the rest and becomes the product state $|{1_{N}}\rangle$.
As a result, we can write the KW-transformed AKLT state as
\begin{equation}
\label{KW AKLT N-site decouple}
    U_{\mathrm{KW}} |\mathrm{AKLT}_{ \mathbf{e}_{\uparrow}, \mathbf{e}_{\uparrow} }\rangle
    = \sum_{ \{s_{j}\} }
    \mathbf{e}_{\uparrow}^{\dagger}
    \widetilde{A}_{1}^{s_{1}}
    \widetilde{A}_{2}^{s_{2}}
    \widetilde{A}_{3}^{s_{3}}
    \cdots \widetilde{A}_{N-2}^{s_{N-2}}
    \widetilde{A}_{N-1}^{s_{N-1}}
    \widetilde{\mathbf{r}}_{\mathrm{R}}
    |s_{1} s_{2} \cdots s_{N-1} \rangle \otimes |{1_{N}}\rangle, \qquad
    \widetilde{\mathbf{r}}_{\mathrm{R}}
    = - \frac{1}{\sqrt{3}} \left( \mathbf{e}_{\uparrow} + \sqrt{2} \mathbf{e}_{\downarrow} \right).
\end{equation}
From Eq.~\eqref{KW AKLT N-site decouple}, we analyze the Clifford disentangling procedure at the two boundaries: between $\mathbf{e}_{\uparrow}^{\dagger} \widetilde{A}_{1}$ and $\widetilde{A}_{2}$ at the left boundary, and between $\widetilde{A}_{N-2}$ and $\widetilde{A}_{N-1} \widetilde{\mathbf{r}}_{\mathrm{R}}$ at the right boundary (the $N$-th site is already fully disentangled).
Using the same method as for the bond between the $j$-th and $(j+1)$-th sites, we compute the singular values for each of the 90 Clifford gates and verify that the identity gate remains optimal.
In conclusion, within the class of Clifford disentanglers, the entanglement cannot be reduced any further.

\section{Symmetries of KW-transformed Spin-1 Heisenberg and BLBQ models}

In this section, we characterize the symmetries of the KW-transformed spin-1 Heisenberg and BLBQ models.
Throughout this section, we denote by $G_{\mathrm{prod}}(H)$ the group of on-site product unitaries that commute with a Hamiltonian $H$:
\begin{align}
    G_{\mathrm{prod}}(H) = \brrr{S = \bigotimes_{j = 1}^{N} U_{j} \,:\, [S,H] = 0, \,\,\, U_{j} \in \mathrm{U}(3)}.
\end{align}
We also define the right-boundary symmetry group associated with the last $Z$-operator by
\begin{align}
    G_{\mathrm{R}}^{\mathrm{bd}} := \brrr{I_{1} \otimes \cdots \otimes I_{N-1} \otimes V_{N} \,:\, [V_{N},Z_{N}] = 0, \,\,\, V_{N} \in \mathrm{U}(3)}.
\end{align}
For a Hamiltonian $H$, let $[H]_{\Sigma}$ denote the sum of all generalized Pauli strings in $H$ whose support is exactly $\Sigma \subset \{1,2,\cdots,N\}$.
Finally, we write $A \sim B$ if $A = e^{i\phi} B$ for some $\phi \in \mathbb{R}$.

\subsection{On-site product symmetries of the KW-transformed Heisenberg model}\label{subsec:hei_product_sym}

Here, we classify $G_{\mathrm{prod}}(\widetilde{H}_{\mathrm{Hei}})$ for the KW-transformed Heisenberg model $\widetilde{H}_{\mathrm{Hei}}$ given in Eq.~\eqref{transformed Hei Ham}.
The following proposition shows that, after quotienting out the arbitrary right-edge unitary commuting with $Z_{N}$, the only remaining on-site product symmetry is the $\mathbb{Z}_{2}$ symmetry generated by $\mathcal{X}$.

\begin{proposition}\label{prop:Hei_symmetry}
    Let $N \geq 4$ and $D \in \mathbb{R}$. 
    Then every on-site product symmetry $S = \bigotimes_{j = 1}^{N} U_{j} \in G_{\mathrm{prod}}(\widetilde{H}_{\mathrm{Hei}})$ can be written as
    \begin{align}
        S = E_{\mathrm{R}} \mathcal{X}^{\varepsilon},
    \end{align}
    where $\varepsilon \in \{0,1\}$, and $E_{\mathrm{R}} \in G_{\mathrm{R}}^{\mathrm{bd}}$. 
    Consequently, $G_{\mathrm{prod}}(\widetilde{H}_{\mathrm{Hei}}) / G_{\mathrm{R}}^{\mathrm{bd}} \cong \mathbb{Z}_{2}$, generated by the class of $\mathcal{X} = \prod_{j = 1}^{N} e^{i \pi S_{j}^{x}}$.
\end{proposition}
\begin{proof}
    The proof proceeds in four steps: (i) we verify that $\mathcal{X}$ is a symmetry; (ii) we show each local unitary is a permutation matrix up to a scalar phase; (iii) we propagate constraints along the chain to show that the bulk sites are uniformly governed by a single reflection label; and (iv) we handle the right-edge site separately.

    Since $e^{i \pi S^{x}} X e^{-i \pi S^{x}} = X^{\dagger}$ and $e^{i \pi S^{x}} Z e^{-i \pi S^{x}} = \omega^{2} Z^{\dagger}$, conjugation by $\mathcal{X}$ sends each local term in $\widetilde{H}_{\mathrm{Hei}}$ to its Hermitian conjugate.
    From the condition $D \in \mathbb{R}$, we can conclude that $\mathcal{X}$ is a symmetry:
    \begin{align}
        \mathcal{X} \widetilde{H}_{\mathrm{Hei}} \mathcal{X}^{\dagger} = \widetilde{H}_{\mathrm{Hei}}.
    \end{align}

    Let $S = \bigotimes_{j = 1}^{N} U_{j}$ be an on-site product symmetry.
    Since conjugation by an on-site product unitary preserves exact supports, and since generalized Pauli strings are linearly independent, we have
    \begin{align}
        S [\widetilde{H}_{\mathrm{Hei}}]_{\Sigma} S^{\dagger} = [\widetilde{H}_{\mathrm{Hei}}]_{\Sigma}, \qquad \forall \Sigma \subset \{1,2,\cdots,N\}.
    \end{align}
    For $2 \leq j \leq N-1$, the support-$\{j-1,j+1\}$ sector is
    \begin{align}
        [\widetilde{H}_{\mathrm{Hei}}]_{\{j-1,j+1\}} = - \frac{1}{3} \br{Z_{j-1} Z_{j+1}^{\dagger} + Z_{j-1}^{\dagger} Z_{j+1}}. \label{eq:j-1,j+1_sector}
    \end{align}
    Therefore, we have
    \begin{align}
        \br{U_{j-1} \otimes U_{j+1}} \br{Z_{j-1} Z_{j+1}^{\dagger} + Z_{j-1}^{\dagger} Z_{j+1}} \br{U_{j-1} \otimes U_{j+1}}^{\dagger} = Z_{j-1} Z_{j+1}^{\dagger} + Z_{j-1}^{\dagger} Z_{j+1}. \label{eq:ZZ_preservation}
    \end{align}
    In the $Z$-basis, $Z_{j-1} Z_{j+1}^{\dagger} + Z_{j-1}^{\dagger} Z_{j+1} = 3 \sum_{m = 0}^{2} \ket{mm} \bra{mm} - I$.
    Hence, $U_{j-1} \otimes U_{j+1}$ preserves the three-dimensional subspace $\mathrm{span}\{\ket{00}, \ket{11}, \ket{22}\}$.
    For each $m$, the state $U_{j-1} \otimes U_{j+1} \ket{mm} = U_{j-1} \ket{m} \otimes U_{j+1} \ket{m}$ is a product state contained in $\mathrm{span}\{\ket{00}, \ket{11}, \ket{22}\}$.
    The only product states contained are vectors proportional to $\ket{00}$, $\ket{11}$, or $\ket{22}$.
    Thus, for each $m$, both $U_{j-1} \ket{m}$ and $U_{j+1} \ket{m}$ are proportional to computational basis states.
    Since $N \geq 4$, every site belongs to at least one pair of the form $\{j-1,j+1\}$ with $2 \leq j \leq N-1$.
    Hence, for any site $j$, there exists a permutation $\pi_{j} \in S_{3}$ such that $U_{j} \ket{m} = e^{i \alpha_{j,m}} \ket{\pi_{j}(m)}$ for some phases $\alpha_{j,m}$.
    Thus, $U_{j}$ has the following form,
    \begin{align}
        U_{j} = \Lambda_{j} P_{j}, \qquad \text{for any }1 \leq j \leq N,
    \end{align}
    where $P_{j}$ denotes the permutation matrix defined by $P_{j} \ket{m} = \ket{\pi_{j}(m)}$ and the phases are collected into a diagonal unitary $\Lambda_{j}$.

    We next use the one-site sectors.
    For $3 \leq j \leq N-1$,
    \begin{align}
        [\widetilde{H}_{\mathrm{Hei}}]_{\{j\}} = \frac{4}{9} \br{X_{j} + X_{j}^{\dagger}}.
    \end{align}
    Using the decomposition $U_{j} = \Lambda_{j} P_{j}$ obtained above, the permutation part does not affect $X_{j} + X_{j}^{\dagger}$, since the matrix has zero diagonal entries and all off-diagonal entries equal to $1$ in the $Z$-basis.
    Thus, $\Lambda_{j}$ itself must fix $X_{j} + X_{j}^{\dagger}$.
    Writing $\Lambda_{j} = \mathrm{diag}(\lambda_{0},\lambda_{1},\lambda_{2})$, the off-diagonal $(m,n)$-entry of $\Lambda_{j} (X_{j} + X_{j}^{\dagger}) \Lambda_{j}^{\dagger}$ is $\lambda_{m} \lambda_{n}^{*}$.
    Equality with $X_{j} + X_{j}^{\dagger}$ gives $\lambda_{m} \lambda_{n}^{*} = 1$ for all $m \neq n$, hence $\lambda_{0} = \lambda_{1} = \lambda_{2}$.
    Therefore, $\Lambda_{j}$ is a scalar.
    Consequently, up to phase, $U_{j}$ is a permutation matrix in the $Z$-basis.
    Since the permutations of the three $Z$-basis vectors are generated by $X_{j}$ and $e^{i \pi S_{j}^{x}}$, we obtain
    \begin{align}
        U_{j} \sim X_{j}^{a_{j}} \br{e^{i \pi S_{j}^{x}}}^{\varepsilon_{j}}, \qquad \text{for any }3 \leq j \leq N-1,
    \end{align}
    for some $a_{j} \in \mathbb{Z}_{3}$ and $\varepsilon_{j} \in \{0,1\}$.
    At the left boundary, the one-site sector at site 1 is
    \begin{align}
        [\widetilde{H}_{\mathrm{Hei}}]_{\{1\}} = \frac{4}{9} X_{1} - \frac{2}{9} \br{\omega^{2} X_{1} + X_{1}^{\dagger}}Z_{1}^{\dagger} - \frac{D}{3} \omega^{2} Z_{1} + \mathrm{H.c.} \doteq \begin{pmatrix}
            D/3 & 2/3 & 0 \\ 2/3 & -2D/3 & 2/3 \\ 0 & 2/3 & D/3
        \end{pmatrix}
    \end{align}
    where $\doteq$ denotes the matrix form in the $Z$-basis.
    The only zero off-diagonal entries are the $(0,2)$ and $(2,0)$ entries.
    Hence, the permutation part of $U_{1} = \Lambda_{1} P_{1}$ must preserve the unordered pair $\{0,2\}$.
    It is therefore either the identity or the exchange $0 \leftrightarrow 2$.
    Writing $\Lambda_{1} = \mathrm{diag}(\lambda_{0},\lambda_{1},\lambda_{2})$, the two nonzero off-diagonal entries $(0,1)$ and $(1,2)$ with $\Lambda_{1} [\widetilde{H}_{\mathrm{Hei}}]_{\{1\}} \Lambda_{1}^{\dagger} = [\widetilde{H}_{\mathrm{Hei}}]_{\{1\}}$ give $\lambda_{0} \lambda_{1}^{*} = 1$ and $\lambda_{1} \lambda_{2}^{*} = 1$.
    Hence, $\lambda_{0} = \lambda_{1} = \lambda_{2}$, so $\Lambda_{1}$ is scalar.
    Consequently, 
    \begin{align}
        U_{1} \sim \br{e^{i \pi S_{1}^{x}}}^{\varepsilon_{1}}, \qquad \varepsilon_{1} \in \{0,1\}.
    \end{align}
    Similarly, at site 2,
    \begin{align}
        [\widetilde{H}_{\mathrm{Hei}}]_{\{2\}} = \frac{4}{9} \br{X_{2} + X_{2}^{\dagger}} - \frac{1}{3} \br{\omega^{2} Z_{2} + \omega Z_{2}^{\dagger}} \doteq \begin{pmatrix}
            1/3 & 4/9 & 4/9 \\ 4/9 & -2/3 & 4/9 \\ 4/9 & 4/9 & 1/3
        \end{pmatrix}.
    \end{align}
    The unequal middle diagonal entry forces the permutation part of $U_{2} = \Lambda_{2} P_{2}$ to fix $\ket{1}$, and hence to be either the identity or the reflection $0 \leftrightarrow 2$.
    Writing $\Lambda_{2} = \mathrm{diag}(\lambda_{0},\lambda_{1},\lambda_{2})$, the nonzero off-diagonal entries with $\Lambda_{2} [\widetilde{H}_{\mathrm{Hei}}]_{\{2\}} \Lambda_{2}^{\dagger} = [\widetilde{H}_{\mathrm{Hei}}]_{\{2\}}$ give $\lambda_{m} \lambda_{n}^{*} = 1$ for all $m \neq n$.
    Hence, $\Lambda_{2}$ is also scalar, and therefore
    \begin{align}
        U_{2} \sim \br{e^{i \pi S_{2}^{x}}}^{\varepsilon_{2}}, \qquad \varepsilon_{2} \in \{0,1\}.
    \end{align}

    We now return to the two-site sectors in order to propagate these local constraints along the chain. 
    From the preceding one-site analysis, for $1 \leq j \leq N-1$, we have $U_{j} \sim X_{j}^{a_{j}} \br{e^{i \pi S_{j}^{x}}}^{\varepsilon_{j}}$ with $a_{1} = a_{2} = 0$.
    Since $X^{a} Z X^{-a} = \omega^{-a} Z$ and $e^{i \pi S^{x}} Z e^{-i \pi S^{x}} = \omega^{2} Z^{\dagger}$, there exist uniquely determined parameters $\eta_{j} \in \{1, \omega, \omega^{2}\}$ and $\sigma_{j} \in \{+1,-1\}$ such that
    \begin{align}
        U_{j} Z_{j} U_{j}^{\dagger} = \eta_{j} Z_{j}^{\sigma_{j}}, \qquad \text{for any }1 \leq j \leq N-1.
    \end{align}
    Moreover, for $1 \leq j \leq N-1$, the pair $(\eta_{j},\sigma_{j})$ determines $(a_{j},\varepsilon_{j})$.
    Using this form in the invariant support-$\{j-1,j+1\}$ sectors gives
    \begin{align}
        \eta_{j-1} \eta_{j+1}^{-1} Z_{j-1}^{\sigma_{j-1}} Z_{j+1}^{-\sigma_{j+1}} + \eta_{j-1}^{-1} \eta_{j+1} Z_{j-1}^{-\sigma_{j-1}} Z_{j+1}^{\sigma_{j+1}} = Z_{j-1} Z_{j+1}^{\dagger} + Z_{j-1}^{\dagger} Z_{j+1}.
    \end{align}
    By linear independence of generalized Pauli strings,
    \begin{align}
        (\eta_{j-1}, \sigma_{j-1}) = (\eta_{j+1}, \sigma_{j+1}) \qquad \text{for any }2 \leq j \leq N-1.
    \end{align}
    Thus, the pair $(\eta_{j},\sigma_{j})$, and hence $(a_{j},\varepsilon_{j})$, is constant on each parity sublattice among the sites $1,\cdots,N-1$.
    Since $a_{1} = a_{2} = 0$, it follows that $a_{j} = 0$ for any $1 \leq j \leq N-1$.
    Consequently, on the first $N-1$ sites, only the reflection labels remain
    \begin{align}
        U_{j} \sim \br{e^{i \pi S_{j}^{x}}}^{\varepsilon_{j}}, \qquad \text{for any }1 \leq j \leq N-1,
    \end{align}
    where $\varepsilon_{j}$ is constant separately on the odd and even sublattices.

    It remains to compare the two parity sublattices.
    We show that the odd and even reflection labels must be equal.
    From the support-$\{1,2\}$ sector, we read off
    \begin{align}
        \bra{00} [\widetilde{H}_{\mathrm{Hei}}]_{\{1,2\}} \ket{10} = \frac{1}{3}, \quad \bra{20} [\widetilde{H}_{\mathrm{Hei}}]_{\{1,2\}} \ket{10} = - \frac{2}{3}, \quad \bra{00} [\widetilde{H}_{\mathrm{Hei}}]_{\{1,2\}} \ket{01} = \frac{2}{9}, \quad \bra{02} [\widetilde{H}_{\mathrm{Hei}}]_{\{1,2\}} \ket{01} = - \frac{4}{9}. \label{eq:1,2sector_matrix_elements}
    \end{align}
    Suppose first that $(\varepsilon_{1}, \varepsilon_{2}) = (1,0)$. 
    Then, on the first two sites, the symmetry acts as $e^{i \pi S_{1}^{x}} \otimes I_{2}$, up to an overall phase.
    This unitary exchanges $\ket{00}$ and $\ket{20}$ while fixing $\ket{10}$.
    Since the support-$\{1,2\}$ sector is invariant, the matrix element $\bra{00} [\widetilde{H}_{\mathrm{Hei}}]_{\{1,2\}} \ket{10}$ must therefore agree with $\bra{20} [\widetilde{H}_{\mathrm{Hei}}]_{\{1,2\}} \ket{10}$.
    This contradicts $1/3 \neq -2/3$ in Eq.~\eqref{eq:1,2sector_matrix_elements}.
    Thus, $(\varepsilon_{1},\varepsilon_{2}) = (1,0)$ is impossible.
    Similarly, suppose that $(\varepsilon_{1},\varepsilon_{2}) = (0,1)$.
    Then the action on the first two sites is $I_{1} \otimes e^{i \pi S_{2}^{x}}$, up to phase.
    This unitary exchanges $\ket{00}$ and $\ket{02}$ while fixing $\ket{01}$.
    Invariance of the support-$\{1,2\}$ sector would imply $\bra{00} [\widetilde{H}_{\mathrm{Hei}}]_{\{1,2\}} \ket{01} = \bra{02} [\widetilde{H}_{\mathrm{Hei}}]_{\{1,2\}} \ket{01}$, which contradicts $2/9 \neq -4/9$ in Eq.~\eqref{eq:1,2sector_matrix_elements}.
    Hence, $(\varepsilon_{1},\varepsilon_{2}) = (0,1)$ is also impossible.
    Therefore, $\varepsilon_{1} = \varepsilon_{2}$.
    Since the reflection labels are already constant separately on the odd and even sublattices, there exists a single $\varepsilon \in \{0,1\}$ such that
    \begin{align}
        U_{j} \sim \br{e^{i \pi S_{j}^{x}}}^{\varepsilon}, \qquad \text{for any }1 \leq j \leq N-1. \label{eq:U_j_form}
    \end{align}

    We finally treat the right edge.
    The support-$\{N-2,N\}$ sector is
    \begin{align}
        [\widetilde{H}_{\mathrm{Hei}}]_{\{N-2,N\}} = - \frac{1}{3} \br{Z_{N-2} Z_{N}^{\dagger} + Z_{N-2}^{\dagger} Z_{N}}.
    \end{align}
    From Eq.~\eqref{eq:U_j_form}, $U_{N-2} \sim \br{e^{i \pi S_{N-2}^{x}}}^{\varepsilon}$.
    Thus,
    \begin{align}
        U_{N-2} Z_{N-2} U_{N-2}^{\dagger} = \br{e^{i \pi S_{N-2}^{x}}}^{\varepsilon} Z_{N-2} \br{e^{i \pi S_{N-2}^{x}}}^{-\varepsilon}.
    \end{align}
    Since the support-$\{N-2,N\}$ sector is invariant under $U_{N-2} \otimes U_{N}$, the $N$-th site must transform compatibly with the $(N-2)$-nd site.
    For $\varepsilon = 0$, the first factor remains $Z_{N-2}$, and invariance of $Z_{N-2} Z_{N}^{\dagger} + Z_{N-2}^{\dagger} Z_{N}$ forces $U_{N}Z_{N} U_{N}^{\dagger} = Z_{N}$.
    For $\varepsilon = 1$, the first factor changes as $e^{i \pi S_{N-2}^{x}} Z_{N-2} e^{-i \pi S_{N-2}^{x}} = \omega^{2} Z_{N-2}^{\dagger}$, and invariance of the same two-site sector forces $U_{N} Z_{N} U_{N}^{\dagger} = \omega^{2} Z_{N}^{\dagger} = e^{i \pi S_{N}^{x}} Z_{N} e^{-i \pi S_{N}^{x}}$.
    In other words, we have
    \begin{align}
        U_{N} Z_{N} U_{N}^{\dagger} = \br{e^{i \pi S_{N}^{x}}}^{\varepsilon} Z_{N} \br{e^{i \pi S_{N}^{x}}}^{-\varepsilon}. \label{eq:N_unitary}
    \end{align}
    Now define
    \begin{align}
        V_{N} := U_{N} \br{e^{i \pi S_{N}^{x}}}^{-\varepsilon}. \label{eq:VN_definition}
    \end{align}
    We now check that $V_{N}$ commutes with $Z_{N}$.
    If $\varepsilon = 0$, the $V_{N} = U_{N}$, and Eq.~\eqref{eq:N_unitary} gives $V_{N} Z_{N} V_{N}^{\dagger} = Z_{N}$.
    If $\varepsilon = 1$, then $V_{N} = U_{N} e^{-i \pi S_{N}^{x}}$, and hence 
    \begin{align}
        V_{N} Z_{N} V_{N}^{\dagger} &= U_{N} e^{-i \pi S_{N}^{x}} Z_{N} e^{i \pi S_{N}^{x}} U_{N}^{\dagger} = U_{N} \br{\omega^{2} Z_{N}^{\dagger}} U_{N}^{\dagger} = \omega^{2} \br{U_{N} Z_{N} U_{N}^{\dagger}}^{\dagger}.
    \end{align}
    By Eq.~\eqref{eq:N_unitary}, in the case $\varepsilon = 1$, we have $U_{N} Z_{N} U_{N}^{\dagger} = e^{i \pi S_{N}^{x}} Z_{N} e^{-i \pi S_{N}^{x}}$.
    Since $e^{i \pi S^{x}} Z e^{-i \pi S^{x}} = \omega^{2} Z^{\dagger}$, it follows that
    \begin{align}
        V_{N} Z_{N} V_{N}^{\dagger} = \omega^{2} \br{e^{i \pi S_{N}^{x}} Z_{N} e^{-i \pi S_{N}^{x}}}^{\dagger} = \omega^{2} \br{\omega^{2} Z_{N}^{\dagger}}^{\dagger} = Z_{N}.
    \end{align}
    Thus, in both cases, 
    \begin{align}
        V_{N} Z_{N} V_{N}^{\dagger} = Z_{N},
    \end{align}
    so $[V_{N},Z_{N}] = 0$. 
    Since $Z_{N}^{\dagger} = Z_{N}^{-1}$, the relation $[V_{N},Z_{N}] = 0$ also implies $[V_{N},Z_{N}^{\dagger}] = 0$.
    Moreover, every term of $\widetilde{H}_{\mathrm{Hei}}$ involving the last site contains only $Z_{N}$, $Z_{N}^{\dagger}$, or $I_{N}$ on that site.
    Hence, $E_{\mathrm{R}} := I_{1} \otimes I_{2} \otimes \cdots \otimes I_{N-1} \otimes V_{N}$ belongs to $G_{\mathrm{R}}^{\mathrm{bd}}$.

    From Eqs.~\eqref{eq:U_j_form} and \eqref{eq:VN_definition}, the local factors of $S$ can be written as
    \begin{align}
        U_{j} = \begin{cases}
            e^{i \phi_{j}} \br{e^{i \pi S_{j}^{x}}}^{\varepsilon} &\quad 1 \leq j \leq N-1, \\
            V_{N} \br{e^{i \pi S_{N}^{x}}}^{\varepsilon} &\quad j = N,
        \end{cases}
    \end{align}
    for some $\phi_{1},\cdots,\phi_{N-1} \in \mathbb{R}$ and $\varepsilon \in \{0,1\}$.
    Therefore, 
    \begin{align}
        S = \bigotimes_{j = 1}^{N} U_{j} = E_{\mathrm{R}} \mathcal{X}^{\varepsilon},
    \end{align}
    where $E_{\mathrm{R}} \in G_{\mathrm{R}}^{\mathrm{bd}}$ and the overall phase $\phi = \sum_{j = 1}^{N-1} \phi_{j} \in \mathbb{R}$ has been absorbed into $E_{\mathrm{R}}$.

    Furthermore, $G_{\mathrm{R}}^{\mathrm{bd}}$ is normal in $G_{\mathrm{prod}}(\widetilde{H}_{\mathrm{Hei}})$.
    We showed that every element of $G_{\mathrm{prod}}(\widetilde{H}_{\mathrm{Hei}})$ has the form $E_{R} \mathcal{X}^{\varepsilon}$ with $E_{R} \in G_{\mathrm{R}}^{\mathrm{bd}}$.
    Hence, it is enough to check that $\mathcal{X}$ preserves $G_{\mathrm{R}}^{\mathrm{bd}}$ under conjugation.
    For $F_{R} = I_{1} \otimes I_{2} \otimes \cdots \otimes I_{N-1} \otimes W_{N} \in G_{\mathrm{R}}^{\mathrm{bd}}$ with $[W_{N},Z_{N}] = 0$, we have 
    \begin{align}
        \mathcal{X} F_{R} \mathcal{X}^{-1} = I_{1} \otimes I_{2} \otimes \cdots \otimes I_{N-1} \otimes \br{e^{i \pi S_{N}^{x}} W_{N} e^{-i \pi S_{N}^{x}}}.
    \end{align}
    Since $e^{-i \pi S^{x}} Z e^{i \pi S^{x}} = \omega^{2} Z^{\dagger}$, the last-site unitary $e^{i \pi S_{N}^{x}} W_{N} e^{-i \pi S_{N}^{x}}$ also commutes with $Z_{N}$.
    Thus, $\mathcal{X} F_{R} \mathcal{X}^{-1} \in G_{\mathrm{R}}^{\mathrm{bd}}$, so $G_{\mathrm{R}}^{\mathrm{bd}} \triangleleft G_{\mathrm{prod}}(\widetilde{H}_{\mathrm{Hei}})$.
    The normal form then gives $G_{\mathrm{prod}}(\widetilde{H}_{\mathrm{Hei}}) / G_{\mathrm{R}}^{\mathrm{bd}} \cong \mathbb{Z}_{2}$.
\end{proof}

\subsection{On-site product symmetries of the KW-transformed BLBQ model}\label{subsec:blbq_product_sym}

We now extend the preceding classification to the KW-transformed BLBQ Hamiltonian $\widetilde{H}_{\mathrm{BLBQ}}$, which is given in Eq.~\eqref{transformed BLBQ Ham}.
Although the local coefficients are modified by the biquadratic terms, the support sectors used in the Heisenberg proof retain the same operator structure for $- \pi / 4 < \theta < \pi / 4$.
Hence, the same support-sector argument gives the following classification.
\begin{proposition}\label{prop:BLBQ_symmetry}
Assume $N \geq 4$ and $- \pi / 4 < \theta < \pi / 4$.
Then every on-site product symmetry $S = \bigotimes_{j = 1}^{N} U_{j} \in G_{\mathrm{prod}}(\widetilde{H}_{\mathrm{BLBQ}})$ can be written as
\begin{align}
    S = E_{\mathrm{R}} \mathcal{X}^{\varepsilon},
\end{align}
where $\varepsilon \in \{0,1\}$, and $E_{\mathrm{R}} \in G_{\mathrm{R}}^{\mathrm{bd}}$. 
Consequently, $G_{\mathrm{prod}}(\widetilde{H}_{\mathrm{BLBQ}}) / G_{\mathrm{R}}^{\mathrm{bd}} \cong \mathbb{Z}_{2}$, generated by the class of $\mathcal{X} = \prod_{j = 1}^{N} e^{i \pi S_{j}^{x}}$.
\end{proposition}
\begin{proof}
    We first note that the support sectors that constrain an on-site product symmetry have the same operator structure as in the KW-transformed Heisenberg case in Proposition~\ref{prop:Hei_symmetry}.
    The biquadratic terms only modify the numerical coefficients of these sectors.
    In the interval $- \pi / 4 < \theta < \pi / 4$, the coefficients used below remain nonzero, so the support-sector constraints are unchanged.

    The global operator $\mathcal{X}$ is a symmetry.
    Since $e^{i \pi S^{x}} X e^{- i \pi S^{x}} = X^{\dagger}$ and $e^{i \pi S^{x}} Z e^{- i \pi S^{x}} = \omega^{2} Z^{\dagger}$, each summand in $\widetilde{T}_{1,2}^{(\mathrm{bd})}$, $\widetilde{T}_{j-1,j,j+1}$, $\widetilde{Q}_{1,2}^{(\mathrm{bd})}$, and $\widetilde{Q}_{j-1,j,j+1}$ is mapped to its Hermitian conjugate. 
    Since $\widetilde{H}_{\mathrm{BLBQ}}$ contains these terms together with their Hermitian conjugates, and since $\cos \theta, \sin \theta \in \mathbb{R}$, we obtain
    \begin{align}
        \mathcal{X} \widetilde{H}_{\mathrm{BLBQ}} \mathcal{X}^{\dagger} = \widetilde{H}_{\mathrm{BLBQ}}.
    \end{align}

    Let $S = \bigotimes_{j = 1}^{N} U_{j} \in G_{\mathrm{prod}}(\widetilde{H}_{\mathrm{BLBQ}})$.
    Since conjugation by an on-site product unitary preserves exact supports, and since generalized Pauli strings are linearly independent, each support sector is separately invariant:
    \begin{align}
        S[\widetilde{H}_{\mathrm{BLBQ}}]_{\Sigma} S^{\dagger} = [\widetilde{H}_{\mathrm{BLBQ}}]_{\Sigma}, \qquad \forall \Sigma \subset \{1,2,\cdots,N\}.
    \end{align}
    For $2 \leq j \leq N-1$, the support-$\{j-1,j+1\}$ sector is proportional to $Z_{j-1} Z_{j+1}^{\dagger} + Z_{j-1}^{\dagger} Z_{j+1}$, with a nonzero proportionality coefficient in the range $- \pi / 4 < \theta < \pi / 4$.
    Therefore, $U_{j-1} \otimes U_{j+1}$ must preserve this two-site operator.
    As in the support-sector argument in the proof of Proposition~\ref{prop:Hei_symmetry}, this implies that each local unitary $U_{j}$ maps computational basis states to computational basis states up to phases.
    Thus, each $U_{j}$ can be written as $U_{j} = \Lambda_{j} P_{j}$, where $\Lambda_{j}$ is diagonal and $P_{j}$ is a permutation matrix in the $Z$-basis.

    We next use the one-site sectors.
    For $3 \leq j \leq N-1$, the bulk one-site sector is 
    \begin{align}
        [\widetilde{H}_{\mathrm{BLBQ}}]_{\{j\}} = \frac{4 \cos \theta - \sin \theta}{9} \br{X_{j} + X_{j}^{\dagger}}. 
    \end{align}
    Since $4 \cos \theta - \sin \theta \neq 0$ for $- \pi / 4 < \theta < \pi / 4$, invariance of this sector forces $\Lambda_{j}$ to commute with $X_{j} + X_{j}^{\dagger}$.
    Hence, $\Lambda_{j}$ is scalar, and therefore 
    \begin{align}
        U_{j} \sim X_{j}^{a_{j}} \br{e^{i \pi S_{j}^{x}}}^{\varepsilon_{j}}, \qquad \text{for any }3 \leq j \leq N-1, \label{eq:Uj_1}
    \end{align}
    where $a_{j} \in \mathbb{Z}_{3}$ and $\varepsilon_{j} \in \{0,1\}$.
    At site $1$, the one-site sector is
    \begin{align}
        [\widetilde{H}_{\mathrm{BLBQ}}]_{\{1\}} \doteq \begin{pmatrix}
            0 & \frac{2 \cos \theta - \sin \theta}{3} & \frac{\sin \theta}{3} \\ 
            \frac{2 \cos \theta - \sin \theta}{3} & 0 & \frac{2 \cos \theta - \sin \theta}{3} \\
            \frac{\sin \theta}{3} & \frac{2 \cos \theta - \sin \theta}{3} & 0
        \end{pmatrix}.
    \end{align}
    Since $2 \cos \theta - \sin \theta \neq 0$ and $2 \cos \theta - \sin \theta \neq \sin \theta$ in the range $- \pi / 4 < \theta < \pi / 4$, the permutation part $P_{1}$ is either the identity or the reflection $0 \leftrightarrow 2$.
    The nonzero $(0,1)$ and $(1,2)$ entries then force $\Lambda_{1}$ to be scalar.
    Hence,
    \begin{align}
        U_{1} \sim \br{e^{i \pi S_{1}^{x}}}^{\varepsilon_{1}}. \label{eq:Uj_2}
    \end{align}
    At site $2$, the one-site sector is
    \begin{align}
        [\widetilde{H}_{\mathrm{BLBQ}}]_{\{2\}} \doteq \begin{pmatrix}
            \frac{\cos \theta - \sin \theta}{3} & \frac{4 \cos \theta - \sin \theta}{9} & \frac{4 \cos \theta - \sin \theta}{9} \\ 
            \frac{4 \cos \theta - \sin \theta}{9} & - \frac{2(\cos \theta - \sin \theta)}{3} & \frac{4 \cos \theta - \sin \theta}{9} \\
            \frac{4 \cos \theta - \sin \theta}{9} & \frac{4 \cos \theta - \sin \theta}{9} & \frac{\cos \theta - \sin \theta}{3}
        \end{pmatrix}.
    \end{align}
    Since $\cos \theta - \sin \theta \neq 0$ and $4 \cos \theta - \sin \theta \neq 0$, the middle diagonal entry is distinguished, and all off-diagonal entries are nonzero.
    Thus, $P_{2}$ is either the identity or the reflection $0 \leftrightarrow 2$, and $\Lambda_{2}$ is scalar.
    Hence,
    \begin{align}
        U_{2} \sim \br{e^{i \pi S_{2}^{x}}}^{\varepsilon_{2}}. \label{eq:Uj_3}
    \end{align}

    We now propagate these constraints along the chain.
    For $1 \leq j \leq N-1$, Eqs.~\eqref{eq:Uj_1}, \eqref{eq:Uj_2}, and \eqref{eq:Uj_3} imply that
    \begin{align}
        U_{j} Z_{j} U_{j}^{\dagger} = \eta_{j} Z_{j}^{\sigma_{j}},
    \end{align}
    for some $\eta_{j} \in \{1,\omega,\omega^{2}\}$ and $\sigma_{j} \in \{+1,-1\}$.
    Invariance of the support-$\{j-1,j+1\}$ sector gives
    \begin{align}
        (\eta_{j-1},\sigma_{j-1}) = (\eta_{j+1},\sigma_{j+1}), \qquad \text{for any }2 \leq j \leq N-1.
    \end{align}
    Hence, the labels are constant on each parity sublattice.
    Because $(\eta_{j},\sigma_{j})$ uniquely fixes $(a_{j},\varepsilon_{j})$, this also implies that $a_{j}$ is constant on each parity sublattice.
    Since the boundary constraints give $a_{1} = a_{2} = 0$, we obtain $a_{j} = 0$ for all $1 \leq j \leq N-1$.

    The equality of the two parity labels follows from the same argument used in the proof of Proposition~\ref{prop:Hei_symmetry}.
    Namely, for $- \pi / 4 < \theta < \pi / 4$, the support-$\{1,2\}$ sector is not invariant under $e^{i \pi S_{1}^{x}} \otimes I_{2}$ or $I_{1} \otimes e^{i \pi S_{2}^{x}}$.
    Hence, the mixed choices $(\varepsilon_{1},\varepsilon_{2}) = (1,0)$ and $(\varepsilon_{1}, \varepsilon_{2}) = (0,1)$ are impossible.
    Therefore,
    \begin{align}
        \varepsilon_{1} = \varepsilon_{2}.
    \end{align}
    Since the labels are already constant on each parity sublattice, there exists a single $\varepsilon \in \{0,1\}$ such that
    \begin{align}
        U_{j} \sim \br{e^{i \pi S_{j}^{x}}}^{\varepsilon}, \qquad 1 \leq j \leq N-1.
    \end{align}

    Finally, the right edge is constrained by the support-$\{N-2,N\}$ sector.
    Since the $(N-2)$-nd site already carries the reflection label $\varepsilon$, invariance of this sector forces
    \begin{align}
        U_{N} Z_{N} U_{N}^{\dagger} = \br{e^{i \pi S_{N}^{x}}}^{\varepsilon} Z_{N} \br{e^{i \pi S_{N}^{x}}}^{-\varepsilon}.
    \end{align}
    Define $V_{N} := U_{N} \br{e^{i \pi S_{N}^{x}}}^{-\varepsilon}$.
    Then $[V_{N},Z_{N}] = 0$ and $[V_{N},Z_{N}^{\dagger}] = 0$ as in the proof of Proposition~\ref{prop:Hei_symmetry}.
    Therefore, $E_{\mathrm{R}} := I_{1} \otimes I_{2} \otimes \cdots \otimes I_{N-1} \otimes V_{N} \in G_{\mathrm{R}}^{\mathrm{bd}}$.
    Combining the local forms of the $U_{j}$, we obtain
    \begin{align}
        S = E_{\mathrm{R}} \mathcal{X}^{\varepsilon}
    \end{align}
    where $\varepsilon \in \{0,1\}$, and $E_{\mathrm{R}} \in G_{\mathrm{R}}^{\mathrm{bd}}$.
    The normality of $G_{\mathrm{R}}^{\mathrm{bd}}$ follows by the same argument as in the proof of Proposition~\ref{prop:Hei_symmetry}.
\end{proof}

\section{Details of Hidden $\mathbb{Z}_{2}$ SSB Revealed by KW Transformation}

In this section, we provide the details of the $\mathbb{Z}_{2}$ SSB phenomenon after applying the KW transformation to the AKLT state.
The MPS description of the AKLT state is given in Eq.~\eqref{def of AKLT A}.
Here, we consider the four types of AKLT states, labeled by $\mathbf{L} \in \{ \mathbf{e}_{\uparrow}, \mathbf{e}_{\downarrow} \} $ and $\mathbf{R} \in \{ \mathbf{e}_{+}, \mathbf{e}_{-} \}$,
where $\mathbf{e}_{+} = \frac{1}{\sqrt{2}}( \mathbf{e}_{\uparrow} + \mathbf{e}_{\downarrow} )$
and $\mathbf{e}_{-} = \frac{1}{\sqrt{2}}( \mathbf{e}_{\uparrow} - \mathbf{e}_{\downarrow} )$.
These four states become mutually orthogonal in the thermodynamic limit, with finite-size overlaps decaying exponentially with $N$.
Note that $\mathbf{e}_{\uparrow}$ and $\mathbf{e}_{\downarrow}$ are the $+1$ and $-1$ eigenstates of the qubit Pauli $Z$ operator $\sigma^{z}$, respectively, and $\mathbf{e}_{+}$ and $\mathbf{e}_{-}$ are the $+1$ and $-1$ eigenstates of the qubit Pauli $X$ operator $\sigma^{x}$, respectively.
The AKLT state corresponding to arbitrary $\mathbf{L}$ and $\mathbf{R}$ can be written as a linear combination of these four states.

\subsection{Expectation value of $S_{j}^{z}$ and the two-point correlation function}

Here, we show that at the thermodynamic limit $N \rightarrow \infty$, the expectation value of $S_{j}^{z}$ for the KW-transformed AKLT state $U_{\mathrm{KW}} | \mathrm{AKLT}_{ \mathbf{L}, \mathbf{R} } \rangle$ is given by
\begin{equation}
\label{AKLT local Sz expectation}
    \lim_{N \to \infty}
    \frac{
    \langle \mathrm{AKLT}_{ \mathbf{L}, \mathbf{R} } |
    (U_{\mathrm{KW}})^{\dagger} S_{j}^{z} U_{\mathrm{KW}}
    | \mathrm{AKLT}_{ \mathbf{L}, \mathbf{R} } \rangle
    }
    {
    \langle \mathrm{AKLT}_{ \mathbf{L}, \mathbf{R} } | \mathrm{AKLT}_{ \mathbf{L}, \mathbf{R} } \rangle
    }
    =
    \begin{cases}
        \frac{1}{2} & \mathbf{L} = \mathbf{e}_{\uparrow} \\
        -\frac{1}{2} & \mathbf{L} = \mathbf{e}_{\downarrow} \\
    \end{cases}.
\end{equation}
As shown in Eq.~\eqref{AKLT local Sz expectation}, it depends only on the left boundary. 
In addition, the two-point correlation function between $S_{j}^{z}$ and $S_{j+r}^{z}$ is given by
\begin{equation}
\label{AKLT two point correlation}
    \lim_{r \rightarrow \infty}
    \lim_{N \to \infty}
    \frac{
    \langle \mathrm{AKLT}_{ \mathbf{L}, \mathbf{R} } |
    (U_{\mathrm{KW}})^{\dagger} S_{j}^{z} S_{j+r}^{z} U_{\mathrm{KW}}
    | \mathrm{AKLT}_{ \mathbf{L}, \mathbf{R} } \rangle
    }
    {
    \langle \mathrm{AKLT}_{ \mathbf{L}, \mathbf{R} } | \mathrm{AKLT}_{ \mathbf{L}, \mathbf{R} } \rangle
    }
    = \frac{1}{4}.
\end{equation}
Since $\acomm{S_{j}^{z}}{\mathcal{X}} = 0$, these results diagnose the $\mathbb{Z}_{2}$ SSB associated $\mathcal{X}$.

Furthermore, these results are clearly distinguished from the corresponding results for the trivial state $|{1}\rangle^{\otimes N}$.
Since $U_{\mathrm{KW}} |{1}\rangle^{\otimes N} = |{1}\rangle^{\otimes N}$, both expectation values are $0$ for the trivial state.
The invariance of $|{1}\rangle^{\otimes N}$ under $U_{\mathrm{KW}}$ follows directly from Eq.~\eqref{KW unitary}; $\mathcal{U}_{j,j+1}$ is defined as the product of $X_{j+1}^{2}$ and the \texttt{SUM} gate.

To compute physical quantities of MPSs, we introduce the transfer matrix $\mathbb{T}_{j}(O)$ associated with an operator $O$.
The definition and the corresponding diagram are given by
\begin{equation}
\label{def of transfer matrix T}
    \mathbb{T}_{j}(O) := \sum_{s_{j}, s_{j}'}
    O_{s_{j}, s_{j}'} A_{j}^{s_{j}'} \otimes (A_{j}^{s_{j}})^{*} = ~~~
    \begin{tikzpicture}[anchor=center, baseline={(0,-0.1)}]
        \def\a{0.35};
    
        \begin{scope}[shift = {(0, 2.5*\a)}]
            \roundedsquare{SkyBlue!40}{\a}{\a}
            \node at (0, 0) {$A_{j}$};    
            
            \draw (1.0*\a, 0) -- (2.0*\a, 0);
            \draw (-1.0*\a, 0) -- (-2.0*\a, 0);
            \draw (0, -1.0*\a) -- (0, -2.0*\a);
        \end{scope}
    
        \begin{scope}[yscale = -1, shift = {(0, 2.5*\a)}]
            \roundedsquare{SkyBlue!40}{\a}{\a}
            \node at (0, 0) {$A_{j}^{*}$};    
            
            \draw (1.0*\a, 0) -- (2.0*\a, 0);
            \draw (-1.0*\a, 0) -- (-2.0*\a, 0);
            \draw (0, -1.0*\a) -- (0, -2.0*\a);
        \end{scope}
    
        \draw[fill = orange!30] (0, 0) circle (1.0*\a);
        \node at (0, 0) {$O$};
    \end{tikzpicture}
\end{equation}
For simplicity, we define $\mathbb{T}_{j} := \mathbb{T}_{j}(I)$, whose explicit form is
\begin{equation}
\label{Tj}
    \mathbb{T}_{j} = \frac{1}{3}
    \begin{pmatrix}
        1 & 0 & 0 & 2 \\
        0 & -1 & 0 & 0 \\
        0 & 0 & -1 & 0 \\
        2 & 0 & 0 & 1
    \end{pmatrix}.
\end{equation}

By using the transfer matrix, the norm of $| \mathrm{AKLT}_{ \mathbf{L}, \mathbf{R} } \rangle$ is expressed as
\begin{equation}
    \langle \mathrm{AKLT}_{ \mathbf{L}, \mathbf{R} } | \mathrm{AKLT}_{ \mathbf{L}, \mathbf{R} } \rangle
    = (\mathbf{L}^{\dagger} \otimes \mathbf{L}^{T})
    \left( \prod_{j=1}^{N} \mathbb{T}_{j} \right)
    (\mathbf{R} \otimes \mathbf{R}^{*}).
\end{equation}
Using the matrix $\mathbb{T}_{j}$ in Eq.~\eqref{Tj}, at the thermodynamic limit $N \rightarrow \infty$, we have
\begin{equation}
    \lim_{N \rightarrow \infty}
    \langle \mathrm{AKLT}_{ \mathbf{L}, \mathbf{R} } | \mathrm{AKLT}_{ \mathbf{L}, \mathbf{R} } \rangle
    = \frac{1}{2}.
\end{equation}

To obtain the quantities in Eqs.~\eqref{AKLT local Sz expectation} and~\eqref{AKLT two point correlation}, we evaluate the following operators:
\begin{equation}
\label{ops Heisenberg picture}
\begin{gathered}
    (U_{\mathrm{KW}})^{\dagger} Z_{j} U_{\mathrm{KW}}
    = \omega \prod_{k=1}^{j} (\omega^{2} Z_{k}) \\
    (U_{\mathrm{KW}})^{\dagger} S_{j}^{z} U_{\mathrm{KW}}
    = \frac{i}{\sqrt{3}} \prod_{k=1}^{j} (\omega^{2} Z_{k})
    - \frac{i}{\sqrt{3}} \prod_{k=1}^{j} (\omega Z_{k}^{\dagger}) \\
    (U_{\mathrm{KW}})^{\dagger} S_{j}^{z} S_{j+r}^{z} U_{\mathrm{KW}}
    = 
    \frac{1}{3} \prod_{k=j+1}^{j+r} (\omega^{2} Z_{k})
    - \frac{1}{3} \prod_{k=1}^{j} (\omega^{2} Z_{k})
    \prod_{k=j+1}^{j+r} (\omega Z_{k}^{\dagger})
    + \text{H.c.}
\end{gathered}
\end{equation}
To obtain Eq.~\eqref{ops Heisenberg picture}, we use Eq.~\eqref{Pauli spinop conversion}.
The expectation values of each term in Eq.~\eqref{ops Heisenberg picture} are expressed as
\begin{equation}
\begin{aligned}
    \langle \mathrm{AKLT}_{ \mathbf{L}, \mathbf{R} } |
    \prod_{k=1}^{j} (\omega^{2} Z_{k})
    | \mathrm{AKLT}_{ \mathbf{L}, \mathbf{R} } \rangle
    =
    (\mathbf{L}^{\dagger} \otimes \mathbf{L}^{T})
    \left( \prod_{k=1}^{j} \mathbb{T}_{k}(\omega^{2} Z_{k}) \right)
    \left( \prod_{k=j+1}^{N} \mathbb{T}_{k} \right)
    (\mathbf{R} \otimes \mathbf{R}^{*}),
\end{aligned}
\end{equation}
\begin{equation}
\begin{aligned}
    & \langle \mathrm{AKLT}_{ \mathbf{L}, \mathbf{R} } |
    \prod_{k=j+1}^{j+r} (\omega^{2} Z_{k})
    | \mathrm{AKLT}_{ \mathbf{L}, \mathbf{R} } \rangle = 
    (\mathbf{L}^{\dagger} \otimes \mathbf{L}^{T})
    \left( \prod_{k=1}^{j} \mathbb{T}_{k} \right)
    \left( \prod_{k=j+1}^{j+r} \mathbb{T}_{k}(\omega^{2} Z_{k}) \right)
    \left( \prod_{k=j+r+1}^{N} \mathbb{T}_{k} \right)
    (\mathbf{R} \otimes \mathbf{R}^{*}),
\end{aligned}
\end{equation}
\begin{equation}
\begin{aligned}
    \langle \mathrm{AKLT}_{ \mathbf{L}, \mathbf{R} } &|
    \prod_{k=1}^{j} (\omega^{2} Z_{k})
    \prod_{k=j+1}^{j+r} (\omega Z_{k}^{\dagger})
    | \mathrm{AKLT}_{ \mathbf{L}, \mathbf{R} } \rangle \\ 
    &\quad = 
    (\mathbf{L}^{\dagger} \otimes \mathbf{L}^{T})
    \left( \prod_{k=1}^{j} \mathbb{T}_{k}(\omega^{2} Z_{k}) \right)
    \left( \prod_{k=j+1}^{j+r} \mathbb{T}_{k}(\omega Z_{k}^{\dagger}) \right)
    \left( \prod_{k=j+r+1}^{N} \mathbb{T}_{k} \right)
    (\mathbf{R} \otimes \mathbf{R}^{*}).
\end{aligned}
\end{equation}
At the thermodynamic limit $N \rightarrow \infty$ and $j \rightarrow \infty$, these quantities are evaluated as
\begin{equation}
\begin{gathered}
    \lim_{N \to \infty}
    \langle \mathrm{AKLT}_{ \mathbf{L}, \mathbf{R} } |
    \prod_{k=1}^{j} (\omega^{2} Z_{k})
    | \mathrm{AKLT}_{ \mathbf{L}, \mathbf{R} } \rangle
    = \begin{cases}
        \frac{1}{4}(1+\omega^{2}) & \mathbf{L} = \mathbf{e}_{\uparrow} \\
        \frac{1}{4}(1+\omega) & \mathbf{L} = \mathbf{e}_{\downarrow}
    \end{cases}, \\
    \lim_{r \to \infty}
    \lim_{N \to \infty}
    \langle \mathrm{AKLT}_{ \mathbf{L}, \mathbf{R} } |
    \prod_{k=j+1}^{j+r} (\omega^{2} Z_{k})
    | \mathrm{AKLT}_{ \mathbf{L}, \mathbf{R} } \rangle
    = \frac{1}{8}, \\
    \lim_{r \rightarrow \infty}
    \lim_{N \to \infty}
    \langle \mathrm{AKLT}_{ \mathbf{L}, \mathbf{R} } |
    \prod_{k=1}^{j} (\omega^{2} Z_{k})
    \prod_{k=j+1}^{j+r} (\omega Z_{k}^{\dagger})
    | \mathrm{AKLT}_{ \mathbf{L}, \mathbf{R} } \rangle
    = \begin{cases}
        \frac{1}{8}\omega & \mathbf{L} = \mathbf{e}_{\uparrow} \\
        \frac{1}{8}\omega^{2} & \mathbf{L} = \mathbf{e}_{\downarrow}
    \end{cases}.
\end{gathered}
\end{equation}
Then, we obtain the desired results in Eqs.~\eqref{AKLT local Sz expectation} and~\eqref{AKLT two point correlation}.

\subsection{Action of symmetry operators on the edge spins}

In this section, we investigate how the symmetry operators $\mathcal{X}$ and $\mathcal{Z}_{\mathrm{R}}$ act on the boundary edge spins. 
Explicitly, we prove Eq.~(12) in the main text, which is given by
\begin{equation}
\label{AKLT X Z edge transformation}
\begin{gathered}
    \mathcal{X} U_{\mathrm{KW}}
    | \mathrm{AKLT}_{ \mathbf{L}, \mathbf{R} } \rangle
    = U_{\mathrm{KW}}
    | \mathrm{AKLT}_{ \sigma^{x} \mathbf{L}, \sigma^{x} \mathbf{R} } \rangle, \\
    \mathcal{Z}_{\mathrm{R}} U_{\mathrm{KW}}
    | \mathrm{AKLT}_{ \mathbf{L}, \mathbf{R} } \rangle
    = U_{\mathrm{KW}}
    | \mathrm{AKLT}_{ \sigma^{z} \mathbf{L}, \sigma^{z} \mathbf{R} } \rangle,
\end{gathered}
\end{equation}
for $\mathbf{L} \in \{ \mathbf{e}_{\uparrow}, \mathbf{e}_{\downarrow} \} $ and $\mathbf{R} \in \{ \mathbf{e}_{+}, \mathbf{e}_{-} \}$.

To verify the first line in Eq.~\eqref{AKLT X Z edge transformation}, we use a diagrammatic identity~\cite{Tasaki2020, Perez2008, Pollmann2010, Pollmann2012, Pollmann2017symmetry}, which is given by
\begin{equation}
\label{X action on A}
    \begin{tikzpicture}[anchor=center, baseline={(0,-0.1)}]
        \def\a{0.35};

        \roundedsquare{SkyBlue!40}{\a}{\a}
        \node at (0, 0) {$A_{j}$};    
        
        \draw (1.0*\a, 0) -- (2.0*\a, 0);
        \draw (-1.0*\a, 0) -- (-2.0*\a, 0);
        \draw (0, -1.0*\a) -- (0, -2.0*\a);

        \draw[fill = orange!30] (0, -3.0*\a) circle (1.15*\a);
        \node at (0, -3.0*\a) {$\mathcal{X}_{j}$};
        \draw (0, -4.15*\a) -- (0, -5.0*\a);
    \end{tikzpicture}
    =
    \begin{tikzpicture}[anchor=center, baseline={(0,-0.1)}]*
        \def\a{0.35};
    
        \roundedsquare{SkyBlue!40}{\a}{\a}
        \node at (0, 0) {$A_{j}$};    
    
        \draw (1.0*\a, 0) -- (2.0*\a, 0);
        \draw (-1.0*\a, 0) -- (-2.0*\a, 0);
        \draw (0, -1.0*\a) -- (0, -2.0*\a);
    
        \begin{scope}[shift = {(2.8*\a, 0)}]
            \diatensor{green!30}{1.3*\a}
            \node at (0, 0) {$\sigma^{x}$};
            \draw (1.3*\a, 0) -- (1.8*\a, 0);
        \end{scope}    
        \begin{scope}[xscale = -1, shift = {(2.8*\a, 0)}]
            \diatensor{green!30}{1.3*\a}
            \node at (0, 0) {$\sigma^{x}$};
            \draw (1.3*\a, 0) -- (1.8*\a, 0);
        \end{scope}
    \end{tikzpicture}, ~~
    \text{where } \mathcal{X}_{j} = e^{i \pi S_{j}^{x}}.
\end{equation}
Additionally, $\mathcal{X}$ commutes with $U_{\mathrm{KW}}$. 
Since $(\sigma^{x})^{2}$ is an identity, all $\sigma^{x}$ operators on the virtual bond cancel out, except at the left and right edges.
Explicitly, we have
\begin{equation}
    (U_{\mathrm{KW}})^{\dagger} \mathcal{X} U_{\mathrm{KW}}
    | \mathrm{AKLT}_{ \mathbf{L}, \mathbf{R} } \rangle
    = \sum_{ \{s_{j}\} }
    (\mathbf{L}^{\dagger} \sigma^{x})
    A_{1}^{s_{1}} A_{2}^{s_{2}} \cdots A_{N}^{s_{N}}
    (\sigma^{x} \mathbf{R})
    |s_{1} s_{2} \cdots s_{N} \rangle.
\end{equation}
Then, the action of $\mathcal{X}$ on $\mathbf{L}$ is $\sigma^{x}$, since $\mathbf{L}^{\dagger} \sigma^{x} = (\sigma^{x} \mathbf{L})^{\dagger}$.
At the right edge, the changed boundary vector $\sigma^{x} \mathbf{R}$ implies that the action of $\mathcal{X}$ on $\mathbf{R}$ is also $\sigma^{x}$.
It establishes the first line of Eq.~\eqref{AKLT X Z edge transformation}.

To derive the second line in Eq.~\eqref{AKLT X Z edge transformation}, we decompose $\mathcal{Z}_{R} = e^{i \pi S_{N}^{z}}$ into a sum of generalized Pauli operators as
\begin{equation}
    \mathcal{Z}_{R} = e^{i \pi S_{N}^{z}}
    = \frac{1}{3} ( 2\omega^{2} Z_{N} + 2\omega Z_{N}^{\dagger} - I).
\end{equation}
Then, using Eq.~\eqref{ops Heisenberg picture}, we obtain
\begin{equation}
\label{ZR KW transformation}
    (U_{\mathrm{KW}})^{\dagger} \mathcal{Z}_{R} U_{\mathrm{KW}}
    = \frac{1}{3}
    \left(
        2 \prod_{j=1}^{N} (\omega^{2} Z_{j}) + 2 \prod_{j=1}^{N} (\omega Z_{j}^{\dagger}) - I
    \right).
\end{equation}
To compute the actions of each term, we derive the following diagrammatic identities:
\begin{equation}
\label{ZR diagram}
    \begin{tikzpicture}[anchor=center, baseline={(0,-0.1)}]
        \def\a{0.35};

        \roundedsquare{SkyBlue!40}{\a}{\a}
        \node at (0, 0) {$A_{j}$};    
        
        \draw (1.0*\a, 0) -- (2.0*\a, 0);
        \draw (-1.0*\a, 0) -- (-2.0*\a, 0);
        \draw (0, -1.0*\a) -- (0, -2.0*\a);

        \draw[fill = orange!30] (0, -3.0*\a) circle (1.15*\a);
        \node at (0, -3.0*\a) {\scriptsize $\omega^{2} Z_{j}$};
        \draw (0, -4.15*\a) -- (0, -5.0*\a);
    \end{tikzpicture}
    =
    \begin{tikzpicture}[anchor=center, baseline={(0,-0.1)}]*
        \def\a{0.35};
    
        \roundedsquare{SkyBlue!40}{\a}{\a}
        \node at (0, 0) {$A_{j}$};    
    
        \draw (1.0*\a, 0) -- (2.0*\a, 0);
        \draw (-1.0*\a, 0) -- (-2.0*\a, 0);
        \draw (0, -1.0*\a) -- (0, -2.0*\a);
    
        \begin{scope}[shift = {(2.8*\a, 0)}]
            \diatensor{green!30}{1.3*\a}
            \node at (0, 0) {$\Omega_{\omega}^{\dagger}$};
            \draw (1.3*\a, 0) -- (1.8*\a, 0);
        \end{scope}    
        \begin{scope}[xscale = -1, shift = {(2.8*\a, 0)}]
            \diatensor{green!30}{1.3*\a}
            \node at (0, 0) {$\Omega_{\omega}$};
            \draw (1.3*\a, 0) -- (1.8*\a, 0);
        \end{scope}
    \end{tikzpicture}, ~~
    \begin{tikzpicture}[anchor=center, baseline={(0,-0.1)}]
        \def\a{0.35};

        \roundedsquare{SkyBlue!40}{\a}{\a}
        \node at (0, 0) {$A_{j}$};    
        
        \draw (1.0*\a, 0) -- (2.0*\a, 0);
        \draw (-1.0*\a, 0) -- (-2.0*\a, 0);
        \draw (0, -1.0*\a) -- (0, -2.0*\a);

        \draw[fill = orange!30] (0, -3.0*\a) circle (1.15*\a);
        \node at (0, -3.0*\a) {\scriptsize $\omega Z_{j}^{\dagger}$};
        \draw (0, -4.15*\a) -- (0, -5.0*\a);
    \end{tikzpicture}
    =
    \begin{tikzpicture}[anchor=center, baseline={(0,-0.1)}]*
        \def\a{0.35};
    
        \roundedsquare{SkyBlue!40}{\a}{\a}
        \node at (0, 0) {$A_{j}$};    
    
        \draw (1.0*\a, 0) -- (2.0*\a, 0);
        \draw (-1.0*\a, 0) -- (-2.0*\a, 0);
        \draw (0, -1.0*\a) -- (0, -2.0*\a);
    
        \begin{scope}[shift = {(2.8*\a, 0)}]
            \diatensor{green!30}{1.3*\a}
            \node at (0, 0) {$\Omega_{\omega}$};
            \draw (1.3*\a, 0) -- (1.8*\a, 0);
        \end{scope}    
        \begin{scope}[xscale = -1, shift = {(2.8*\a, 0)}]
            \diatensor{green!30}{1.3*\a}
            \node at (0, 0) {$\Omega_{\omega}^{\dagger}$};
            \draw (1.3*\a, 0) -- (1.8*\a, 0);
        \end{scope}
    \end{tikzpicture}, ~~
    \text{where } \Omega_{\omega} = \begin{pmatrix} 1 & 0 \\ 0 & \omega \end{pmatrix}.
\end{equation}
Since $\Omega_{\omega}$ is unitary, $(U_{\mathrm{KW}})^{\dagger} \mathcal{Z}_{R} U_{\mathrm{KW}}$ acts only on both edges, by analogy with $\mathcal{X}$.
Thus, we obtain 
\begin{equation}
\label{KW ZR apply onto AKLT}
\begin{aligned}
    (U_{\mathrm{KW}})^{\dagger} \mathcal{Z}_{\mathrm{R}} U_{\mathrm{KW}}
    | \mathrm{AKLT}_{ \mathbf{L}, \mathbf{R} } \rangle
    & = \frac{2}{3} \sum_{ \{s_{j}\} }
    \mathbf{L}^{\dagger} \Omega_{\omega}
    A_{1}^{s_{1}} A_{2}^{s_{2}} \cdots A_{N}^{s_{N}}
    (\Omega_{\omega})^{\dagger}
    \mathbf{R}
    |s_{1} s_{2} \cdots s_{N} \rangle \\
    &\qquad + \frac{2}{3} \sum_{ \{s_{j}\} }
    \mathbf{L}^{\dagger} (\Omega_{\omega})^{\dagger}
    A_{1}^{s_{1}} A_{2}^{s_{2}} \cdots A_{N}^{s_{N}}
    \Omega_{\omega}
    \mathbf{R}
    |s_{1} s_{2} \cdots s_{N} \rangle \\
    &\qquad - \frac{1}{3} \sum_{ \{s_{j}\} }
    \mathbf{L}^{\dagger}
    A_{1}^{s_{1}} A_{2}^{s_{2}} \cdots A_{N}^{s_{N}}
    \mathbf{R}
    |s_{1} s_{2} \cdots s_{N} \rangle.
\end{aligned}
\end{equation}
At the left edge, for $\mathbf{L} = \mathbf{e}_{\uparrow}$, we have $\mathbf{e}_{\uparrow}^{\dagger} \Omega_{\omega} = \mathbf{e}_{\uparrow}^{\dagger} \Omega_{\omega}^{\dagger} = \mathbf{e}_{\uparrow}^{\dagger} = (\sigma^{z} \mathbf{e}_{\uparrow})^{\dagger}$.
Consequently, we find
\begin{equation}
\begin{aligned}
    (U_{\mathrm{KW}})^{\dagger} \mathcal{Z}_{\mathrm{R}} U_{\mathrm{KW}}
    | \mathrm{AKLT}_{ \mathbf{e}_{\uparrow}, \mathbf{R} } \rangle
    & = \sum_{ \{s_{j}\} }
    \mathbf{e}_{\uparrow}^{\dagger}
    A_{1}^{s_{1}} A_{2}^{s_{2}} \cdots A_{N}^{s_{N}}
    \times
    \frac{1}{3}(2 \Omega_{\omega}^{\dagger} + 2 \Omega_{\omega} - I)
    \times 
    \mathbf{R}
    |s_{1} s_{2} \cdots s_{N} \rangle \\
    & = \sum_{ \{s_{j}\} }
    (\sigma^{z} \mathbf{e}_{\uparrow})^{\dagger}
    A_{1}^{s_{1}} A_{2}^{s_{2}} \cdots A_{N}^{s_{N}}
    (\sigma^{z} \mathbf{R})
    |s_{1} s_{2} \cdots s_{N} \rangle,
\end{aligned}
\end{equation}
where we use the identity $\frac{1}{3}(2 \Omega_{\omega}^{\dagger} + 2 \Omega_{\omega} - I) = \sigma^{z}$.
For the case where $\mathbf{L} = \mathbf{e}_{\downarrow}$, we use the fact that $\mathbf{e}_{\downarrow}^{\dagger} \Omega_{\omega} = \omega \mathbf{e}_{\downarrow}^{\dagger}$, and $\mathbf{e}_{\downarrow}^{\dagger} \Omega_{\omega}^{\dagger} = \omega^{2} \mathbf{e}_{\downarrow}^{\dagger}$.
Hence, we obtain
\begin{equation}
\begin{aligned}
    (U_{\mathrm{KW}})^{\dagger} \mathcal{Z}_{\mathrm{R}} U_{\mathrm{KW}}
    | \mathrm{AKLT}_{ \mathbf{e}_{\downarrow}, \mathbf{R} } \rangle
    & = \sum_{ \{s_{j}\} }
    \mathbf{e}_{\downarrow}^{\dagger}
    A_{1}^{s_{1}} A_{2}^{s_{2}} \cdots A_{N}^{s_{N}}
    \times
    \frac{1}{3}(2 \omega \Omega_{\omega}^{\dagger} + 2 \omega^{2} \Omega_{\omega} - I)
    \times 
    \mathbf{R}
    |s_{1} s_{2} \cdots s_{N} \rangle \\
    & = \sum_{ \{s_{j}\} }
    (\sigma^{z} \mathbf{e}_{\downarrow})^{\dagger}    
    A_{1}^{s_{1}} A_{2}^{s_{2}} \cdots A_{N}^{s_{N}}
    (\sigma^{z} \mathbf{R})
    |s_{1} s_{2} \cdots s_{N} \rangle,
\end{aligned}
\end{equation}
Therefore, we confirm the second line of Eq.~\eqref{AKLT X Z edge transformation}.

\subsection{Comparison with Kennedy-Tasaki transformation}

Here, we compare the KW transformation with the Kennedy-Tasaki (KT) transformation~\cite{Kennedy1992CMP, Kennedy1992PRB}.
The KT transformation unitary operator $U_{\mathrm{KT}}$ is given by
\begin{equation}
    U_{\mathrm{KT}}
    = \prod_{1 \leq u < v \leq N} e^{ i \pi S_{u}^{z} S_{v}^{x} }.
\end{equation}
First, we show that $U_{\mathrm{KT}}$ is not a Clifford unitary, while $U_{\mathrm{KW}}$ is.
For a two-qutrit system, a generalized Pauli string $Z_{2}$ is transformed as
\begin{equation}
    U_{\mathrm{KT}} Z_{2} (U_{\mathrm{KT}})^{\dagger}
    = Z_{2} + i \frac{\sqrt{3}}{9} \omega
    ( 2I - \omega^{2} Z_{1} - \omega Z_{1}^{\dagger} )
    ( (1-\omega) Z_{2} + (1-\omega^{2}) Z_{2}^{\dagger} )
\end{equation}
This resulting operator is not a Pauli string.

After the KT transformation, the Heisenberg and BLBQ models exhibit a $\mathbb{Z}_{2} \times \mathbb{Z}_{2}$ symmetry generated by $\{ \mathcal{X}, \mathcal{Z} \}$, where $\mathcal{Z} = \prod_{j=1}^{N} e^{i \pi S_{j}^{z}}$.
This follows from the fact that both $\mathcal{X}$ and $\mathcal{Z}$ commute with $U_{\mathrm{KT}}$.
In contrast to the KW-transformed Hamiltonians, which possess a $\mathcal{Z}_{\mathrm{R}}$, the KT-transformed Hamiltonians have a ``global'' symmetry operator $\mathcal{Z}$.
Moreover, the actions of each symmetry operator are computed as
\begin{equation}
\label{KT X Z edge transformation}
\begin{gathered}
    \mathcal{X} U_{\mathrm{KT}}
    | \mathrm{AKLT}_{ \mathbf{L}, \mathbf{R} } \rangle
    = U_{\mathrm{KT}}
    | \mathrm{AKLT}_{ \sigma^{x} \mathbf{L}, \sigma^{x} \mathbf{R} } \rangle, \\
    \mathcal{Z} U_{\mathrm{KT}}
    | \mathrm{AKLT}_{ \mathbf{L}, \mathbf{R} } \rangle
    = U_{\mathrm{KT}}
    | \mathrm{AKLT}_{ \sigma^{z} \mathbf{L}, \sigma^{z} \mathbf{R} } \rangle,
\end{gathered}
\end{equation}
for $\mathbf{L} \in \{ \mathbf{e}_{\uparrow}, \mathbf{e}_{\downarrow} \} $ and $\mathbf{R} \in \{ \mathbf{e}_{+}, \mathbf{e}_{-} \}$.
The first line follows from the commutation of $\mathcal{X}$ with $U_{\mathrm{KT}}$ and Eq.~\eqref{X action on A}.
The second line is obtained using the commutation of $\mathcal{Z}$ with $U_{\mathrm{KT}}$, together with the following diagrammatic identity analogous to Eq.~\eqref{ZR diagram}~\cite{Tasaki2020, Perez2008, Pollmann2010, Pollmann2012, Pollmann2017symmetry}:
\begin{equation}
\label{Z action on A}
    \begin{tikzpicture}[anchor=center, baseline={(0,-0.1)}]
        \def\a{0.35};

        \roundedsquare{SkyBlue!40}{\a}{\a}
        \node at (0, 0) {$A_{j}$};    
        
        \draw (1.0*\a, 0) -- (2.0*\a, 0);
        \draw (-1.0*\a, 0) -- (-2.0*\a, 0);
        \draw (0, -1.0*\a) -- (0, -2.0*\a);

        \draw[fill = orange!30] (0, -3.0*\a) circle (1.15*\a);
        \node at (0, -3.0*\a) {$\mathcal{Z}_{j}$};
        \draw (0, -4.15*\a) -- (0, -5.0*\a);
    \end{tikzpicture}
    =
    \begin{tikzpicture}[anchor=center, baseline={(0,-0.1)}]*
        \def\a{0.35};
    
        \roundedsquare{SkyBlue!40}{\a}{\a}
        \node at (0, 0) {$A_{j}$};    
    
        \draw (1.0*\a, 0) -- (2.0*\a, 0);
        \draw (-1.0*\a, 0) -- (-2.0*\a, 0);
        \draw (0, -1.0*\a) -- (0, -2.0*\a);
    
        \begin{scope}[shift = {(2.8*\a, 0)}]
            \diatensor{green!30}{1.3*\a}
            \node at (0, 0) {$\sigma^{z}$};
            \draw (1.3*\a, 0) -- (1.8*\a, 0);
        \end{scope}    
        \begin{scope}[xscale = -1, shift = {(2.8*\a, 0)}]
            \diatensor{green!30}{1.3*\a}
            \node at (0, 0) {$\sigma^{z}$};
            \draw (1.3*\a, 0) -- (1.8*\a, 0);
        \end{scope}
    \end{tikzpicture}, ~~
    \text{where } \mathcal{Z}_{j} = e^{i \pi S_{j}^{z}}.
\end{equation}
As compared with Eq.~\eqref{AKLT X Z edge transformation}, the global operator $\mathcal{Z}$ plays the same role as the local operator $\mathcal{Z}_{\mathrm{R}}$ in the KW transformation.
Therefore, the two states $U_{\mathrm{KT}} | \mathrm{AKLT}_{ \mathbf{L}, \mathbf{e}_{+} } \rangle$ and $U_{\mathrm{KT}} | \mathrm{AKLT}_{ \mathbf{L}, \mathbf{e}_{-} } \rangle$ are related by the \emph{global} unitary $\mathcal{Z}$ and belong to distinct broken-symmetry sectors characterized by the order parameter $\langle S_{j}^{x} \rangle$.
Consequently, the KT transformation reveals the hidden $\mathbb{Z}_{2} \times \mathbb{Z}_{2}$ SSB.

\bibliography{reference_supp}